\documentclass[reprint,superscriptaddress,showpacs]{revtex4-2}
\usepackage{amsmath}
\usepackage{amssymb}
\usepackage{graphicx}

\begin{document}

\title{Skew-Gaussian model of small-photon-number coherent Ising machines}

\author{Yoshitaka Inui}
\email[]{yoshitaka.inui@ntt-research.com}

\affiliation{Physics \& Informatics Laboratories, NTT Research, Inc., Sunnyvale, California 94085, USA}

\author{Edwin Ng}

\affiliation{Physics \& Informatics Laboratories, NTT Research, Inc., Sunnyvale, California 94085, USA}
\affiliation{E. L. Ginzton Laboratory, Stanford University, Stanford, California 94305, USA}

\author{Yoshihisa Yamamoto}

\affiliation{Physics \& Informatics Laboratories, NTT Research, Inc., Sunnyvale, California 94085, USA}
\affiliation{E. L. Ginzton Laboratory, Stanford University, Stanford, California 94305, USA}

\date{\today}

\begin{abstract}

A Gaussian quantum theory of bosonic modes has been widely used 
to describe quantum optical systems, including coherent Ising machines (CIMs) 
that consist of $\chi^{(2)}$ degenerate optical parametric oscillators (DOPOs) as nonlinear elements. 
However, Gaussian models have been thought to be invalid 
in the extremely strong-gain-saturation limit. 
Here, we develop an extended Gaussian model including two third-order fluctuation products, 
$\langle \delta \hat{X}^3\rangle$ and $\langle \delta \hat{X}\delta \hat{P}^2\rangle$, 
which we call self-skewness and cross-skewness, respectively. 
This new model which we call skew-Gaussian model more precisely replicates 
the success probability predicted by the quantum master equation (QME), relative to Gaussian models. 
We also discuss the impact of skew variables on the performance of CIMs. 

\end{abstract}


\maketitle

\section{Introduction}

Optical neural network machines \cite{Caulfield10,Shen17,Lin18,Wetzstein20,Shastri21,Feldmann21,Xu21,Marandi14,McMahon16,Inagaki16,Hamerly19,Pierangeli19,Okawachi20,Honjo21,Inagaki21,Hamerly19x,Wang22,Kalinin23} 
have been expected to reduce the energy costs of solving hard computational problems 
involving matrix-vector-multiplication and combinatorial optimization, such as 
quadratic unconstrained binary optimization (QUBO), and Boolean satisfiability problem (SAT). 
Recently, several studies have addressed the question of whether such machines can work with small numbers of photons, 
for example a few photons, or even less than one photon per multiplication\cite{Hamerly19x,Wang22}. 
Several machines (e.g., coherent Ising machines (CIMs) \cite{Wang13,McMahon16,Inagaki16,Hamerly19,Kako20,Honjo21,Reifenstein21,Inagaki21,Ng22,Inui22}) 
use optical nonlinear interactions. 
The study of small-photon-number operations in such machines entails 
the design and fabrication of systems in which the nonlinear coefficient per photon is as large as the linear loss coefficient\cite{Lu20,Zhao22,Yanagimoto22}, 
and an evaluation involving a numerical simulation of its nonlinear quantum optical physics\cite{Kako20,Ng22,Inui22}, 
which is the main focus of this study. 

Gaussian models\cite{Corney03,Olivares12} have been used in the numerical simulation of CIMs\cite{Kako20,Ng22,Inui22}.  
However, strong nonlinear effects are expected to produce non-Gaussian states of light, 
where Gaussian models, $c$-number Heisenberg-Langevin equations\cite{Wang13,Lax66}, and truncated-Wigner $c$-number stochastic differential equations (CSDEs) \cite{Maruo16,Inui20} might be inaccurate\cite{Plimak01}, 
and simulations using the positive-$P$ CSDEs \cite{Drummond80,Takata15,Shoji17} would be unstable\cite{Gilchrist97}. 
We developed a non-Gaussian model of CIMs as an extended Gaussian model including the third-order fluctuation products, 
that is expected to be more accurate than Gaussian models\cite{Kako20,Ng22,Inui22}. 
By using it, we evaluate the performance of small-photon-number CIM, operated with less than three photons per pulse on average. 

A CIM in this paper is an optical and electronic hybrid system and 
uses degenerate optical parametric oscillators (DOPOs). 
It is designed to solve the Ising problem, 
which is to find the ground state of the Ising Hamiltonian, 
\begin{equation}
H=-\frac{1}{2}\sum_{r=1}^N\sum_{r'=1}^N \tilde{J}_{rr'}\sigma_r\sigma_{r'}. 
\end{equation}
Here, $\sigma_r=\{\pm 1\}$ and $N$ is the problem size (number of lattice sites). 
We denote the $r$-th optical pulse generated by the DOPO as $\hat{a}_r (r=1,\cdots,N)$. 
Full optoelectronic connection of up to $10^5$ optical pulses\cite{Honjo21} has been demonstrated 
through homodyne measurement of partly outcoupled pulses (indirect measurement\cite{Braginsky95}) and coherent injection feedback \cite{McMahon16,Inagaki16}. 
The Gaussian model of CIM uses three variables to represent each optical pulse\cite{Kako20,Inui22}, i.e., 
the mean amplitude $\langle \hat{X}_r\rangle$, the variance of the canonical coordinate $\langle \delta \hat{X}_r^2\rangle$, 
and the variance of the canonical momentum $\langle \delta \hat{P}_r^2\rangle$, 
to approximate the full quantum description which is called the quantum master equation (QME). 
Here, $\hat{X}_r=\frac{\hat{a}_r+\hat{a}_r^{\dagger}}{\sqrt{2}}$, $\hat{P}_r=\frac{\hat{a}_r-\hat{a}_r^{\dagger}}{\sqrt{2} i}$, 
$\delta \hat{X}_r=\hat{X}_r-\langle \hat{X}_r\rangle$, and $\delta \hat{P}_r=\hat{P}_r-\langle \hat{P}_r\rangle$, 
where the mean amplitudes of the canonical momenta $\langle \hat{P}_r\rangle$ are assumed to be always zero. 
As a lowest-order correction to the Gaussian model, we can use the third-order fluctuation products for each pulse. 
For CIMs, the self-skewness $\langle \delta \hat{X}_r^3\rangle$ and 
cross-skewness $\langle \delta \hat{X}_r\delta \hat{P}_r^2\rangle$ can have non-zero values. 
We call this model, using five variables to represent each pulse, a skew-Gaussian model of CIM. 
Here, the other third-order fluctuation products, $\langle \delta \hat{X}_r^2\delta \hat{P}_r\rangle$ and $\langle \delta \hat{P}_r^3\rangle$, 
are zero due to the assumption of $\langle \hat{P}_r\rangle=0$. 

The previous studies on the CIM in Refs.\cite{Kako20,Inui22} used the density-operator master equation under the continuous-time evolution model, 
in which the changes in the field variables in a single round trip 
that are caused by the DOPO, extraction beam splitter, indirect homodyne measurement\cite{Wiseman93}, 
and injection beam splitter are all sufficiently small 
so that all the effects of these optical components in a single round trip 
can be packed into one differential equation of the density-matrix\cite{Kako20,Inui22,Takata15,Shoji17}, 
instead of considering them separately in order \cite{Clements17,Yamamura17,Ng22} (which we call the discrete-component model). 
In the first part of this paper, we derive the skew-Gaussian model as a continuous-time model. 
However, when there are only a few photons inside the cavity, 
the information extracted by indirect homodyne measurement signals can be easily buried 
in the large amount of quantum noise 
particularly when the reflection coefficient of the extraction beam splitter is small. 
In the latter part of the paper, we will use the discrete-component model, 
to evaluate the performance when the beam splitters have a large reflection rate. 
Throughout the paper, we squeeze the vacuum state input of the extraction beam splitter \cite{Maruo16} 
to improve the signal-to-noise ratio in the homodyne measurement signals, 
and use chaotic amplitude control (CAC) \cite{Leleu19} to solve the problem of amplitude inhomogeneity\cite{Leleu17}. 
The CAC introduces an auxiliary variable $e_r$ to stabilize the squared measured amplitudes to the target value. 
Using the skew-Gaussian model, we evaluated the success probability ($P_s$) and 
the minimum number of matrix-vector-multiplications needed to reach a solution ($\min {\rm MVMTS}$) 
for small-photon-number, strongly nonlinear CIMs up to $N=200$. 
The $\min {\rm MVMTS}$ was at most only $33\%$ larger than that for a large-photon-number CIM when solving $\alpha_{WP}=0.8$ Wishart planted instances \cite{Hamze20}. 
Here, $\alpha_{WP}N^2$ is the number of independent Gaussian random numbers used to generate an instance. 

This paper is structured as follows. 
Section II describes the system and introduces two continuous-time models of the CIM. 
In particular, we introduce the quantum master equation (QME) for the density-matrix 
using Wiseman-Milburn's description\cite{Wiseman93,Wiseman93a,Wiseman09} of the homodyne measurement. 
Then, we introduce the skew-Gaussian model to approximate it. 
Appendix A provides information about the continuous-time QME. 
Appendix B derives the equations of the continuous-time skew-Gaussian model. 
Section III presents the non-Gaussian Wigner functions with negative and positive sign-adjusted self-skewness 
${\rm sgn}(\langle \hat{X}_r\rangle)\langle \delta \hat{X}_r^3\rangle$, obtained by solving the mean-field coupled QME. 
Appendix C provides analytical details on the skew variables in the mean-field coupled model. 
Section IV compares the success probability ($P_s$) and mean photon number 
of continuous-time models when the saturation coefficient $g^2$ is increased. 
It is found that the skew-Gaussian model provides closer results to the QME than the Gaussian models do. 
Section V introduces the discrete-component models (detailed in Appendix D) 
and compares the success probability $P_s^{(s)}$ obtained by the skew-Gaussian model 
with the success probability $P_s^{(g)}$ obtained by the Gaussian model. 
It is found that, when the spin state $\sigma_r$ is 
determined by individual indirect homodyne measurement results\cite{Braginsky95}, 
the ratio of success probabilities $\frac{P_s^{(s)}}{P_s^{(g)}}-1$ 
is governed by the sign-adjusted weighted sum of skew variables, ${\rm sgn}(\langle \hat{X}_r\rangle)(3\langle \delta \hat{X}_r^3\rangle+\langle \delta \hat{X}_r\delta \hat{P}_r^2\rangle)$. 
Moreover, when the spin state $\sigma_r$ is determined by homodyne measurement results of 
totally outcoupled pulses (direct measurement), 
the sign-adjusted self-skewness ${\rm sgn}(\langle \hat{X}_r\rangle)\langle \delta \hat{X}_r^3\rangle$ also contributes to the success probability. 
Section VI provides the site-number-dependent success probability and 
$\min {\rm MVMTS}$ under the optimized parameters. 
The parameter optimization for large success probability and smaller $\min {\rm MVMTS}$ leads to a large beam splitter loss, 
under which the difference between the skew-Gaussian model and Gaussian model becomes almost negligible. 
The instance-averaged success probability and $\min {\rm MVMTS}$ of the small-photon-number CIM turns out to 
be comparable to that of the large-photon-number CIM. 
Appendix E provides information about the parameter optimization and site-number dependence of small-photon-number CIMs. 
Appendix F provides information about large-photon-number CIMs. 
Section VII summarizes the paper. 

\section{Model}

\subsection{Hybrid optical-digital system}

Figure 1(a) shows the system architecture of the hybrid optical-digital CIM. 
It consists of $\chi^{(2)}$ degenerate optical parametric oscillators (DOPOs), 
a probe field in a squeezed vacuum state at the extraction beam splitter (BS1), 
homodyne measurements of the probe field, 
computation of the feedback signal amplitude by using a digital processing unit 
consisting of an analog-to-digital converter (ADC), field programmable gate array (FPGA), 
and digital-to-analog converter (DAC), 
and coherent signal injection at the second beam splitter (BS2) \cite{Inagaki16,McMahon16}. 
The $\chi^{(2)}$ DOPOs are coherently and synchronously excited by a sequence of pump pulses, 
which generates $N$ signal pulses that work as analog (soft) spins. 
These signal pulses are denoted as $\hat{a}_r (r=1,\cdots,N)$. 
The blue letters (b-g) in Figure 1(a) show the points where Wigner functions in Figures 1(b-g) are obtained. 
In the first round trip, the pulses are prepared in a vacuum state $\hat{\rho}^{(r)}=|0\rangle \langle 0|$ (Figure 1(b)), 
and when the normalized parametric gain $p$ is positive, 
$\hat{X}$-anti-squeezed ($\hat{P}$-squeezed) states are obtained after the DOPO (Figure 1(c)). 
We assume there are two linear loss coefficients for a signal pulse, $\hat{a}_r\propto e^{-(\gamma_s+J)t}$, 
where $J$ describes the linear loss at the two beam splitters\cite{Takata15,Maruo16} and $\gamma_s$ describes the background linear loss. 
We will suppose that the lifetime $1/\gamma_s$ due to the background loss is normalized to be $1$, 
and that it is sufficiently longer than the round trip time $\Delta t$. 
The normalized coupling loss $j=J/\gamma_s$ is decomposed as $j=\frac{j_1}{2}+\frac{j_2}{2}$. 
Here, $R_1=j_1\Delta t$, is the intensity reflection rate of BS1, 
with which the internal state is indirectly measured \cite{Braginsky95}. 
The squeezed probe states $\hat{b}_r$ at BS1 are generated by 
phase-sensitive deamplification of the vacuum state\cite{Caves82,Maruo16}, 
i.e., $|\psi_{B,r}\rangle=e^{-\frac{\ln\sqrt{G_j}}{2}(\hat{b}_r^{\dagger 2}-\hat{b}_r^2)}|0\rangle$. 
The $\hat{X}$ and $\hat{P}$ variances of the state $|\psi_{B,r}\rangle$ 
are prepared to be $\frac{1}{2G_j}$ and $\frac{G_j}{2}$, respectively. 
The Wigner function of mode $\hat{b}_r$ when $G_j>1$ is shown in Figure 1(d). 
The two-mode mixing by BS1 can be described in the Heisenberg picture as, 
$\begin{bmatrix} \hat{a}_{r}\\ \hat{b}_{r}\end{bmatrix}\rightarrow \begin{bmatrix} \sqrt{1-R_1} & -\sqrt{R_1} \\ \sqrt{R_1} & \sqrt{1-R_1} \end{bmatrix} \begin{bmatrix} \hat{a}_{r}\\ \hat{b}_{r}\end{bmatrix}$. 
Here, $\hat{a}_r$ and $\hat{b}_r$ describe the $r$-th internal pulse and 
the probe field which interacts with it, respectively. 
In the Schr{\"o}dinger picture, this effect of mode mixing is calculated as, 
$\hat{\rho}_{AB}^{(r)}\rightarrow e^{\theta_1(\hat{a}_{r}^{\dagger}\hat{b}_{r}-\hat{b}_{r}^{\dagger}\hat{a}_{r})}\hat{\rho}_{AB}^{(r)}e^{\theta_1(\hat{b}_{r}^{\dagger}\hat{a}_{r}-\hat{a}_{r}^{\dagger}\hat{b}_{r})}$\cite{Olivares12}, 
where $\hat{\rho}_{AB}^{(r)}$ is the two-mode density-matrix of $\hat{a}_r$ and $\hat{b}_r$, and $\theta_1=\sin^{-1} (-\sqrt{R_1})$. 
The $\hat{X}$ component of BS1's output is measured to be $X_{B,r}$ by the homodyne measurement, 
which affects the transmitted internal state due to the quantum correlation between the signal and probe fields after BS1. 
Figure 1(e) shows the state after the homodyne measurement. 
It is an $\hat{X}$-squeezed state because of the mixing with the $G_j>1$ input state, 
and the mean amplitude is shifted to a negative value 
due to the negative value of a homodyne measurement in this case. 
By inferring the internal amplitudes $\tilde{\mu}_r=X_{B,r}/\sqrt{2R_1}$ from the measured values $X_{B,r}$, 
the FPGA calculates the feedback amplitudes $\zeta_r=j\Delta t\mu_{F,r}/\sqrt{R_2}$, 
where $\mu_{F,r}=e_r \sum_{r'} \tilde{J}_{rr'}\tilde{\mu}_{r'}$ 
and $R_2=j_2\Delta t$ is the intensity reflection of BS2. 
The input mode $\hat{z}_r$ of the injection beam splitter (BS2) is set to 
a coherent field $|\zeta_r\rangle$ by using the intensity modulator and phase modulator (IM/PM) (Figure 1(f)). 
The two-mode mixing at BS2 is described in the Heisenberg picture as 
$\begin{bmatrix} \hat{a}_{r}\\ \hat{z}_{r}\end{bmatrix}\rightarrow \begin{bmatrix} \sqrt{1-R_2} & \sqrt{R_2} \\ -\sqrt{R_2} & \sqrt{1-R_2} \end{bmatrix} \begin{bmatrix} \hat{a}_{r}\\ \hat{z}_{r}\end{bmatrix}$, 
which is equivalent to 
\begin{equation}
\label{rhotf}
\hat{\rho}_{AZ}^{(r)}\rightarrow e^{\theta_2(\hat{a}_{r}^{\dagger}\hat{z}_{r}-\hat{z}_{r}^{\dagger}\hat{a}_{r})}\hat{\rho}_{AZ}^{(r)}e^{\theta_2(\hat{z}_{r}^{\dagger}\hat{a}_{r}-\hat{a}_{r}^{\dagger}\hat{z}_{r})}, 
\end{equation}
where $\theta_2=\sin^{-1} (\sqrt{R_2})$. 
Here, $\hat{\rho}_{AZ}$ is the two-mode density-matrix of $\hat{a}_r$ and $\hat{z}_r$. 
The state after BS2 (Figure 1(g)) obtained by tracing out the feedback input mode $\hat{z}_r$ 
is the signal state input to the DOPO in the next round trip. 
The auxiliary variable $e_r$ for the chaotic amplitude control\cite{Leleu19} develops as 
\begin{equation}
\label{tdeve}
\frac{de_r}{dt}=-\beta(\tilde{\mu}_r^2-\tau)e_r, 
\end{equation}
where $\beta$ is the strength of the amplitude control and $\tau$ is the target value. 
We define the effective computation time $t$ as the number of round trips multiplied by $\Delta t$. 
We introduce the ratio of reflection coefficients $R:=R_1/R_2$, 
with which $j_1$ and $j_2$ are written as $j_1=\frac{2R}{1+R}j$, and $j_2=\frac{2}{1+R}j$. 

\begin{figure*}
\begin{center}
\includegraphics[width=15.0cm]{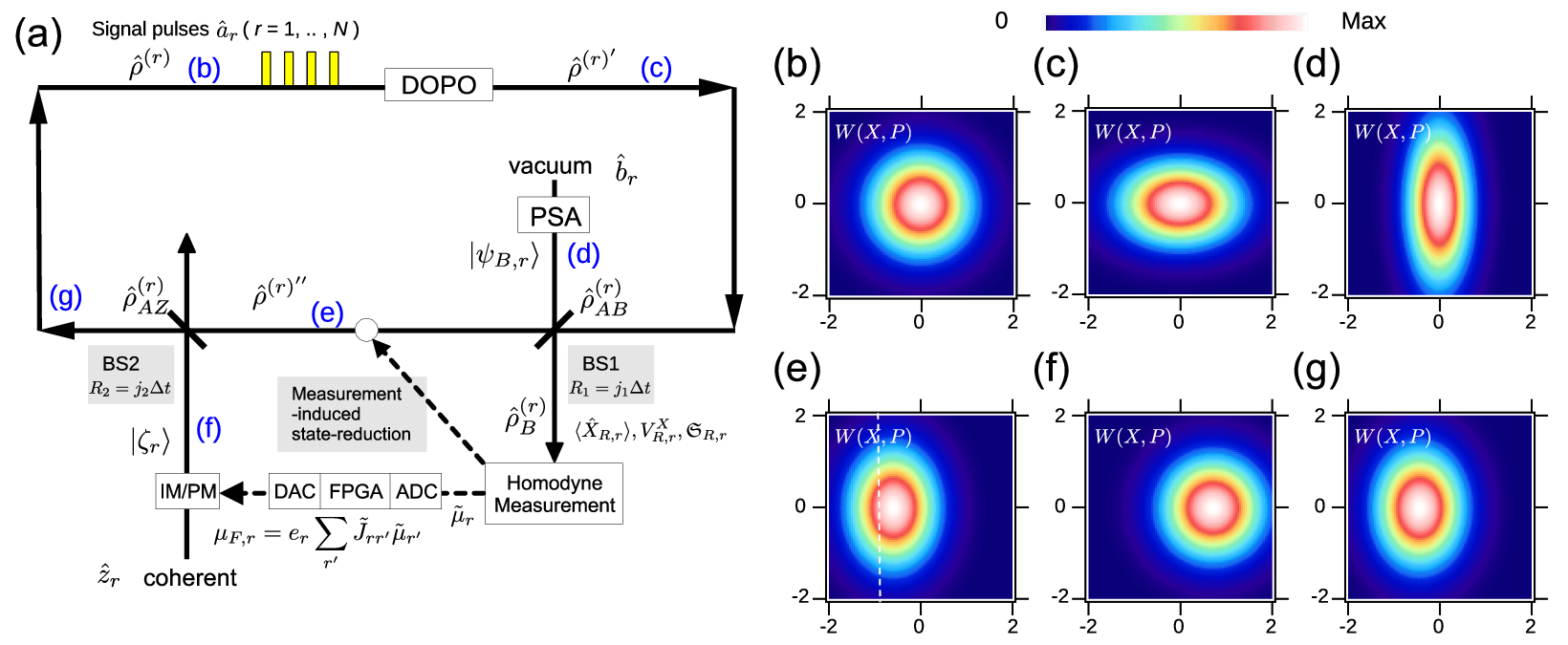}
\caption{Operation of a small-photon-number CIM. (a) System architecture: 
DOPO, PSA, ADC, FPGA, DAC, IM and PM mean degenerate optical parametric oscillator, phase-sensitive amplifier, 
analog-to-digital converter, field programmable gate array, digital-to-analog converter, intensity modulator and phase modulator, respectively. 
(b-g) Example of Wigner functions for the first round trip (for the same site index $r$). 
(b) Initial vacuum state $\hat{\rho}^{(r)}=|0\rangle \langle 0|$. 
(c) State $\hat{\rho}^{(r)'}$ after the DOPO ($p>0$). 
(d) Input state $|\psi_{B,r}\rangle \langle \psi_{B,r}|$ of BS1 ($G_j>1$). 
(e) Internal state $\hat{\rho}^{(r)''}$ after the homodyne measurement. 
The white dashed line shows the measured value $X_{B,r}$. 
(f) Input state $|\zeta_r\rangle \langle \zeta_r|$ of BS2 ($\zeta_r>0$). 
(g) Internal state $\hat{\rho}^{(r)}$ after BS2. 
}
\label{model}
\end{center}
\end{figure*}

\subsection{Continuous-time quantum master equation}

Here, we present the quantum master equation for the density-matrix in the continuous-time limit, where $\Delta t,R_1,R_2 \ll 1(=\gamma_s^{-1})$. 
We use the Wiseman-Milburn model to describe an optical homodyne measurement in such a high-$Q$ case \cite{Wiseman93,Wiseman93a,Wiseman09}. 
The master equation of a measurement-feedback CIM consists of three parts 
describing the DOPO, extraction beam splitter (BS1), and injection beam splitter (BS2), 
\begin{equation}
\label{qmect}
\frac{\partial \hat{\rho}}{\partial t}=\Bigl(\frac{\partial \hat{\rho}}{\partial t}\Bigr)_{DOPO}+\Bigl(\frac{\partial \hat{\rho}}{\partial t}\Bigr)_{BS1}+\Bigl(\frac{\partial \hat{\rho}}{\partial t}\Bigr)_{BS2}.
\end{equation} 
By assuming the linear decay rate of the pump mode $\gamma_p$ is much larger than $\gamma_s$, 
we can adiabatically eliminate the pump mode\cite{Kinsler91}. The effect of the DOPO is described as 
\begin{widetext}
\begin{equation}
\label{qmedopo}
\Bigl(\frac{\partial \hat{\rho}}{\partial t}\Bigr)_{DOPO}=\sum_r([\hat{a}_r,\hat{\rho}\hat{a}_r^{\dagger}]+{\rm H.c.})+\frac{g^2}{2}\sum_r([\hat{a}_r^2,\hat{\rho}\hat{a}_r^{\dagger 2}]+{\rm H.c.})+\frac{p}{2}\sum_r[\hat{a}_r^{\dagger 2}-\hat{a}_r^2,\hat{\rho}]. 
\end{equation}
\end{widetext}
Here, $\hat{a}_r$ is the annihilation operator of the $r$-th signal pulse, $[\hat{A},\hat{B}]:=\hat{A}\hat{B}-\hat{B}\hat{A}$, 
and ${\rm H.c.}$ indicates the Hermitian conjugate. 
The first term on the right hand side (R.H.S.) represents the linear dissipation of the signal mode with the normalized amplitude decay rate of $\gamma_s=1$. 
The second and third terms are related to the nonlinear $\chi^{(2)}$-interaction Hamiltonian $\hat{H}=i\hbar\frac{\chi}{2}(\hat{a}_s^{\dagger 2}\hat{a}_p-{\rm H.c.})$. 
The second term on the R.H.S. represents the gain saturation, which is equivalent to the two-photon absorption loss of the signal mode, 
due to the transition to the pump mode. 
That rate follows the electromagnetic perturbation theory \cite{Winn99} 
where the final state of the transition is the pump cavity mode whose density of states is $\propto \frac{1}{\gamma_p}$\cite{Purcell46}. 
The saturation coefficient $g^2$ is represented as $g^2=\frac{\chi^2}{2\gamma_p \gamma_s}$. 
The third term on the R.H.S. is the parametric gain 
depending on the normalized pump rate $p=\frac{\chi \varepsilon_p}{\gamma_p \gamma_s}$, 
where $\varepsilon_p$ is the strength of coherent excitation of the pump mode\cite{Takata15}. 

The extraction beam splitter (BS1) affects the system 
via the linear loss, injection of the squeezed probe light \cite{Gardiner85,Milburn87}, 
and the measurement-induced state-reduction \cite{Wiseman09} (Appendix A.1): 
\begin{widetext}
\begin{eqnarray}
\label{qmesr}
\Bigl(\frac{\partial \hat{\rho}}{\partial t}\Bigr)_{BS1}&=&\frac{j_1}{2} \sum_{r}([\hat{a}_r,\hat{\rho}\hat{a}_r^{\dagger}]+{\rm H.c.})+j_1n_j\sum_r[\hat{a}_r,[\hat{\rho},\hat{a}_r^{\dagger}]]-\frac{j_1}{2}m_j\sum_r([\hat{a}_r,[\hat{a}_r,\hat{\rho}]]+{\rm H.c.}) \\
&+& \sqrt{j_1 G_j}\frac{1+G_j^{-1}}{2}\sum_{r}(\hat{a}_{r}\hat{\rho}+\hat{\rho}\hat{a}^{\dagger}_{r}-\langle \hat{a}_r+\hat{a}^{\dagger}_r\rangle\hat{\rho})\dot{W}_r+ \sqrt{j_1 G_j}\frac{1-G_j^{-1}}{2}\sum_{r}(\hat{a}_{r}^{\dagger}\hat{\rho}+\hat{\rho}\hat{a}_{r}-\langle \hat{a}_r+\hat{a}^{\dagger}_r\rangle\hat{\rho})\dot{W}_r. \nonumber 
\end{eqnarray}
\end{widetext}
$\dot{W}_r$ are real random numbers obeying a zero-mean normal distribution, 
which originate from the quantum noise incident on the open port of BS1, and they satisfy 
$\overline{\dot{W}_r(t)\dot{W}_{r'}(t')}=\delta_{rr'}\delta(t-t')$. 
Here, $n_j=\frac{1}{4}(G_j+G_j^{-1})-\frac{1}{2}$ and $m_j=\frac{1}{4}(G_j-G_j^{-1})$ satisfy $n_j(1+n_j)=m_j^2$\cite{Gardiner86}. 
The injection beam splitter (BS2) affects the system via the following terms, representing the linear loss and the coherent injection \cite{Wiseman93,Shoji17,Leleu19}: 
\begin{equation}
\label{qmefb}
\Bigl(\frac{\partial \hat{\rho}}{\partial t}\Bigr)_{BS2}=\frac{j_2}{2} \sum_{r}([\hat{a}_r,\hat{\rho}\hat{a}_r^{\dagger}]+{\rm H.c.}) + \sum_r \varepsilon_r [\hat{a}_r^{\dagger}-\hat{a}_r,\hat{\rho}]. 
\end{equation}
Here, 
\begin{equation}
\label{eps_r}
\varepsilon_r= j \mu_{F,r}=je_r\sum_{r'}\tilde{J}_{r,r'}\tilde{\mu}_{r'}, 
\end{equation}
and $\tilde{\mu}_r=\frac{\langle \hat{a}_{r}+\hat{a}^{\dagger}_{r}\rangle}{2}+\frac{\dot{W}_r}{\sqrt{4j_1 G_j}}$\cite{Wiseman93}. 
The auxiliary variable $e_r$ for amplitude control follows Eq.(\ref{tdeve}).

\subsection{Continuous-time skew-Gaussian model}

The full numerical simulation of the quantum master equation for the density-matrix requires many equations to describe even a single DOPO. 
When we set the maximum photon number to $N_M$ and neglect photon-number states higher than $N_M$, 
$(N_M+1)^2$ equations are required for each DOPO (Appendix A.2). 
Instead, the Gaussian models\cite{Kako20,Inui22,Ng22} require only three variables, $\mu_r=\langle \hat{a}_r\rangle$, $m_r=\langle \delta \hat{a}_r^2\rangle$, 
and $n_r=\langle \delta \hat{a}_r^{\dagger}\delta \hat{a}_r\rangle$, to describe each DOPO. 
Here, we can introduce two additional variables, 
$\gamma_r=\langle \delta \hat{a}_r^3\rangle$ and $\kappa_r=\langle \delta \hat{a}_r^{\dagger}\delta \hat{a}_r^2\rangle$ 
to describe the lowest-order non-Gaussian correction. 
Following the derivations shown in Appendix B, 
we obtain a continuous-time skew-Gaussian model using five variables $\mu_r$, $m_r$, $n_r$, $\gamma_r$, and $\kappa_r$ to describe each pulse, 
which approximates the continuous-time quantum master equation (Eq.(\ref{qmect})). 
\begin{widetext}
\begin{equation}
\label{ctsk1}
\frac{d\mu_r}{dt}=-(1-p+j)\mu_r-g^2(\mu_r^2+2n_r+m_r)\mu_r-g^2\kappa_r+\sqrt{j_1G_j}V_r' \dot{W}_r+\varepsilon_r, 
\end{equation}
\begin{eqnarray}
\label{ctsk2}
\frac{d m_r}{dt}&=&-2(1+j)m_r+2p n_r-2g^2 \mu_r^2 (2m_r+n_r)+p-g^2(\mu_r^2+m_r)-j_1G_jV_r'^2+\frac{j_1}{4}\Bigl(\frac{1}{G_j}-G_j\Bigr) \nonumber \\ 
&-& 6g^2n_rm_r-2g^2\mu_r(\gamma_r+2\kappa_r) +\sqrt{j_1 G_j} \dot{W}_r (\gamma_r+\kappa_r),
\end{eqnarray}
\begin{eqnarray}
\label{ctsk3}
\frac{d n_r}{dt}&=&-2(1+j)n_r+2 pm_r-2g^2\mu_r^2(2n_r+m_r)- j_1G_j V_r'^2+\frac{j_1}{4}\Bigl(G_j+\frac{1}{G_j}-2\Bigr) \nonumber \\
&-& 2g^2(m_r^2+2n_r^2)-6g^2\mu_r \kappa_r +2 \sqrt{j_1 G_j} \dot{W}_r \kappa_r,
\end{eqnarray}
\begin{eqnarray}
\label{ctsk4}
\frac{d\gamma_r}{dt}&=&-3(1+j+g^2)\gamma_r+3p \kappa_r-3g^2\mu_r^2 (2\gamma_r+\kappa_r) \nonumber \\
&-&6g^2\mu_r ( m_r+ m_r^2+2 m_rn_r )-3g^2(4n_r\gamma_r+5m_r\kappa_r)-3j_1 G_j V_r'(\kappa_r+\gamma_r), 
\end{eqnarray}
\begin{eqnarray}
\label{ctsk5}
\frac{d\kappa_r}{dt}&=&-(3+3j-2p+g^2)\kappa_r+p\gamma_r -g^2\mu_r^2 (\gamma_r +8\kappa_r) \nonumber \\
&-& 2g^2\mu_r (n_r +2m_r^2+4n_rm_r+3n_r^2)-g^2(3m_r\gamma_r+8m_r\kappa_r+16n_r\kappa_r)-j_1 G_j V_r'(5\kappa_r+\gamma_r), 
\end{eqnarray}
\end{widetext}
where $V_r'=n_r+m_r+\frac{1}{2}-\frac{1}{2G_j}$ represents the strength of the mean amplitude shift in Eq.(\ref{ctsk1}) 
and the $-j_1 G_j V_r^{'2}$ term in Eqs.(\ref{ctsk2}) and (\ref{ctsk3}) represents the fluctuation reduction. 
The strength of the coherent excitation $\varepsilon_r$ is given by Eq.(\ref{eps_r}), 
where $\tilde{\mu}_r=\mu_r+\frac{\dot{W}_r}{\sqrt{4j_1G_j}}$. 
The characteristics of the canonical coordinates and momenta are obtained from these five variables as 
$\langle \hat{X}_r\rangle=\sqrt{2}\mu_r$, $\langle \delta \hat{X}_r^2\rangle=n_r+m_r+\frac{1}{2}$, 
$\langle \delta \hat{P}_r^2\rangle=n_r-m_r+\frac{1}{2}$, 
$\langle \delta \hat{X}_r^3\rangle=\frac{\gamma_r+3\kappa_r}{\sqrt{2}}$, 
and $\langle \delta \hat{X}_r\delta \hat{P}_r^2\rangle=\frac{-\gamma_r+\kappa_r}{\sqrt{2}}$. 
If we assume that $\gamma_r$ and $\kappa_r$ in Eqs.(\ref{ctsk1})(\ref{ctsk2}) and (\ref{ctsk3}) are each zero, 
these equations become identical to those of the Gaussian model in Ref.\cite{Ng22}. 
We call that model 'Gaussian model' in this paper. 
Another model using Gaussian variables in Ref.\cite{Inui22} was obtained 
by the linearization of positive-$P$ CSDEs. 
That model which we call the Gaussian-approximated positive-$P$ (GAPP) model\cite{Inui22}, 
does not have terms containing the products of Gaussian variables, i.e., 
the $-6g^2n_rm_r$ term in Eq.(\ref{ctsk2}), and the $-2g^2(m_r^2+2n_r^2)$ term in Eq.(\ref{ctsk3}). 

\section{Wigner function for negative and positive self-skewness. }

In this section, we discuss two types of Wigner function, 
where the signs of self-skewness $\langle \delta\hat{X}_r^3\rangle$ are different from and the same as 
those of the mean amplitude $\langle \hat{X}_r\rangle$, respectively. 
In other words, these two have the negative and positive 
sign-adjusted self-skewness ${\rm sgn}(\langle \hat{X}_r\rangle) \langle \delta \hat{X}_r^3\rangle$, respectively. 
Appendix C discusses how the sign of ${\rm sgn}(\langle \hat{X}_r\rangle) \langle \delta \hat{X}_r^3\rangle$ 
depends on the normally ordered variance $\langle :\delta \hat{X}_r^2:\rangle$. 
To simplify the discussion, we will use the mean field coupling model, 
which provides the averaged characteristics over measurement results. 
We simplify the coupling terms in Eq.(\ref{qmect}) 
to a sum of the coupling-related loss part and the coherent injection part, 
\begin{widetext}
\begin{equation}
\label{qmemfc}
\Bigl(\frac{\partial \hat{\rho}}{\partial t}\Bigr)_{BS1}+\Bigl(\frac{\partial \hat{\rho}}{\partial t}\Bigr)_{BS2} \sim j \sum_{r}([\hat{a}_r,\hat{\rho}\hat{a}_r^{\dagger}]+{\rm H.c.}) + \sum_r \varepsilon_r [\hat{a}_r^{\dagger}-\hat{a}_r,\hat{\rho}], 
\end{equation}
\end{widetext}
where the coefficient of coherent injection depends on the mean amplitude 
\begin{equation}
\varepsilon_r=j G_F \mu_r (\mu_r=\langle \hat{a}_r\rangle). 
\end{equation}
Here, $G_F$ is the effective gain coefficient that 
represents the ratio of coherent injection $\varepsilon_r$ to the coupling loss $j\mu_r$. 
The equation of $\mu_r$ is $\frac{d\mu_r}{dt}=-(1-p+j-jG_F)\mu_r+O(g^2)$. 
From this equation, when $G_F=1$, the parametric gain $p$ at the oscillation threshold (denoted as $p_{thr}$) is $p_{thr}=1$, 
which is the threshold of an isolated DOPO. 
Generally, when $G_F\ne 1$, this threshold can be represented as, 
\begin{equation}
p_{thr}=1+j-jG_F. 
\end{equation}
It can be negative, which means that the effective linear amplification due to the coherent injection 
balances the linear deamplification only when there is a parametric deamplification, $p<0$. 

Here, we obtained the steady-state characteristics of an $N=1$ mean-field-coupled CIM by calculating the QME, 
developing from a vacuum state $\hat{\rho}=|0\rangle \langle 0|$ at $t=0$. 
We changed the oscillation threshold $p_{thr}$ (which is equivalent to changing $G_F=\frac{1-p_{thr}+j}{j}$) 
and set the parametric gain as $p=p_{thr}+10$. 
The initial coherent feedback at the first time step was $\varepsilon_r=jG_F$ and after that 
the coherent feedback depended on the mean field $\varepsilon_r=jG_F\langle \hat{a}_r\rangle=jG_F {\rm Tr}\hat{\rho}\hat{a}_r$. 
The time step $\Delta t$, nonlinear saturation coefficient, and coupling coefficient were $\Delta t=5\times 10^{-4}$, $g^2=5$, and $j=5$, respectively. 
The maximum photon number $N_M$ in the QME was 20. 
The steady-state characteristics determined from the QME at $t=10$ are shown as lines in Figure 2. 
The black line in Figure 2(a) represents the obtained mean amplitudes $\langle \hat{X}_r\rangle$ plotted versus $p_{thr}$. 
The values are close to the theoretically predicted one, $\langle \hat{X}_r\rangle\sim \sqrt{\frac{2(p-p_{thr})}{g^2}}\sim 2$. 
The steady-state mean field $\mu_r$ was also obtained by solving the self-consistent equation Eq.(\ref{scl}), starting from $\mu_r=1$. 
The results from the Eqs.(\ref{scl}) and (\ref{scl2}) in Appendix C are shown with circles in Figure 2. 
Here, the number of iterations of Eq.(\ref{scl}) was $2\times 10^3$. 
As can be seen, the circles follow the QME results. 
Figure 2(b) plots the $p_{thr}$-dependent fluctuation of the canonical coordinates $\langle \delta \hat{X}_r^2\rangle$. 
When $G_F$ is small ($p_{thr}\sim 1+j$), the CIM primarily operates as a system 
where the parametric gain $p$ and the two-photon absorption $g^2$ compete. 
In that case, the variance $\langle \delta \hat{X}_r^2\rangle$ has a larger value than that of the vacuum fluctuation (gray dashed line). 
In Figures 2(b)(c), when $p_{thr}$ has a positive value ($G_F<\frac{1+j}{j}$), the normally ordered variance $\langle :\delta \hat{X}_r^2:\rangle$ is positive, 
and the sign-adjusted self-skewness ${\rm sgn}(\langle \hat{X}_r\rangle) \langle \delta \hat{X}_r^3\rangle$ has a negative value\cite{Shoji17}. 
On the other hand, when $G_F$ is large ($p_{thr}\ll 0$), mainly the coherent excitation $\varepsilon_r$ and the two-photon absorption $g^2$ compete, 
producing a photon-number squeezed state as discussed in Ref.\cite{Chaturvedi77}. 
As shown in Figure 2(b), variance $\langle \delta \hat{X}_r^2\rangle$ is squeezed and below the vacuum level. 
In Figures 2(b)(c), the negative value of $p_{thr}$ ($G_F>\frac{1+j}{j}$) is related to the negative $\langle :\delta \hat{X}_r^2:\rangle$ and 
positive sign-adjusted self-skewness (${\rm sgn}(\langle \hat{X}_r\rangle) \langle \delta \hat{X}_r^3\rangle>0$). 
The normally ordered variance $\langle :\delta \hat{P}_r^2:\rangle$ and cross-skewness $\langle \delta \hat{X}_r\delta \hat{P}_r^2\rangle$ 
also change signs at $p_{thr}=0$, having the opposite signs from those of $\langle :\delta \hat{X}_r^2:\rangle$ and $\langle \delta \hat{X}_r^3\rangle$, respectively. 

The Wigner function can be obtained from a density-matrix simulation of the mean-field coupled CIM. 
Figure 3(a) shows the two-dimensional Wigner function $W(X_1,P_1)=\frac{1}{\pi}\int \langle X_1+y|\hat{\rho}|X_1-y\rangle e^{-2iP_1y}dy$ at $t=10$ 
for the case of $p_{thr}=2 (p=12)$. 
It shows a squeezed state with an anti-squeezed $\hat{X}$ (squeezed $\hat{P}$), 
slightly modified so that the distribution decreases more gradually on the side facing $\hat{X}_1=0$. 
At $t=10$, the mean amplitude is $\langle \hat{X}_1\rangle\sim 1.96$. 
The self-skewness $\langle \delta \hat{X}_1^3\rangle\sim -0.081$ and the cross-skewness $\langle \delta \hat{X}_1\delta \hat{P}_1^2\rangle \sim 0.020$ 
have the opposite sign from and same sign as the mean amplitude, respectively. 
Figure 3(b) shows the Wigner function at $t=10$, for the case of $p_{thr}=-5 (p=5)$. 
This distribution appears to be one of a slightly modified $\hat{X}$-squeezed state. 
This type of distribution is often seen in photon-number squeezed states and is generated by, 
for example, a coherently excited two-photon absorber\cite{Chaturvedi77} or an optical Kerr medium\cite{Kitagawa86,WilsonGordon91}. 
Its contour is flattened on the side facing $\hat{X}_1=0$. 
At $t=10$, the mean amplitude is $\langle \hat{X}_1\rangle\sim 2.03$. 
The self-skewness $\langle \delta \hat{X}_1^3\rangle\sim 0.011$ has the same sign as that of $\langle \hat{X}_1\rangle$, 
and the cross-skewness $\langle \delta \hat{X}_1\delta \hat{P}_1^2\rangle \sim -0.044$ has the opposite sign from it. 
The reduced Wigner functions are obtained after integrating over the momentum $\hat{P}_1$, $W_X(X_1)=\int W(X_1,P_1) dP_1$. 
The normalized values $W_X/W_X^{(g)}-1$ for $p_{thr}=2,-5$, 
using the Gaussian distribution $W_X^{(g)}(X_1)=\frac{1}{\sqrt{2\pi \langle \delta \hat{X}_1^2\rangle}}e^{-\frac{(X_1-\langle \hat{X}_1\rangle)^2}{2\langle \delta \hat{X}_1^2\rangle}}$ 
are shown in Figure 3(c). 
When $p_{thr}=2$ (${\rm sgn}(\langle \hat{X}_1\rangle)\langle \delta \hat{X}_1^3\rangle<0$), 
the skewed distribution $W_X(X_1)$ has a larger value than the Gaussian distribution $W_X^{(g)}$ at $X_1=0$. 
On the other hand, when $p_{thr}=-5$, the skewed distribution has a smaller value at $X_1=0$ than the Gaussian distribution. 
For the case of $p_{thr}=-5$, Figure 3(d) shows the $W_X$ (red solid line) and the Gaussian distribution $W_X^{(g)}$ (black dashed line). 
The up-arrow (down-arrow) shows where the skewed distribution has a larger (smaller) value than the Gaussian distribution. 
This skewed distribution suffices $\langle \delta \hat{X}_1\rangle \sim 0$ and 
the skew-Gaussian model was constructed in that manner, 
where these differ from the skew-normal distribution using the cumulative distribution function \cite{Azzalini96}. 

\begin{figure*}
\begin{center}
\includegraphics[width=15.0cm]{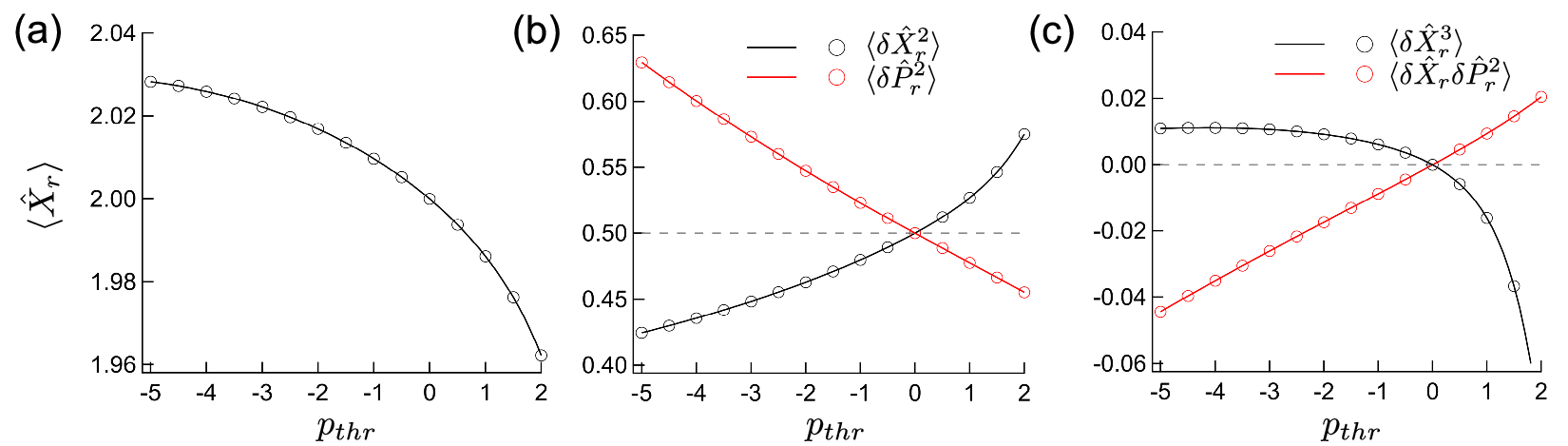}
\caption{Steady-state characteristics of mean-field coupled CIM versus the threshold value $p_{thr}=1+j-jG_F$. 
(a) Mean canonical coordinate $\langle \hat{X}_r\rangle$. 
(b) Fluctuation variance of canonical coordinate $\langle \delta \hat{X}_r^2\rangle$ (black) 
and momentum $\langle \delta \hat{P}_r^2\rangle$ (red). 
(c) Self-skewness $\langle \delta \hat{X}_r^3\rangle$ (black) and cross-skewness $\langle \delta \hat{X}_r\delta \hat{P}_r^2\rangle$ (red). 
Lines are obtained by time developing the QME at $t=10$. 
Open circles are obtained by solving the self-consistent equation Eq.(\ref{scl}) in Appendix C. }
\end{center}
\end{figure*}

\begin{figure}
\begin{center}
\includegraphics[width=9.0cm]{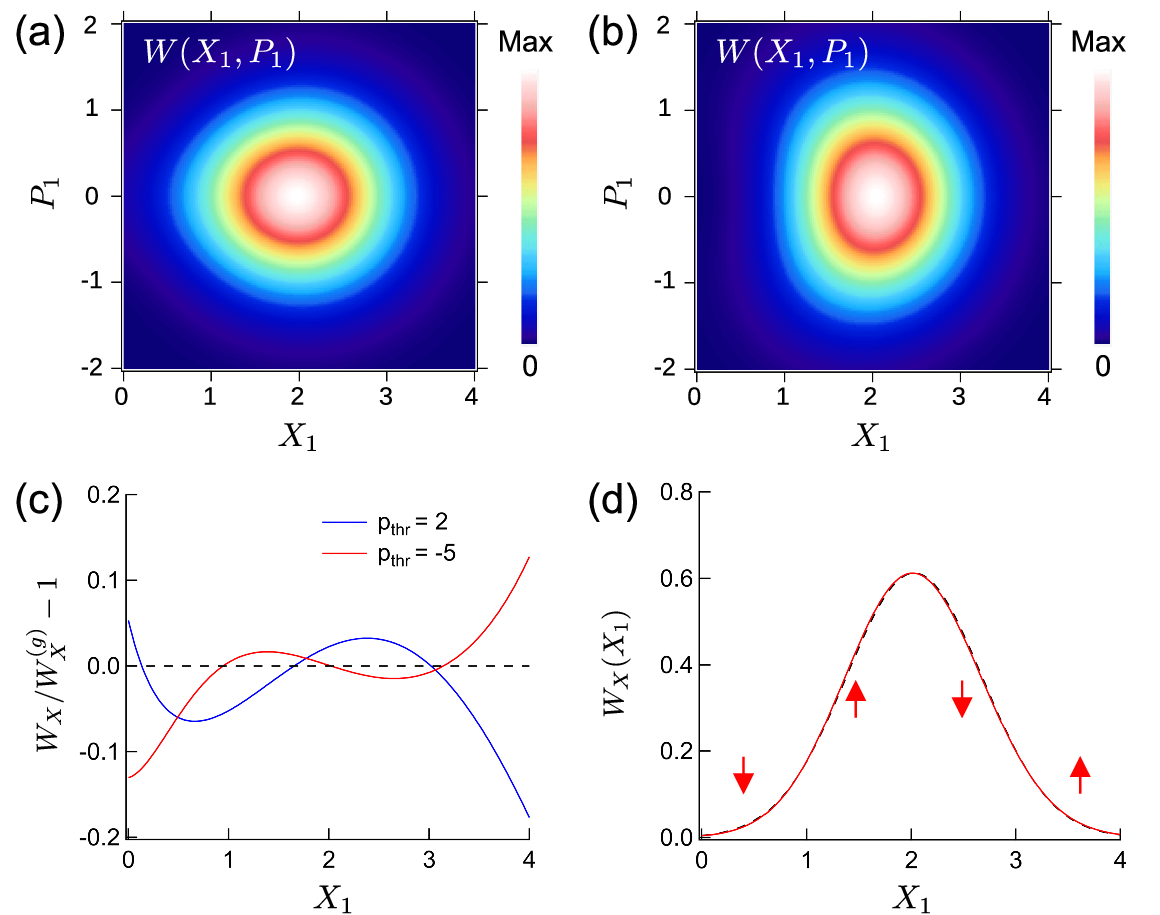}
\caption{Wigner function of mean-field-coupled CIM obtained from the continuous-time QME. 
(a)(b) Wigner function $W(X_1,P_1)$ at $t=10$, with (a) $p_{thr}=2$ 
( ${\rm sgn}(\langle \hat{X}_1\rangle )\langle \delta \hat{X}_1^3\rangle<0$) 
and (b) $p_{thr}=-5$ 
( ${\rm sgn}(\langle \hat{X}_1\rangle )\langle \delta \hat{X}_1^3\rangle>0$). 
(c) Reduced Wigner function $W_X(X_1)$ as a deviation from Gaussian distribution $W_X^{(g)}(X_1)$. 
(d) Reduced Wigner function $W_X$ (red solid line) and Gaussian distribution $W_X^{(g)}$ (black dashed line) for $p_{thr}=-5$. }
\end{center}
\end{figure}

\section{Comparison of continuous-time models}

In this section, we compare the simulation results obtained by various continuous-time models, 
including the quantum master equation and the skew-Gaussian model. 
In particular, we study an amplitude-controlled $N=6$ CIM where each independent component of $\tilde{J}_{rr'}$ is randomly chosen from 
21 discrete values, $\{-1,-0.9,\cdots,1\}$. 
Here, $\tilde{J}_{rr}=0$, and $\tilde{J}_{rr'}=\tilde{J}_{r'r}$. 
The auxiliary variables for the amplitude control $e_r$ follow Eq.(\ref{tdeve}), 
which stabilizes the squared measured values $\tilde{\mu}_r^2$ to $\tau$. 
Here, $\tilde{\mu}_r^2\sim \mu_r^2+\frac{1}{4j_1G_j\Delta t}$, 
and $\frac{1}{4j_1G_j\Delta t}$ represents the contribution from input noise to BS1. 
In a small-photon-number CIM, that noise part can be comparable to or greater than the squared actual internal amplitudes $\mu_r^2$.  
We represent the target value $\tau$ in the following way, 
\begin{equation}
\label{tau0def2}
\tau=\tau_0+\frac{1}{4j_1 G_j \Delta t}. 
\end{equation}
Here, $\tau_0$ is the real target value to which the squared intracavity amplitudes $\mu_r^2$ are stabilized. 
Figures 4(a-c) show the time-dependent mean amplitude $\langle \hat{X}_1\rangle$, 
self-skewness $\langle \delta \hat{X}_1^3\rangle$, and cross-skewness $\langle \delta \hat{X}_1\delta \hat{P}_1^2\rangle$, 
obtained by simulations using the skew-Gaussian model (Section II.C) and QME (Appendix A.2). 
The parameters were $\Delta t=2.5\times 10^{-4}, p=0.8, g^2=4, G_j=2, j=5, R=9, \beta=0.1, \tau_0=0.5$, and $e_r(t=0)=0.4$. 
The maximum photon number in the QME simulation was truncated at $N_M=16$. 
The self-skewness and cross-skewness of the skew-Gaussian model mostly follow 
those of the QME simulation, including at $t=5$ where the Wigner function is shown in Figure 4(d). 
However, around $t=1.22$ and $t=3.04$, the self-skewness $\langle \delta \hat{X}_1^3\rangle$ calculated by the QME has spikes to negative values. 
The Wigner functions at these time frames are shown in Figures 4(e) and (f). 
In Figures 4(e) and (f), they have concave contours for the sides of smaller $\hat{X}_1$ 
in contrast to the flat contours shown in Figures 3(b) and 4(d). 
Moreover in Figure 4(e), the Wigner function has negative values\cite{Mabuchi12}. 

Figures 5(a) and (b) show the success probability $P_s$ and mean photon number at $t=5$ for an $N=6$ CIM. 
Since we are discussing the small-photon-number continuous-time model with $\frac{1}{\sqrt{4j_1 G_j \Delta t}}\sim 7.5$, 
for the experimentally measurable value $\tilde{\mu}_r$, 
the ${\rm sgn}(\tilde{\mu}_r)$ takes $+1$ or $-1$ almost randomly. 
To validate the skew-Gaussian model, we define the spins as $\sigma_r={\rm sgn}(\langle \hat{a}_r\rangle)$, 
and $P_s$ is the probability that 
$E=-\frac{1}{2}\sum_{rr'}\tilde{J}_{rr'}\sigma_r\sigma_{r'}$ at $t=5$ is identical to the ground-state energy. 
The exact ground-state energy was independently obtained by making a brute-force search. 
We generated $10^6$ instances and obtained the instance-averaged success probability $P_s$ by running a time-development simulation for each instance only once. 
Figure 5(a) shows the $g^2$-dependent success probability $P_s$ obtained by five continuous-time models (skew-Gaussian, QME, Gaussian, GAPP, and GATW). 
Here, the parameters except $g^2$ were the same as in Figure 4. 
The success probability $P_s$ of the skew-Gaussian model follows that of the QME even at $g^2\sim 3$, 
although the spiking behaviors of the skew variables (Figures 4(b) and (c)) at $g^2=4$ 
cannot be reproduced by the skew-Gaussian model. 
The success probability of the Gaussian model \cite{Ng22} (red line) which assumes $\gamma_r=\kappa_r=0$ in the skew-Gaussian model 
deviates from that of the QME and skew-Gaussian model even at $g^2 \sim 1$. 
Two other approximate Gaussian models of the CIM have been derived \cite{Kako20,Inui22}. 
The GAPP formulation in Ref.\cite{Inui22} derived from a Gaussian approximation of the positive-$P$\cite{Drummond80} stochastic differential equation 
provides a more accurate success probability $P_s$ than 
the Gaussian-approximated truncated-Wigner (GATW) model in Refs.\cite{Kako20,Inui22}. 
However, the equations of the GAPP neglect the fourth-order fluctuation product terms 
($\langle \delta \alpha_r^+ \delta \alpha_r^3\rangle$, $\langle \delta \alpha_r^{+2} \delta \alpha_r^2\rangle$ in Eqs.(\ref{da2mf}) and (\ref{daTdamf})) of Appendix B.2. 
The Gaussian model in Ref.\cite{Ng22}, where the fourth-order fluctuation products are decomposed into products of Gaussian variances 
provides more accurate results than the GAPP does. 
For the mean photon number per pulse shown in Figure 5(b), the skew-Gaussian model closely follows the results of the QME even at $g^2\sim 10$. 
For $N=12$ CIM, the success probability $P_s$ and photon number per pulse at $t=5$ are shown in Figures 5(c) and (d), respectively. 
The parameters and the cutoff photon number $N_M$ were the same as in the $N=6$ simulations. 
The success probability $P_s$ from the skew-Gaussian model stays close to the QME values until $g^2\sim 3$, 
while $P_s$ of the Gaussian model (red line) deviates from those of the skew-Gaussian model and QME at $g^2\sim 1$. 

\begin{figure*}
\begin{center}
\includegraphics[width=15.0cm]{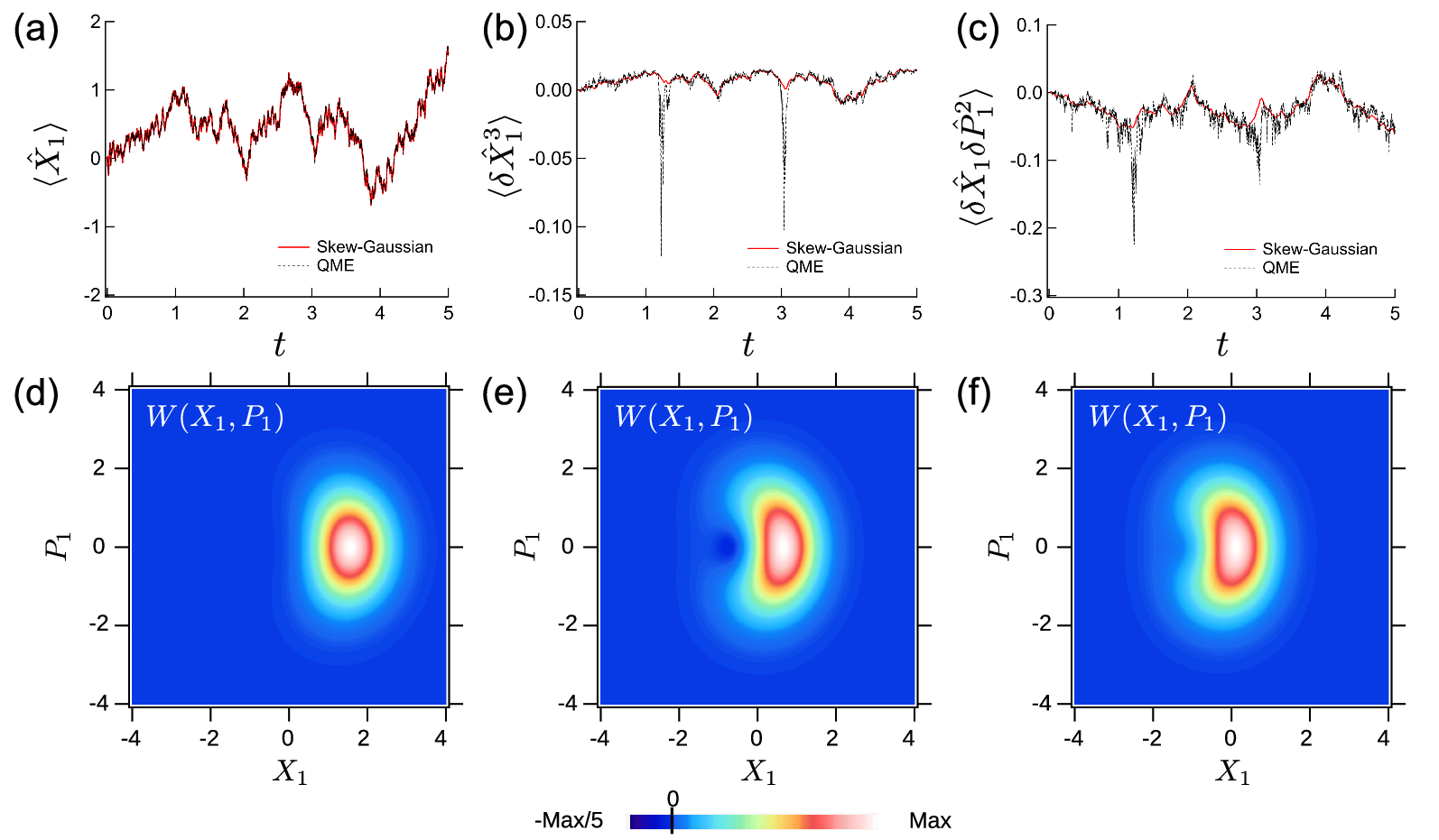}
\caption{Simulated time-development of $N=6$ CIM using two continuous-time models, 
skew-Gaussian model and QME. 
(a) Time-dependent mean canonical coordinate $\langle \hat{X}_1\rangle$. 
(b) Time-dependent self-skewness $\langle \delta \hat{X}_1^3\rangle$. 
(c) Time-dependent cross-skewness $\langle \delta \hat{X}_1 \delta \hat{P}_1^2\rangle$. 
(d)(e)(f) Wigner function $W(X_1,P_1)$ at $t=5$, $t=1.22$, and $t=3.04$, respectively. }
\end{center}
\end{figure*}

\begin{figure}
\begin{center}
\includegraphics[width=9.0cm]{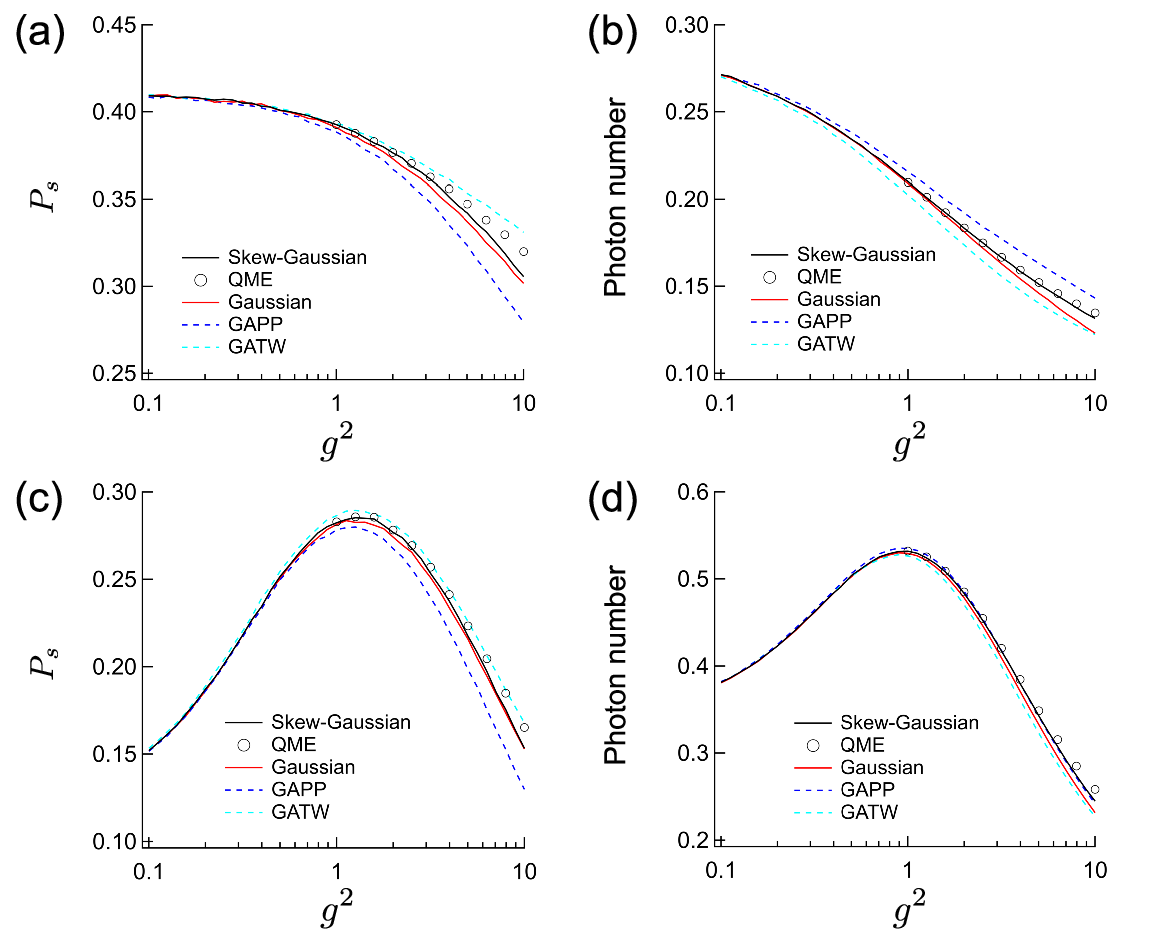}
\caption{$g^2$-dependent characteristics of $N=6,12$ CIMs at $t=5$ using continuous-time models. 
For $N=6$ CIM, (a) success probability $P_s$, and (b) mean photon number per pulse. 
For $N=12$ CIM, (c) success probability $P_s$, and (d) mean photon number per pulse. }
\end{center}
\end{figure}

\section{Correction by skew variables in discrete-component model}

In the previous section, we determined whether the simulation run was success or not by spin $\sigma_r$ 
defined as $\sigma_r={\rm sgn}(\langle \hat{a}_r\rangle)$. 
In this section, we define the spin in an experimentally measurable form $\sigma_r={\rm sgn}(\tilde{\mu}_r)$ \cite{Kako20}, 
while keeping the type of instance ($\tilde{J}_{rr'}\in \{-1,-0.9,\cdots,1\}$) the same. 
To realize a sufficiently large success probability, we simulated the CIM with a larger $\Delta t$, $R_1$ and $R_2$ 
by using the discrete-component models \cite{Clements17,Ng22}, instead of the continuous-time models. 
The details of discrete-component QME (Figure 1(a)) and skew-Gaussian model are presented in Appendix D.1 and Appendix D.2, respectively, 
and we compare these two models in Appendix D.3. 
Figure 6 shows the discrete-component skew-Gaussian model. 
It uses a set of five variables $(\langle \hat{X}_r\rangle, V_r^X, V_r^P, \mathfrak{S}_r, \mathfrak{C}_r)$ 
to represent the initial and final state $\hat{\rho}^{(r)}$. 
Here, $V_r^X$, $V_r^P$, $\mathfrak{S}_r$, and $\mathfrak{C}_r$ 
represent $\langle \delta \hat{X}_r^2\rangle$, $\langle \delta \hat{P}_r^2\rangle$, $\langle \delta \hat{X}_r^3\rangle$, and $\langle \delta \hat{X}_r\delta \hat{P}_r^2\rangle$, respectively. 
There are the other six variations of them with the prime symbol $'$, subscript $B$, subscript $T$, subscript $R$, double prime $''$, and subscript $Z$, 
representing the state after the DOPO, probe input state of BS1, 
transmitted state of BS1, reflected state of BS1, the internal state after the measurement-induced state-reduction, and external input state of the BS2. 
In the discrete-component model, the squared measured amplitude has the value $X_{B,r}^2\sim R_1\langle \hat{X}_r\rangle^{'2}+(1-R_1)V_{B,r}^X+R_1 V_r^{X'}$. 
For the inferred internal amplitude $\tilde{\mu}_r=X_{B,r}/\sqrt{2R_1}$, it is represented as, 
\begin{equation}
\tilde{\mu}_r^2\sim \frac{\langle \hat{X}_r\rangle^{'2}}{2}+\frac{V_r^{X'}}{2}+\frac{1-j_1\Delta t}{4j_1 G_j\Delta t}. 
\end{equation}
We approximate the variance as $V_r^{X'}\sim \frac{1+\frac{j_1}{2G_j}+\frac{j_2}{2}}{2(1-p+j)}$, 
by neglecting the $g^2$-dependent terms and the measurement-induced state-reduction in Eqs. (\ref{ctsk2}) and (\ref{ctsk3}). 
We thus introduce an approximated actual target value $\tau_0$ 
for the discrete-component model as follows, 
\begin{equation}
\label{tau0def}
\tau=\tau_0+\frac{1-j_1 \Delta t}{4G_j j_1\Delta t}+\frac{1+\frac{j_1}{2G_j}+\frac{j_2}{2}}{4(1-p+j)}. 
\end{equation}
Because of the huge computational cost of discrete-component QME, 
we used the skew-Gaussian model in the study reported below and compared it with the Gaussian model. 

Figure 7(a) shows the $g^2$-dependent success probability $P_s$ of $N=6$ CIM, 
obtained by the discrete-component skew-Gaussian and Gaussian models. 
The parameters were $\Delta t=3.2\times 10^{-3},  p=-10, G_j=1.1, j=10, R=9, \beta=1, \tau_0=1$, and $e_r(t=0)=0.4$, 
and the number of instances was $10^7$, each of which was solved only once. 
The $P_s$ is the probability that the energy $E=-\frac{1}{2}\sum_{rr'}\tilde{J}_{rr'}\sigma_r\sigma_{r'}$ 
calculated from $\sigma_r={\rm sgn}(\tilde{\mu}_r)$ at $t=10$ is equal to the ground-state energy 
determined by a brute-force search. 
Here, $\tilde{\mu}_r=X_{B,r}/\sqrt{2R_1}$ is the indirect homodyne measurement result (Eq.(\ref{XRmeas})). 
When $g^2$ is increased, the skew-Gaussian model has a slightly smaller success probability than that of the Gaussian model. 
Figure 7(b) shows the distribution of mean amplitudes $\langle \hat{X}_r\rangle'$ at $t=10$, 
obtained by $6\times 10^7$ samples ($10^7$ simulation runs and six sites), where the saturation coefficient was $g^2=5$. 
The skew-Gaussian model has a slightly larger value at $\langle \hat{X}_r\rangle'=0$ and slightly smaller peak values. 
The larger values at $\langle \hat{X}_r\rangle'\sim 0$ seem to have contributed to the smaller success probability, 
since smaller $|\langle \hat{X}_r\rangle'|$ values are easily flipped by the noise from the open port of BS1, 
even when ${\rm sgn}(\langle \hat{X}_r\rangle')$ have one of the ground-state configurations. 
Figure 7(c) shows the $p$-dependent success probability and photon number per pulse of the $N=6$ CIM 
obtained by the discrete-component skew-Gaussian and Gaussian models 
for $\Delta t=3.2\times 10^{-3}, g^2=5, G_j=1.1, j=10, R=9, \beta=1, \tau_0=1$, and $e_r(t=0)=0.4$. 
The number of simulation runs was $10^8$. 
Since the difference in success probability between Gaussian and skew-Gaussian models is small, we took the deviation 
$D_1=P_s^{(s)}/P_s^{(g)}-1$, where $P_s^{(s)}$, $P_s^{(g)}$ are the success probabilities obtained by the skew-Gaussian and Gaussian models, respectively. 
This value is plotted as the black line in Figure 7(d). 
As we saw in Figure 7(a), it is smaller than 0 ($P_s^{(g)}>P_s^{(s)}$) when $p$ is negative. 
When $p$ is increased, it crosses 0 at $p\sim 0$, but again turns negative between $p\sim5$ and $p\sim10$. 
This behavior is similar to that of ${\rm sgn}(\langle \hat{X}_r\rangle)(3\langle \delta \hat{X}_r^3\rangle+\langle \delta \hat{X}_r\delta \hat{P}_r^2\rangle)$, 
shown by the blue line in Figure 7(d). 
The plotted values for the blue line were obtained using the amplitudes and skew variables after BS2 at $t=10$, and are averaged over instances and sites. 
It is equivalent to $\mu_r(\gamma_r+5\kappa_r)$ appearing in $\frac{d(n_r+m_r)}{dt}$ of Eqs.(\ref{ctsk2}) and (\ref{ctsk3}) 
and represents the skew-induced negative correction to the fluctuation variance $\langle \delta \hat{X}_r^2\rangle(\sim V_r^{X'})$ . 
Because the amplitude control fixes the value 
$\tilde{\mu}_r^2=\frac{\langle \hat{X}_r\rangle^{'2}}{2}+\frac{V_r^{X'}}{2}+\frac{1-j_1\Delta t}{4j_1G_j\Delta t}$ to the target value, 
when $V_r^{X'}$ is smaller, the squared mean amplitude $\langle \hat{X}_r\rangle^{'2}$ can be greater, 
which results in a larger success probability $P_s$. 

Figure 8(a) shows the success probabilities $P_s$ of $N=12$ CIM 
when we read out spins $\sigma_r$ through a direct homodyne measurement at $t=10$ by extracting all internal pulses. 
The number of simulation runs was $2\times 10^7$. 
For the skew-Gaussian model, the spin is defined as the sign of the direct measurement \cite{Plimak01}, 
$\sigma_r={\rm sgn}\Bigl(\langle \hat{X}_{r}\rangle'-\frac{\mathfrak{S}_{r}'}{6V_{r}^{X'}}+\sqrt{V_{r}^{X'}+\frac{\mathfrak{S}_{r}'}{3\sqrt{V_{r}^{X'}}}\mathcal{N}_r}\mathcal{N}_r\Bigr)$, 
where $\langle \hat{X}_r\rangle'$, $V_r^{X'}$, and $\mathfrak{S}_r'$ are mean amplitude, variance, and self-skewness after the DOPO (Figure 1(a)). 
Here, the normal random number $\mathcal{N}_r$ outside the square root is identical to that inside the square root. 
On the other hand, for the Gaussian model, the spin does not depend on the self-skewness, $\sigma_r={\rm sgn}(\langle \hat{X}_{r}\rangle'+\sqrt{V_{r}^{X'}}\mathcal{N}_r)$. 
The success probabilities $P_s$ are higher than those of $N=6$ CIM in Figure 7(c) because of the direct measurement. 
In Figure 8(b), the black line is the skew-induced deviation of the success probability $D_1=P_s^{(s)}/P_s^{(g)}-1$, 
while the blue line is the skew-induced negative correction to the variance 
${\rm sgn}(\langle \hat{X}_r\rangle)(3\langle \delta \hat{X}_r^3\rangle+\langle \delta \hat{X}_r\delta \hat{P}_r^2\rangle)$. 
The black line crosses zero at the smaller $p\sim -2.7$ than the blue line does ($p\sim -0.5$). 
Figure 8(b) also shows the deviation $D_2=P_s^{(s')}/P_s^{(g)}-1$ (the red line), 
where $P_s^{(s')}$ is the success probability of the skew-Gaussian model 
where the definition of the spin is modified to be that of a Gaussian model $\sigma_r={\rm sgn}(\langle \hat{X}_{r}\rangle'+\sqrt{V_{r}^{X'}}\mathcal{N}_r)$. 
That deviation $D_2$ crosses zero at $p\sim 0$ which is closer to where 
${\rm sgn}(\langle \hat{X}_r\rangle)(3\langle \delta \hat{X}_r^3\rangle+\langle \delta \hat{X}_r\delta \hat{P}_r^2\rangle)$ changed sign. 
We introduce the deviation $D_3=P_s^{(s)}/P_s^{(s')}-1$ of success probabilities 
between the skew-Gaussian model ($P_s^{(s)}$) and 
the one with the modified definition of spin ($P_s^{(s')}$) and present it in Figure 8(c). 
$D_3$ depends on the sign-adjusted self-skewness, ${\rm sgn}(\langle \hat{X}_r\rangle)(\langle \delta \hat{X}_r^3\rangle)$ 
shown with a blue line in Figure 8(c). 
Overall, Figure 8 shows that the success probability $P_s$ obtained by the direct measurement results 
is improved by the positive sign-adjusted self-skewness, 
since as shown in Figure 3(b) the skewed distribution at $X_r=0$ has a smaller value compared with that of the Gaussian model 
and the signs of the measured values are less likely to be flipped from ${\rm sgn}(\langle \hat{X}_r\rangle)$ 
when compared to the Gaussian distribution. 

\begin{figure}
\begin{center}
\includegraphics[width=7.0cm]{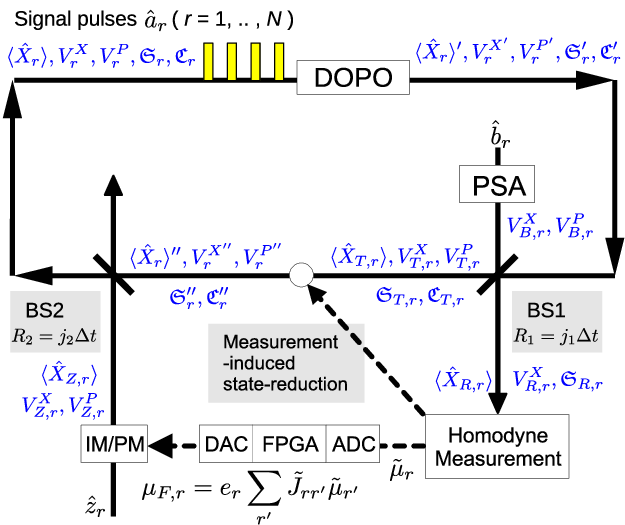}
\caption{Discrete-component skew-Gaussian model. }
\end{center}
\end{figure}

\begin{figure}
\begin{center}
\includegraphics[width=9.0cm]{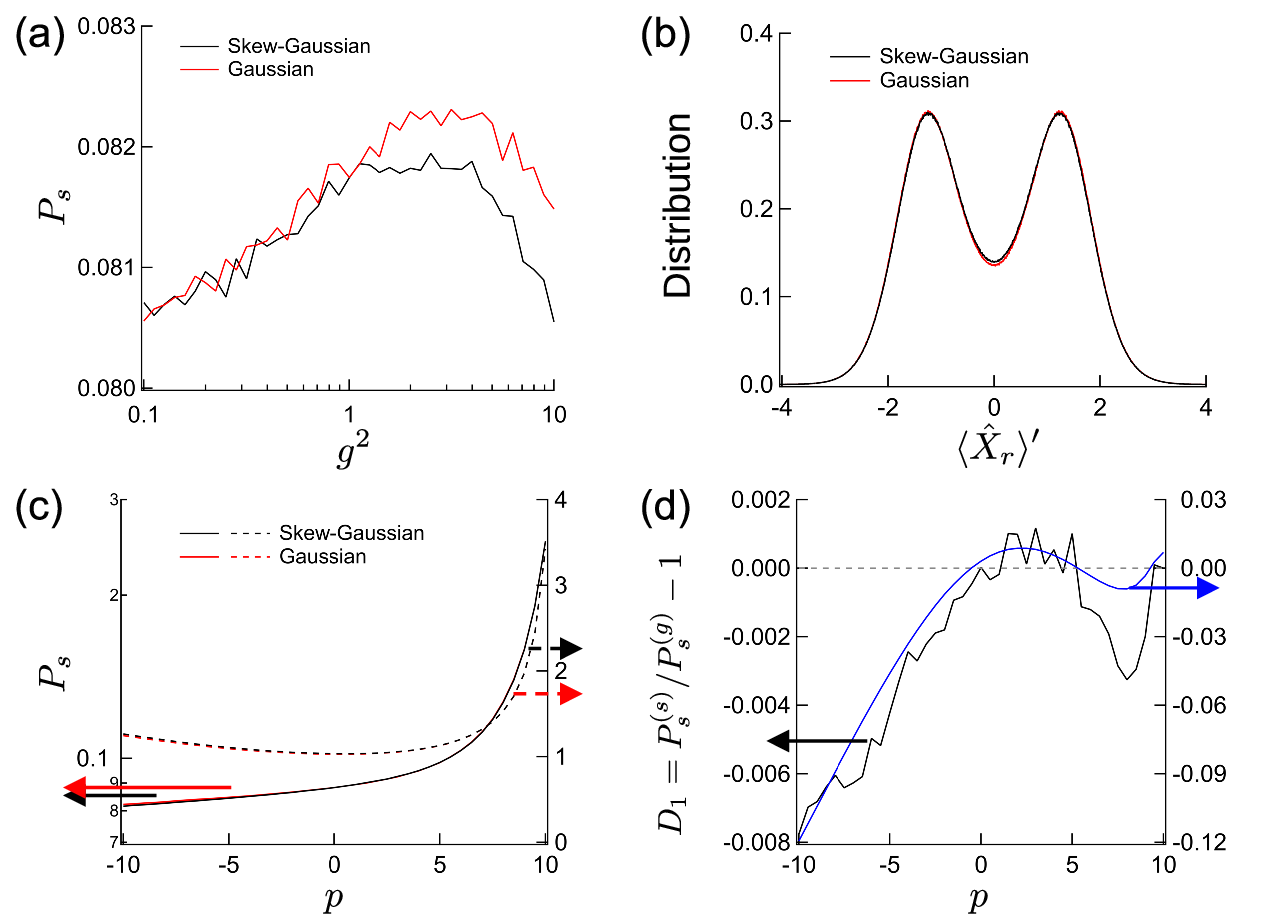}
\caption{Skew-induced correction of $N=6$ CIM assuming indirect measurement. 
(a) $g^2$-dependent success probability $P_s$ by discrete-component skew-Gaussian and Gaussian models. 
(b) The distribution of mean canonical coordinate $\langle \hat{X}_r\rangle'$ for skew-Gaussian and Gaussian models. 
(c) $p$-dependent success probability (left axis) and photon number per site (right axis). 
(d) Deviation $D_1$ of success probabilities (left axis) and sign-adjusted $3\langle \delta \hat{X}_r^3\rangle+\langle \delta \hat{X}_r\delta \hat{P}_r^2\rangle$ (right axis). }
\end{center}
\end{figure}

\begin{figure*}
\begin{center}
\includegraphics[width=15.0cm]{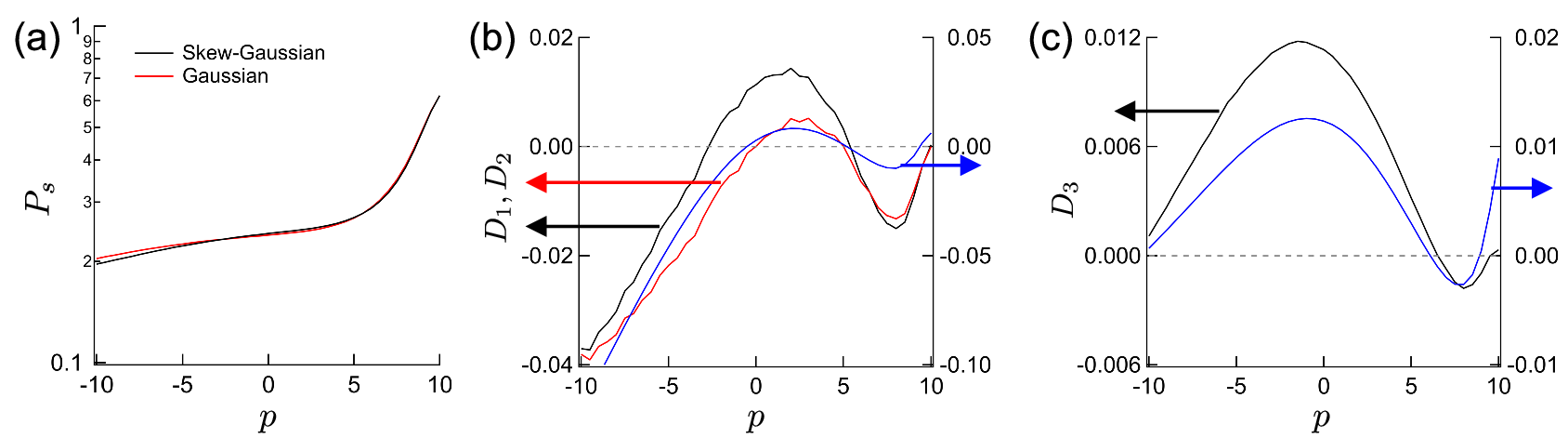}
\caption{Skew-induced correction of $N=12$ CIM assuming direct measurement. 
(a) $p$-dependent success probability $P_s$ by discrete-component skew-Gaussian and Gaussian models. 
(b) Deviation $D_1$ (black line), $D_2$ (red line) of success probabilities (left axis), 
and sign-adjusted $3\langle \delta \hat{X}_r^3\rangle+\langle \delta \hat{X}_r\delta \hat{P}_r^2\rangle$ (right axis). 
(c) Deviation $D_3$ of success probabilities (left axis) and sign-adjusted self-skewness $\langle \delta \hat{X}_r^3\rangle$ (right axis). }
\end{center}
\end{figure*}

\section{Site-number dependence}

Here, we discuss whether skew variables affect the performance of CIMs, 
after the parameters are optimized to maximize the success probability. 
In this context, we studied the site-number- ($N$-) dependent success probability and 
the minimum number of matrix-vector-multiplications needed to reach a solution ($\min {\rm MVMTS}$) 
for up to $N=200$ Wishart planted instances \cite{Hamze20}. 
With a discrete-component model of small-photon-number CIM, we solved the $\alpha_{WP}=0.8$ Wishart planted instance, 
for which a CIM-inspired heuristic method has already been applied\cite{Leleu21}. 
We redefined the success probability (denoted as $P_s'(t)$) as the cumulative probability of finding the ground state at least once until time $t$. 
We used the signs of indirect homodyne measurement values as spins $\sigma_r={\rm sgn}(\tilde{\mu}_r)$, 
and for each time step, calculated the energy $E=-\frac{1}{2}\sum_{rr'}\tilde{J}_{rr'}\sigma_r\sigma_{r'}$ 
and checked whether it equaled the ground-state energy of the Wishart planted instance. 
Figure 9(a) shows the success probability $P_s'$ at $t=10$ for $N=100, \alpha_{WP}=0.8$ Wishart planted instances, 
solved with the discrete-component skew-Gaussian and Gaussian models. 
The parameters were $\Delta t=3.2\times 10^{-3}, p=0, g^2=15, G_j=2, R=9, \beta=0.3, \tau_0=1.5$, and $e_r(t=0)=2$. 
We generated $5\times 10^5$ instances and ran the time-development simulation for each instance only once. 
The percent deviation $(P_s^{'(s)}/P_s^{'(g)}-1)\times 100$ is shown on the right axis. 
Here, $P_s^{'(s)}$ and $P_s^{'(g)}$ are the cumulative success probabilities for the skew-Gaussian and Gaussian models, respectively. 
Figure 9(b) shows the mean photon number per site averaged over all time steps, 
sites ($N=100$) and $5\times 10^5$ simulation runs. 
The success probability and photon number per site increased with the coupling coefficient $j$. 
The percent deviations between the skew-Gaussian and Gaussian models (right axis) 
became smaller as $j$ increased. 
Figure 9(c) shows the sign-adjusted skew variables obtained at $t=10$ by averaging over the sites and $5\times 10^5$ simulation runs. 
The negative value of ${\rm sgn}(\langle \hat{X}_r\rangle)(3\langle \delta \hat{X}_r^3\rangle+\langle \delta \hat{X}_r \delta \hat{P}_r^2\rangle)$ 
resulted in the skew-Gaussian model having a smaller success probability than the Gaussian model, as we saw in Figure 7. 
However, the self-skewness and cross-skewness became closer to zero as $j$ increased. 

Figure 10(a) shows the site-number-dependent success probability $P_s'$ at $t=10$ 
calculated with the discrete-component skew-Gaussian and Gaussian models. 
The coupling coefficient was optimized to be $j=100$, 
and two parameters were optimized as functions of the site number: $g^2=1.5\sqrt{N}$, and $\beta=3000/N^2$. 
The other parameters, $\Delta t=3.2\times 10^{-3}, p=0, G_j=2, R=9, \tau_0=1.5$, and $e_r(t=0)=2$, were independent of the site number. 
Appendix E presents the success probability $P_s'$ at $t=10$ as functions of $j$, $g^2$, and $\beta$. 
In Figure 10(a), the number of instances is $5\times 10^5$ and we solved each instance only once. 
The results for the large-photon-number CIM \cite{Inui22} are plotted as the blue dashed line. 
The parameters of the large-photon-number CIM are detailed in Appendix F. 
The success probability of the small-photon-number CIM calculated with the skew-Gaussian model 
is more than 78\% that of the large-photon-number CIM, although 
the mean photon number per site is more than $10^{3}$ times smaller, as shown in Figure 10(b). 
We also calculated the $\min_t {\rm MVMTS}(t)$ to find the ground state energy with 99\% probability, 
\begin{equation}
\label{mvmts_def}
\min_t {\rm MVMTS}(t)=\min_t \frac{t}{\Delta t}\frac{\ln (0.01)}{\ln (1-P_s'(t))}. 
\end{equation}
Figure 10(c) plots the $\min_{t\le 10} {\rm MVMTS}(t)$ for the simulation run until $t=10$. 
The values obtained by the small-photon-number CIM were comparable to those obtained by 
the CIM-inspired heuristic method \cite{Leleu21}. 
Such a level of performance is expected to be achievable in an optical machine 
using indirect homodyne measurement results, with only a few photons per pulse, 
when a large nonlinear coefficient $g^2$ is introduced. 
Although $g^2$ is large, the difference between the skew-Gaussian and Gaussian models 
become negligible because of the use of large $j$ to achieve a high success probability. 

Instead of the similar instance-averaged performances in Figure 10, 
we show the performances for each Wishart planted instance in Figure 11. 
Figure 11(a) shows the success probabilities $P_s'$ at $t=10$ of the small- and large-photon-number CIMs 
for $N=100$, $\alpha_{WP}=0.8$ Wishart planted instances. 
We used the parameters at $N=100$ of Figure 10. 
We generated 100 instances and ran the simulation $10^5$ times for each instance. 
The $P_s'$ of the small-photon-number CIM simulated 
by the skew-Gaussian and Gaussian models differ by at most 2.6\% (instance 86). 
The instance dependence of the success probabilities is smaller in the small-photon-number CIM 
than in the large-photon-number CIM. 
The large-photon-number CIM had a larger $P_s'$ than that of the small-photon-number CIM (skew-Gaussian model) 
for 85 of 100 instances, while for four of the 100 instances the success probabilities of the large-photon-number CIM are less than 2\%. 
Figures 11(b) and (c) show scatter plots of the success probability $P_s'$ at $t=10$ and 
minimum number of matrix-vector-multiplications needed to reach a solution $\min_{t\le 10} {\rm MVMTS}(t)$ for 100 instances. 
The horizontal axis shows the values obtained by the small-photon-number CIM (skew-Gaussian model) 
and the vertical axis shows those obtained by the large-photon-number CIM. 
For three instances (instance 1,16,96), the $\min {\rm MVMTS}$ obtained by the large-photon-number CIM is 
more than ten times larger than that obtained by the small-photon-number CIM. 
Because of these instances, the averaged $\min {\rm MVMTS}$ over 100 instances is 
$2.3\times 10^5$ for large-photon-number CIM, 
which is much larger than $4.8\times 10^4$ in Figure 10 (c) 
(obtained from the averaged $P_s'$ for $5\times 10^5$ instances), 
and the median value of 100 instances ($4.4\times 10^4$). 
For small-photon-number CIM, these values are $6.8\times 10^4$, $6.3\times 10^4$, and $7.0\times 10^4$, respectively. 

\begin{figure*}
\begin{center}
\includegraphics[width=15.0cm]{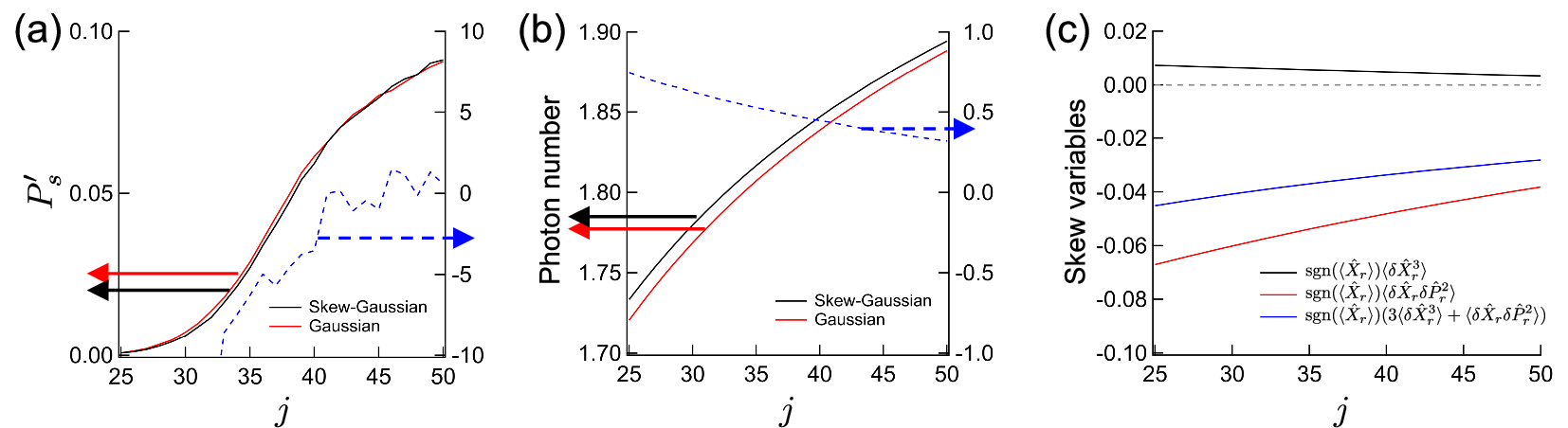}
\caption{Coupling- ($j$-) dependent characteristics of small-photon-number CIM with skew-Gaussian and Gaussian models. 
(a) Success probability $P_s'$ at $t=10$ (left axis) and percent deviation (right axis). 
(b) Mean photon number per site (left axis) and percent deviation (right axis). 
(c) Sign-adjusted self-skewness $\langle \delta \hat{X}_r^3\rangle$, cross-skewness 
$\langle \delta \hat{X}_r \delta \hat{P}_r^2\rangle$, and skew-induced negative correction to the $\hat{X}$-variance, 
$3\langle \delta \hat{X}_r^3\rangle+\langle \delta \hat{X}_r \delta \hat{P}_r^2\rangle$, at $t=10$. }
\end{center}
\end{figure*}

\begin{figure*}
\begin{center}
\includegraphics[width=15.0cm]{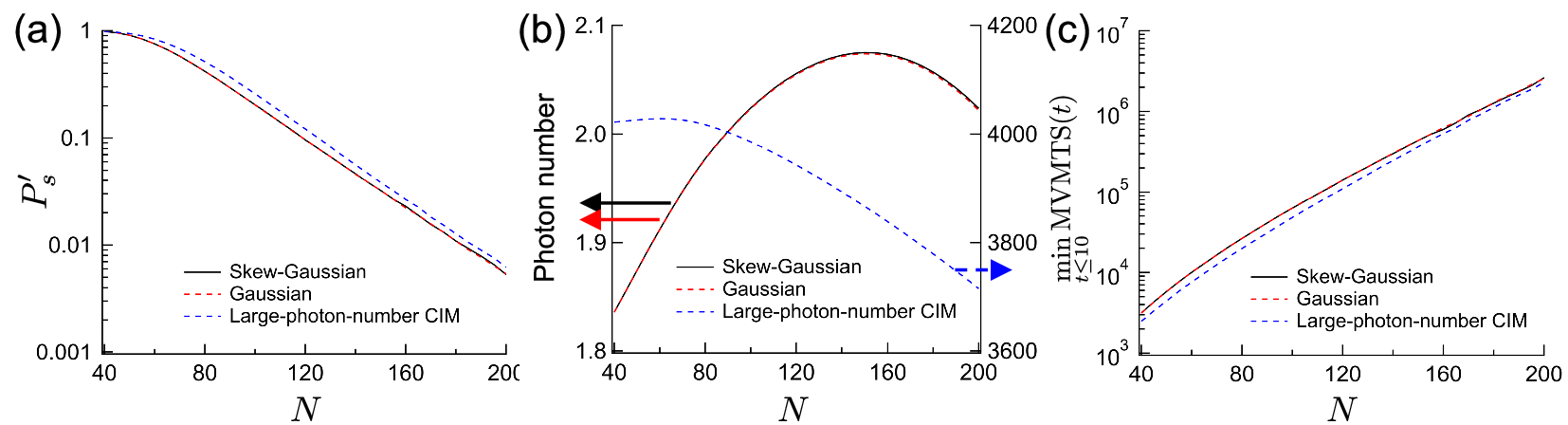}
\caption{Site-number- ($N$-) dependent characteristics of small-photon-number CIM (skew-Gaussian and Gaussian models) and large-photon-number CIM. 
(a) Success probability $P_s'$ at $t=10$. 
(b) Mean photon number per site of small-photon-number CIM (left axis) and large-photon-number CIM (right axis). 
(c) Minimum number of matrix-vector-multiplications needed to reach a solution, $\min_{t\le 10} {\rm MVMTS}(t)$. }
\end{center}
\end{figure*}

\begin{figure*}
\begin{center}
\includegraphics[width=15.0cm]{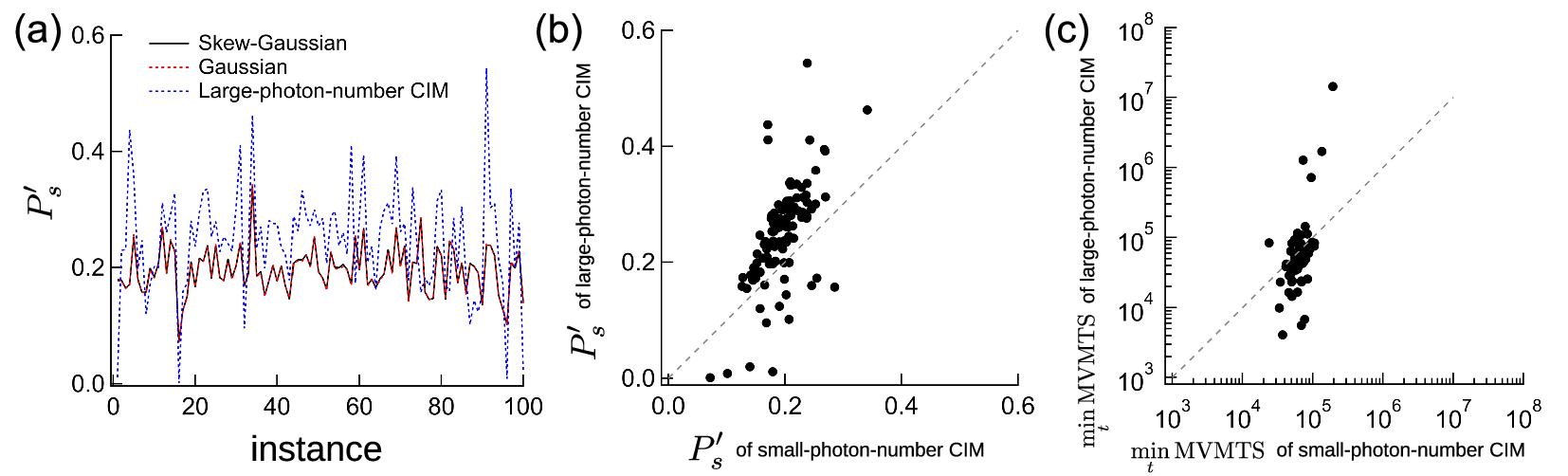}
\caption{Instance-dependent characteristics. (a) Success probability $P_s'$ at $t=10$. 
(b) Scatter plot of $P_s'$ at $t=10$ between small-photon-number CIM (skew-Gaussian model) and large-photon-number CIM. 
(c) Scatter plot of $\min_{t\le 10} {\rm MVMTS}(t)$ between small-photon-number CIM (skew-Gaussian model) and large-photon-number CIM. }
\end{center}
\end{figure*}

\section{Summary}

In this paper, we introduced the skew-Gaussian model to describe coherent Ising machines (CIMs). 
It is an extended Gaussian model using third-order fluctuation products, 
$\langle \delta \hat{X}_r^3\rangle$ and $\langle \delta \hat{X}_r\delta \hat{P}_r^2\rangle$. 
We summarized the impact of skew variables on the characteristics of amplitude-controlled CIMs operating 
with a few photons per pulse and a large nonlinear saturation coefficient $g^2$. 
First, we described the skew-Gaussian model under the continuous-time description 
where we used Wiseman-Milburn's model to describe homodyne measurement. 
The skew-Gaussian model provides a more accurate success probability $P_s$ than the Gaussian model when $g^2>1$. 
Second, we studied the CIM performance using the discrete-component description. 
The deviation in the success probabilities due to the skew variables can be explained by the skew-induced negative correction to $\langle \delta \hat{X}_r^2\rangle$, 
depending on ${\rm sgn}(\langle \hat{X}_r\rangle)(3\langle \delta \hat{X}_r^3\rangle+\langle \delta \hat{X}_r\delta \hat{P}_r^2\rangle)$. 
Because of the amplitude control, the smaller fluctuation $\langle \delta \hat{X}_r^2\rangle$ resulted in larger squared mean amplitudes $\langle \hat{X}_r\rangle^2$, 
which improved the success probability. 
In practice, the increase in the coupling-related linear loss $j$ effectively improved the success probability and decreased the $\min {\rm MVMTS}$. 
However, since skew variables become smaller for large $j$, 
there are an almost negligible difference between the skew-Gaussian and the Gaussian models, when we tuned parameters to maximize the success probability. 
We evaluated the $\min {\rm MVMTS}$ of a small-photon-number CIM ($<3$ photons per pulse) for up to $N=200$ Wishart planted instances, 
which was comparable to that of a large-photon-number CIM ($>10^3$ photons per pulse). 

The skew-Gaussian model can take into account non-Gaussian effects and be simulated with much 
fewer computational resources compared with directly solving quantum master equations, 
which in a simulation entails solving  many equations due to the photon-number state expansion. 
The discussion of skew variables in this paper to some extent might be applicable to other systems with dissipative boson modes, 
particularly those related to two photon absorption\cite{Chaturvedi77} and the optical Kerr effect\cite{Kitagawa86,WilsonGordon91}, 
including non-Gaussian entanglement \cite{Olsen13} in such systems. 
However, the negative Wigner function and concave structure of the Wigner function would not be represented accurately as we saw in Figure 4. 
In addition to simple extensions of our approach, for example, to take into account fourth-order fluctuation products, 
different approaches might be required to treat such a Wigner function, 
for example those related to the Pegg-Barnett phase eigenstates\cite{Pegg89,Ezaki99} or 
photon-added coherent states\cite{Genoni13}.

\appendix
\setcounter{equation}{0}
\renewcommand{\theequation}{A\arabic{equation}}

\section{Continuous-time quantum master equation}

\subsection{Derivation of Eq.(\ref{qmesr})}

In this Appendix, we provide supplementary information about the continuous-time quantum master equation (QME). 
Here, we derive the second line of Eq.(\ref{qmesr}), following Ref.\cite{Wiseman09}. 
That part represents the measurement-induced state-reduction, when the probe field is 
in a squeezed pure state $|\psi_{B,r}\rangle=e^{-\frac{\ln\sqrt{G_j}}{2}(\hat{b}_r^{\dagger 2}-\hat{b}_r^2)}|0\rangle$, 
whose uncertainty product is identical to that of the vacuum state.  
The probe state satisfies, 
\begin{equation}
\hat{Q}_r=(1+G_j^{-1})\hat{b}_r+(1-G_j^{-1})\hat{b}_r^{\dagger}, 
\end{equation}
\begin{equation}
\hat{Q}_r|\psi_{B,r}\rangle=0. 
\end{equation}
The moments $\langle \psi_{B,r}|\hat{b}_r^2|\psi_{B,r}\rangle=-\frac{1}{4}(G_j-G_j^{-1})$ and 
$\langle \psi_{B,r}|\hat{b}_r^{\dagger}\hat{b}_r|\psi_{B,r}\rangle=\frac{1}{4}(G_j+G_j^{-1}-2)$, 
which are identical to $-m_j$ and $n_j$ respectively, are obtained from 
$\langle \psi_{B,r}|\hat{Q}_r^2|\psi_{B,r}\rangle=\langle \psi_{B,r}|\hat{Q}_r^{\dagger}\hat{Q}_r|\psi_{B,r}\rangle=0$. 
When we represent the internal state of the $r$-th pulse as $|\psi_r\rangle$, 
the state after the measurement is 
\begin{equation}
|\psi_r\rangle'=\frac{1}{\sqrt{P(X_{B,r})}}\langle X_{B,r}|e^{\theta_1(\hat{a}_r^{\dagger}\hat{b}_r-\hat{b}_r^{\dagger}\hat{a}_r)}|\psi_r\rangle |\psi_{B,r}\rangle. 
\end{equation}
Here, when $\theta_1$ is small, $P(X_{B,r})\sim |\langle X_{B,r}|\psi_{B,r}\rangle|^2$ 
is the probability to obtain the value $X_{B,r}$ by the homodyne measurement. 
We calculate the following increment $(\Delta |\psi_r\rangle)_{WM}$ due to the 
beam-splitter coupling and the homodyne measurement, 
\begin{equation}
(\Delta |\psi_r\rangle)_{WM}\sim \frac{\langle X_{B,r}|e^{\theta_1(\hat{a}_r^{\dagger}\hat{b}_r-\hat{b}_r^{\dagger}\hat{a}_r)}|\psi_r\rangle |\psi_{B,r}\rangle}{\langle X_{B,r}|\psi_{B,r}\rangle}-|\psi_r\rangle. 
\end{equation}
Assuming small $\Delta t$, we expand the exponential up to $O(\theta_1)$.  Here, $\theta_1 \sim -\sqrt{j_1\Delta t}$. 
\begin{equation}
\label{psi_wm1}
(\Delta |\psi_r\rangle)_{WM}\sim - \sqrt{j_1\Delta t}\frac{\langle X_{B,r}|(\hat{a}_r^{\dagger}\hat{b}_r-\hat{b}_r^{\dagger}\hat{a}_r)|\psi_r\rangle |\psi_{B,r}\rangle}{\langle X_{B,r}|\psi_{B,r}\rangle}. 
\end{equation}
Since $\hat{Q}_r$ projects $|\psi_{B,r}\rangle$ to zero, 
$\hat{b}_r$ in Eq.(\ref{psi_wm1}) can be replaced by $\hat{b}_r-\frac{G_j}{2}\hat{Q}_r$, 
\begin{equation}
\hat{b}_r\rightarrow -G_j\frac{1-G_j^{-1}}{2}(\hat{b}_r+\hat{b}_r^{\dagger}). 
\end{equation}
Next, replacing $\hat{b}_r^{\dagger}$ in Eq.(\ref{psi_wm1}) by $\hat{b}_r^{\dagger}+\frac{G_j}{2}\hat{Q}_r$, 
\begin{equation}
\hat{b}_r^{\dagger} \rightarrow G_j\frac{1+G_j^{-1}}{2}(\hat{b}_r+\hat{b}_r^{\dagger}). 
\end{equation}
Hence, we obtain, 
\begin{eqnarray}
\label{psi_wm2}
(\Delta |\psi_r\rangle)_{WM}& \sim &\sqrt{j_1\Delta t}G_j\Bigl(\frac{1-G_j^{-1}}{2}\hat{a}_r^{\dagger}+\frac{1+G_j^{-1}}{2}\hat{a}_r\Bigr)|\psi_r\rangle \nonumber \\
&&\times \frac{\langle X_{B,r}|(\hat{b}_r+\hat{b}_r^{\dagger})|\psi_{B,r}\rangle}{\langle X_{B,r}|\psi_{B,r}\rangle}. 
\end{eqnarray}
Assuming small $j_1\Delta t$, the measured value of $\hat{b}_r+\hat{b}_r^{\dagger}$ is a random number 
following the normal distribution with the standard deviation $\frac{1}{\sqrt{G_j}}$. 
Therefore, the second line of Eq.(\ref{psi_wm2}) can be replaced by $\sqrt{\frac{\Delta t}{G_j}} \dot{W}_r$. 
Here, $\dot{W}_r$ are real random numbers satisfying $\overline{\dot{W}_r(t)\dot{W}_{r'}(t')}=\delta_{rr'}\delta (t-t')$.  
Therefore, 
\begin{equation}
(\Delta |\psi_r\rangle)_{WM}\sim \sqrt{j_1G_j}\dot{W}_r \Bigl(\frac{1-G_j^{-1}}{2}\hat{a}_r^{\dagger}+\frac{1+G_j^{-1}}{2}\hat{a}_r\Bigr)|\psi_r\rangle \Delta t. 
\end{equation}
By turning this to the equation of the density-matrix $|\psi_r\rangle \langle \psi_r|$, and then normalizing it 
we obtain the second line of Eq.(\ref{qmesr}). 

\subsection{Photon-number state expansion of quantum master equation}

Here, we show the photon-number state expansion for the continuous-time QME (Eq.(\ref{qmect})). 
In the numerical simulation of the density matrix, 
we added the additional term depending on $j_{f,r}=\frac{\varepsilon_r^2}{2}\Delta t$ to Eq.(\ref{qmect})\cite{Wiseman93,Inui22,Kiesewetter22}. 
\begin{widetext}
\begin{equation}
\frac{\partial \hat{\rho}}{\partial t}= \Bigl(\frac{\partial \hat{\rho}}{\partial t}\Bigr)_{DOPO}+\Bigl(\frac{\partial \hat{\rho}}{\partial t}\Bigr)_{BS1}+\Bigl(\frac{\partial \hat{\rho}}{\partial t}\Bigr)_{BS2}+\sum_{r=1}^N j_{f,r}[\hat{a}^{\dagger}_r-\hat{a}_r,[\hat{a}^{\dagger}_r-\hat{a}_r,\hat{\rho}]]. 
\end{equation}
\end{widetext}
The last term is given by $O(\Delta t^2)$ expansion of Eq.(\ref{rhotf}), 
where we approximated the feedback mode $\hat{z}_{r}$ to be a constant coherent amplitude $\zeta_{r}$, 
$\theta_2 \hat{z}_{r}\rightarrow \theta_2 \zeta_{r}\sim \varepsilon_r \Delta t$\cite{Paris96}. 
The density-matrix of the $r$-th DOPO $\hat{\rho}^{(r)}$ ($\hat{\rho}=\otimes_{r=1}^N \hat{\rho}^{(r)}$) 
can be expanded using photon-number states $|N_r\rangle (N_r=0,1,\cdots)$, as $\hat{\rho}^{(r)}=\sum_{N_r,N_r'}\rho_{N_r,N_r'}^{(r)}|N_r\rangle\langle N_r'|$. 
The density-matrix element $\rho^{(r)}_{N_r,N_r'}$ follows: 
\begin{widetext}
\begin{eqnarray}
\label{qmectpns}
\frac{\partial \rho^{(r)}_{N_r,N_r'}}{\partial t}&=& 2(1+j+j_{fr}')\sqrt{(N_r+1)(N_r'+1)}\rho^{(r)}_{N_r+1,N_r'+1}-(1+j+j_{fr}')(N_r+N_r')\rho^{(r)}_{N_r,N_r'} \nonumber \\
&+&2j_{fr}'\sqrt{N_rN_r'}\rho^{(r)}_{N_r-1,N_r'-1}-j_{fr}'(N_r+N_r'+2)\rho^{(r)}_{N_r,N_r'}\nonumber \\
&-&2j_{fr}''\sqrt{(N_r+1)N_r'}\rho^{(r)}_{N_r+1,N_r'-1}+j_{fr}''\sqrt{(N_r+1)(N_r+2)}\rho^{(r)}_{N_r+2,N_r'}+j_{fr}''\sqrt{N_r'(N_r'-1)}\rho^{(r)}_{N_r,N_r'-2}\nonumber \\
&-&2j_{fr}''\sqrt{N_r(N_r'+1)}\rho^{(r)}_{N_r-1,N_r'+1}+j_{fr}''\sqrt{(N_r'+1)(N_r'+2)}\rho^{(r)}_{N_r,N_r'+2}+j_{fr}''\sqrt{N_r(N_r-1)}\rho^{(r)}_{N_r-2,N_r'}\nonumber \\
&+& g^2 \sqrt{(N_r+1)(N_r+2)(N_r'+1)(N_r'+2)}\rho^{(r)}_{N_r+2,N_r'+2}-\frac{g^2}{2}[N_r(N_r-1)+N_r'(N_r'-1)]\rho^{(r)}_{N_r,N_r'} \nonumber \\
&+& \varepsilon_r(\sqrt{N_r}\rho^{(r)}_{N_r-1,N_r'}+\sqrt{N_r'}\rho^{(r)}_{N_r,N_r'-1}-\sqrt{N_r+1}\rho^{(r)}_{N_r+1,N_r'}-\sqrt{N_r'+1}\rho^{(r)}_{N_r,N_r'+1}) \nonumber \\
&+& \frac{p}{2}(\sqrt{N_r(N_r-1)}\rho^{(r)}_{N_r-2,N_r'}+\sqrt{N_r'(N_r'-1)}\rho^{(r)}_{N_r,N_r'-2}) \nonumber \\
&-&\frac{p}{2}(\sqrt{(N_r+1)(N_r+2)}\rho^{(r)}_{N_r+2,N_r'}+\sqrt{(N_r'+1)(N_r'+2)}\rho^{(r)}_{N_r,N_r'+2}) \nonumber \\
&+& \sqrt{j_1G_j}\frac{1+G_j^{-1}}{2}\dot{W}_r(\sqrt{N_r+1}\rho^{(r)}_{N_r+1,N_r'}+\sqrt{N_r'+1}\rho^{(r)}_{N_r,N_r'+1})\nonumber \\
&+& \sqrt{j_1G_j}\frac{1-G_j^{-1}}{2}\dot{W}_r(\sqrt{N_r}\rho^{(r)}_{N_r-1,N_r'}+\sqrt{N_r'}\rho^{(r)}_{N_r,N_r'-1})- \sqrt{j_1G_j}\dot{W}_r\langle \hat{a}_r+\hat{a}_r^{\dagger}\rangle\rho^{(r)}_{N_r,N_r'}.
\end{eqnarray}
Here, $j_{fr}'=j_{fr}+\frac{j_1}{2} n_j$, $j_{fr}''=j_{fr}-\frac{j_1}{2} m_j$. 
In the numerical simulation, we set the maximum photon number $N_M$ to $|N_r\rangle(N_r=0,1,\cdots,N_M)$. 
We neglected the components on the R.H.S. of Eq.(\ref{qmectpns}) that have negative indices or positive indices greater than $N_M$. 

\setcounter{equation}{0}
\renewcommand{\theequation}{B\arabic{equation}}

\section{Continuous-time skew-Gaussian model}

Here, we derive the continuous-time skew-Gaussian model. 
The quantum master equation (QME) for the density matrix $\hat{\rho}$ consists of three parts 
$(\frac{\partial \hat{\rho}}{\partial t})_{DOPO}+(\frac{\partial \hat{\rho}}{\partial t})_{BS1}+(\frac{\partial \hat{\rho}}{\partial t})_{BS2}$ (Eq.(\ref{qmect})). 
The first one $(\frac{\partial \hat{\rho}}{\partial t})_{DOPO}$, detailed as Eq.(\ref{qmedopo}), describes the effects of the DOPO, 
and the others represent the coupling. 
In Appendix B.1, we derive the equations of the skew-Gaussian model for the DOPO part. 
The equations derived are summarized below as Eqs.(\ref{sk1}-\ref{sk5}), 
\begin{equation}
\label{sk1}
\Bigl(\frac{d\mu_r}{dt}\Bigr)_{DOPO}=-(1-p)\mu_r-g^2(\mu_r^2+2n_r+m_r)\mu_r-g^2\kappa_r , 
\end{equation}
\begin{eqnarray}
\label{sk2}
\Bigl(\frac{d m_r}{dt}\Bigr)_{DOPO}&=&-2m_r+2p n_r-2g^2 \mu_r^2 (2m_r+n_r)+p-g^2(\mu_r^2+m_r) \nonumber \\ 
&-& 6g^2n_rm_r-2g^2\mu_r(\gamma_r+2\kappa_r) ,
\end{eqnarray}
\begin{eqnarray}
\label{sk3}
\Bigl(\frac{d n_r}{dt}\Bigr)_{DOPO}=-2n_r+2 pm_r-2g^2\mu_r^2(2n_r+m_r)- 2g^2(m_r^2+2n_r^2)-6g^2\mu_r \kappa_r ,
\end{eqnarray}
\begin{eqnarray}
\label{sk4}
\Bigl(\frac{d\gamma_r}{dt}\Bigr)_{DOPO}&=&-3(1+g^2)\gamma_r+3p \kappa_r-3g^2\mu_r^2 (2\gamma_r+\kappa_r) \nonumber \\
&-&6g^2\mu_r ( m_r+ m_r^2+2 m_rn_r )-3g^2(4n_r\gamma_r+5m_r\kappa_r), 
\end{eqnarray}
\begin{eqnarray}
\label{sk5}
\Bigl(\frac{d\kappa_r}{dt}\Bigr)_{DOPO}&=&-(3-2p+g^2)\kappa_r+p\gamma_r -g^2\mu_r^2 (\gamma_r +8\kappa_r) \nonumber \\
&-& 2g^2\mu_r (n_r +2m_r^2+4n_rm_r+3n_r^2)-g^2(3m_r\gamma_r+8m_r\kappa_r+16n_r\kappa_r). 
\end{eqnarray} 
Next, in Appendix B.2, we derive the equations of the skew-Gaussian model for the coupling part, 
$(\frac{\partial \hat{\rho}}{\partial t})_{BS1}+(\frac{\partial \hat{\rho}}{\partial t})_{BS2}$. 
The full equations Eqs.(\ref{ctsk1}-\ref{ctsk5}) of the continuous-time skew-Gaussian model are obtained by combining the DOPO part and the coupling part. 

\subsection{For an isolated DOPO}

We start with the expansion of the density-matrix $\hat{\rho}^{(r)}$ for the $r$-th optical pulse 
by using the positive-$P$ quasi-distribution function $P^{(r)}(\alpha_r,\alpha_r^+)$ \cite{Drummond80,Takata15}, 
\begin{equation}
\hat{\rho}^{(r)}=\int P^{(r)}(\alpha_r,\alpha^{+}_r) \frac{|\alpha_r \rangle \langle \alpha_r^{+ *}|}{\langle \alpha_r^{+ *}|\alpha_r\rangle}d^2\alpha_r d^2 \alpha^{+}_r. 
\end{equation} 
From the quantum master equation (\ref{qmedopo}) for the DOPO part, the Fokker-Planck equation of the quasi-distribution function $P^{(r)}(\alpha_r,\alpha^{+}_r)$ is obtained as, 
\begin{equation}
\Bigl(\frac{\partial P^{(r)}}{\partial t}\Bigr)_{DOPO}=\frac{\partial}{\partial \alpha_r}(\alpha_r-p\alpha_r^{+}+g^2\alpha_r^{+}\alpha_r^2) P^{(r)}+\frac{1}{2}\frac{\partial^2}{\partial \alpha_r^2}(p-g^2\alpha_r^2)P^{(r)}+{\rm H.c.}.
\end{equation}
Here, ${\rm H.c.}$ means the first and second terms on the R.H.S. with the replacement 
$(\alpha_r,\alpha_r^+)\rightarrow (\alpha_r^+,\alpha_r)$. 
This Fokker-Planck equation can be rewritten as equivalent $c$-number stochastic differential equations, 
\begin{equation}
\label{a1mf}
\Bigl(\frac{d\alpha_r}{dt}\Bigr)_{DOPO}=-\alpha_r+p\alpha_r^{+}-g^2\alpha_r^{+}\alpha_r^2+\sqrt{p-g^2\alpha_r^2}\xi_{R,r}, 
\end{equation}
\begin{equation}
\Bigl(\frac{d\alpha_r^{+}}{dt}\Bigr)_{DOPO}=-\alpha_r^{+}+p\alpha_r-g^2\alpha_r^{+ 2}\alpha_r+\sqrt{p-g^2\alpha_r^{+ 2}}\xi^{+}_{R,r}.
\end{equation}
Here, $\xi_{R,r}$ and $\xi_{R,r}^{+}$ are independent real random numbers satisfying 
$\langle \xi_{R,r}(t)\xi_{R,r'}(t')\rangle =\delta_{rr'}\delta (t-t')$, $\langle \xi_{R,r}^{+}(t)\xi_{R,r'}^{+}(t')\rangle=\delta_{rr'}\delta(t-t')$. 
We take the average $\langle \alpha_r\rangle=\int \alpha_rP^{(r)}(\alpha_r,\alpha_r^+)d^2\alpha_rd^2\alpha_r^+$ 
of Eq.(\ref{a1mf}) and separate the fluctuation part, $\delta \alpha_r=\alpha_r-\langle \alpha_r\rangle$ from the mean field part. 
We also introduce $\delta \alpha_r^{+}=\alpha_r^{+}-\langle \alpha_r^{+}\rangle$ for its conjugate. 
Assuming $\langle \delta \alpha_r\rangle=\langle \delta \alpha_r^+\rangle=0$, we obtain 
\begin{eqnarray}
\label{meandadt}
\Bigl(\frac{d \langle \alpha_r\rangle}{dt}\Bigr)_{DOPO}&=&-\langle \alpha_r\rangle +p \langle \alpha_r^{+}\rangle-g^2\langle \alpha_r^{+}\rangle \langle \alpha_r\rangle^2 \nonumber \\ 
&-& g^2 \langle \delta \alpha^{+}_r\delta \alpha_r^2\rangle-2g^2\mu_r \langle \delta \alpha^{+}_r\delta \alpha_r\rangle-g^2\mu_r \langle \delta \alpha_r^2\rangle. 
\end{eqnarray}
This equation is equivalent to Eq.(\ref{sk1}) since 
$\mu_r=\langle \alpha_r\rangle=\langle \alpha_r^{+}\rangle$, 
$m_r=\langle \delta \alpha_r^2\rangle=\langle \delta \alpha_r^{+ 2}\rangle$, 
$n_r=\langle \delta \alpha_r^{+}\delta \alpha_r\rangle$, 
$\gamma_r=\langle \delta \alpha_r^3\rangle=\langle \delta \alpha_r^{+ 3}\rangle$, and 
$\kappa_r=\langle \delta \alpha_r^{+} \delta \alpha_r^2\rangle=\langle \delta \alpha_r^{+ 2} \delta \alpha_r\rangle$. 

The fluctuation parts, $\delta \alpha_r$ and $\delta \alpha_r^{+}$ follow the $c$-number stochastic differential equations, 
\begin{eqnarray}
\label{da1}
\Bigl(\frac{d \delta \alpha_r}{dt}\Bigr)_{DOPO}&=&-\delta \alpha_r+p \delta \alpha_r^{+}-2g^2 \mu_r^2\delta \alpha_r-g^2 \mu_r^2 \delta \alpha_r^{+}+\sqrt{p-g^2\mu_r^2-2g^2\mu_r\delta \alpha_r-g^2\delta \alpha_r^2}\xi_{R,r} \nonumber \\ 
&-& g^2(\delta \alpha^{+}_r\delta \alpha_r^2- \langle \delta \alpha^{+}_r\delta \alpha_r^2\rangle)-2g^2\mu_r(\delta \alpha^{+}_r\delta \alpha_r- \langle \delta \alpha^{+}_r\delta \alpha_r\rangle)-g^2\mu_r(\delta \alpha_r^2- \langle \delta \alpha_r^2\rangle), 
\end{eqnarray}
\begin{eqnarray}
\label{daT1}
\Bigl(\frac{d \delta \alpha_r^{+}}{dt}\Bigr)_{DOPO}&=&-\delta \alpha_r^{+}+p \delta \alpha_r-2g^2 \mu_r^2\delta \alpha_r^{+}-g^2 \mu_r^2 \delta \alpha_r+\sqrt{p-g^2\mu_r^2-2g^2\mu_r\delta \alpha_r^{+}-g^2\delta \alpha_r^{+ 2}}\xi_{R,r}^{+} \nonumber \\ 
&-& g^2(\delta \alpha^{+ 2}_r\delta \alpha_r- \langle \delta \alpha^{+ 2}_r\delta \alpha_r\rangle)-2g^2\mu_r(\delta \alpha^{+}_r\delta \alpha_r- \langle \delta \alpha^{+}_r\delta \alpha_r\rangle)-g^2\mu_r(\delta \alpha_r^{+ 2}- \langle \delta \alpha_r^{+ 2}\rangle). 
\end{eqnarray}
From Eq.(\ref{da1}), the second-order fluctuation $\delta \alpha_r^2$ obeys, 
\begin{eqnarray}
\label{da2}
\Bigl(\frac{d \delta \alpha_r^2}{dt}\Bigr)_{DOPO}&=&-2\delta \alpha_r^2+2p \delta \alpha_r^{+}\delta \alpha_r-4g^2 \mu_r^2\delta \alpha_r^2-2g^2 \mu_r^2 \delta \alpha_r^{+}\delta \alpha_r+p-g^2\mu_r^2-2g^2\mu_r\delta \alpha_r-g^2\delta \alpha_r^2 \nonumber \\ 
&+& 2\delta \alpha_r \sqrt{p-g^2\mu_r^2-2g^2\mu_r\delta \alpha_r-g^2\delta \alpha_r^2}\xi_{R,r} \\ 
&-& 2g^2(\delta \alpha^{+}_r\delta \alpha_r^3- \delta \alpha_r \langle \delta \alpha^{+}_r\delta \alpha_r^2\rangle)-4g^2\mu_r(\delta \alpha^{+}_r\delta \alpha_r^2- \delta \alpha_r \langle \delta \alpha^{+}_r\delta \alpha_r\rangle)-2g^2\mu_r(\delta \alpha_r^3- \delta \alpha_r \langle \delta \alpha_r^2\rangle). \nonumber 
\end{eqnarray}
By taking the mean value of this equation, we obtain 
\begin{eqnarray}
\label{da2mf}
\Bigl(\frac{d \langle \delta \alpha_r^2\rangle }{dt}\Bigr)_{DOPO}&=&-2 \langle \delta \alpha_r^2\rangle+2p \langle \delta \alpha_r^{+}\delta \alpha_r\rangle -4g^2 \mu_r^2\langle \delta \alpha_r^2\rangle-2g^2 \mu_r^2 \langle \delta \alpha_r^{+}\delta \alpha_r\rangle+p-g^2\mu_r^2-g^2\langle \delta \alpha_r^2\rangle \nonumber \\ 
&-& 2g^2 \langle \delta \alpha^{+}_r\delta \alpha_r^3 \rangle -4g^2\mu_r \langle \delta \alpha^{+}_r\delta \alpha_r^2 \rangle -2g^2\mu_r \langle \delta \alpha_r^3\rangle. 
\end{eqnarray}
In a similar way, we obtain the equation for the mean fluctuation product $\langle \delta \alpha_r^{+}\delta \alpha_r\rangle$, 
\begin{eqnarray}
\label{daTdamf}
\Bigl(\frac{d \langle \delta \alpha_r^{+}\delta \alpha_r\rangle }{dt}\Bigr)_{DOPO}&=&-2 \langle \delta \alpha_r^{+}\delta \alpha_r \rangle+p \langle \delta \alpha_r^{+ 2}\rangle+p\langle \delta \alpha_r^2 \rangle -4g^2 \mu_r^2\langle \delta \alpha_r^{+}\delta \alpha_r \rangle-g^2 \mu_r^2 \langle \delta \alpha_r^{+ 2}\rangle -g^2\mu_r^2 \langle \delta \alpha_r^2\rangle \nonumber \\ 
&-& 2g^2 \langle \delta \alpha^{+ 2}_r \delta \alpha_r^2 \rangle -3g^2\mu_r \langle \delta \alpha^{+ 2}_r\delta \alpha_r \rangle -3g^2\mu_r \langle \delta \alpha_r^{+} \delta \alpha_r^2\rangle. 
\end{eqnarray}
For these two equations, we decompose the fourth-order moments into $\langle \delta \alpha_r^{+} \delta \alpha_r^3\rangle=3\langle \delta \alpha_r^{+}\delta \alpha_r \rangle \langle \delta \alpha_r^2\rangle$, and 
$\langle \delta \alpha_r^{+ 2}\delta \alpha_r^2\rangle=2\langle \delta \alpha_r^{+}\delta \alpha_r\rangle^2+\langle \delta \alpha_r^{+ 2}\rangle \langle \delta \alpha_r^2\rangle$. 
The coefficients show the degrees of freedom for decomposition\cite{Ng23}. 
After this procedure, these equations Eqs.(\ref{da2mf}) and (\ref{daTdamf}) are equivalent to Eqs.(\ref{sk2}) and (\ref{sk3}). 
Next, from Eq.(\ref{da1}) and Eq.(\ref{da2}), the mean of the third-order fluctuation product $\langle \delta \alpha_r^3\rangle$ follows 
\begin{eqnarray}
\Bigl(\frac{d \langle \delta \alpha_r^3\rangle}{dt}\Bigr)_{DOPO}&=&-3(1+g^2)\langle \delta \alpha_r^3\rangle+3p \langle \delta \alpha_r^{+}\delta \alpha_r^2\rangle-6g^2 \mu_r^2\langle \delta \alpha_r^3\rangle-3g^2 \mu_r^2 \langle \delta \alpha_r^{+}\delta \alpha_r^2\rangle-6g^2\mu_r\langle \delta \alpha_r^2\rangle \\ 
&-& 3g^2(\langle \delta \alpha^{+}_r\delta \alpha_r^4\rangle- \langle \delta \alpha^{+}_r\delta \alpha_r^2\rangle\langle \delta \alpha_r^2\rangle)-6g^2\mu_r(\langle \delta \alpha^{+}_r\delta \alpha_r^3\rangle- \langle \delta \alpha^{+}_r\delta \alpha_r\rangle\langle \delta \alpha^2_r\rangle)-3g^2\mu_r(\langle \delta \alpha_r^4\rangle- \langle \delta \alpha_r^2\rangle^2). \nonumber
\end{eqnarray}
Here, we decompose the fourth-order and fifth-order moments into 
$\langle \delta \alpha_r^4\rangle=3\langle \delta \alpha_r^2\rangle^2$, and 
$\langle \delta \alpha_r^{+} \delta \alpha_r^4\rangle=6\langle \delta \alpha_r^{+}\delta \alpha_r^2 \rangle \langle \delta \alpha_r^2\rangle+4\langle \delta \alpha_r^{+}\delta \alpha_r \rangle \langle \delta \alpha_r^3 \rangle$, 
and obtain Eq.(\ref{sk4}). 
From Eq.(\ref{daT1}) and Eq.(\ref{da2}), the mean of the third-order fluctuation product $\langle \delta \alpha_r^{+}\delta \alpha_r^2\rangle$ follows 
\begin{eqnarray}
& &\Bigl(\frac{d \langle \delta \alpha_r^{+} \delta \alpha_r^2\rangle}{dt}\Bigr)_{DOPO} \nonumber \\
&=&-(3+g^2)\langle \delta \alpha_r^{+}\delta \alpha_r^2\rangle+2p \langle \delta \alpha_r^{+ 2}\delta \alpha_r\rangle+p\langle \delta \alpha_r^3\rangle-6g^2 \mu_r^2\langle \delta \alpha_r^{+}\delta \alpha_r^2\rangle-2g^2 \mu_r^2 \langle \delta \alpha_r^{+ 2}\delta \alpha_r\rangle-g^2\mu_r^2\langle \delta \alpha_r^3\rangle-2g^2\mu_r\langle \delta \alpha_r^{+} \delta \alpha_r\rangle \nonumber \\ 
&-& g^2(\langle \delta \alpha^{+ 2}_r\delta \alpha_r^3\rangle- \langle \delta \alpha^{+ 2}_r\delta \alpha_r\rangle\langle \delta \alpha_r^2\rangle)-4g^2\mu_r(\langle \delta \alpha^{+}_r\delta \alpha_r^3\rangle- \langle \delta \alpha^{+}_r\delta \alpha_r\rangle\langle \delta \alpha^2_r\rangle)-g^2\mu_r(\langle \delta \alpha_r^{+ 2}\delta \alpha_r^2\rangle- \langle \delta \alpha_r^{+ 2}\rangle \langle \delta \alpha_r^2\rangle) \nonumber \\ 
&-& 2g^2(\langle \delta \alpha^{+ 2}_r\delta \alpha_r^3\rangle- \langle \delta \alpha^{+}_r\delta \alpha_r^2\rangle\langle\delta \alpha_r^{+} \delta \alpha_r\rangle)-4g^2\mu_r(\langle \delta \alpha^{+ 2}_r \delta \alpha_r^2\rangle- \langle \delta \alpha_r^{+} \delta \alpha_r\rangle^2). 
\end{eqnarray}
\end{widetext}
Here, we decompose the fourth- and fifth-order moments into $\langle \delta \alpha_r^{+ 2}\delta \alpha_r^3\rangle=6\langle \delta \alpha_r^{+}\delta \alpha_r^2\rangle \langle \delta \alpha_r^{+}\delta \alpha_r\rangle+3\langle \delta \alpha_r^{+ 2}\delta \alpha_r\rangle\langle \delta \alpha_r^2\rangle+\langle \delta \alpha_r^{+ 2}\rangle \langle \delta \alpha_r^3\rangle$, 
and obtain Eq.(\ref{sk5}). 

\subsection{Non-Gaussian correction to mutual coupling}

\renewcommand{\theequation}{B\arabic{equation}}

Here, we derive the coupling part of the skew-Gaussian equations. 
We rewrite the mutual coupling terms of the quantum master equation (\ref{qmect}), 
$(\frac{\partial \hat{\rho}}{\partial t})_{BS1}+(\frac{\partial \hat{\rho}}{\partial t})_{BS2}$, 
into a Gaussian part $(\frac{\partial \hat{\rho}}{\partial t})_{G}$ consisting of the linear loss, coherent injection and squeezed light injection 
and a non-Gaussian part $(\frac{\partial \hat{\rho}}{\partial t})_{WM}$ representing Wiseman-Milburn's measurement-induced state-reduction, 
\begin{equation}
\label{qmebs12}
\Bigl(\frac{\partial \hat{\rho}}{\partial t}\Bigr)_{BS1}+\Bigl(\frac{\partial \hat{\rho}}{\partial t}\Bigr)_{BS2}=\Bigl(\frac{\partial \hat{\rho}}{\partial t}\Bigr)_{G}+\Bigl(\frac{\partial \hat{\rho}}{\partial t}\Bigr)_{WM}, 
\end{equation}
\begin{widetext}
\begin{eqnarray}
\Bigl(\frac{\partial \hat{\rho}}{\partial t}\Bigr)_{G}&=&j\sum_r([\hat{a}_r,\hat{\rho}\hat{a}_r^{\dagger}]+{\rm H.c.})+\sum_r \varepsilon_r[\hat{a}_r^{\dagger}-\hat{a}_r,\hat{\rho}], \nonumber \\
&+&j_1n_j\sum_r[\hat{a}_r,[\hat{\rho},\hat{a}_r^{\dagger}]]-\frac{j_1}{2}m_j\sum_r([\hat{a}_r,[\hat{a}_r,\hat{\rho}]]+{\rm H.c.}), 
\end{eqnarray}
\begin{eqnarray}
\label{wmqme}
\Bigl(\frac{\partial \hat{\rho}}{\partial t}\Bigr)_{WM} &=& \sqrt{j_1G_j}\frac{1+G_j^{-1}}{2}\sum_{r}(\hat{a}_{r}\hat{\rho}+\hat{\rho}\hat{a}^{\dagger}_{r}-\langle \hat{a}_r+\hat{a}^{\dagger}_r\rangle\hat{\rho})\dot{W}_r\nonumber \\
&+& \sqrt{j_1G_j}\frac{1-G_j^{-1}}{2}\sum_{r}(\hat{a}_{r}^{\dagger}\hat{\rho}+\hat{\rho}\hat{a}_{r}-\langle \hat{a}_r+\hat{a}^{\dagger}_r\rangle\hat{\rho})\dot{W}_r. 
\end{eqnarray}
\end{widetext}
The skew-Gaussian equations related to $(\frac{\partial \hat{\rho}}{\partial t})_{G}$ are relatively simple. 
Following the same method as was used to derive the skew-Gaussian equations for the DOPO part, 
we obtain $c$-number stochastic differential equations, 
which is equivalent to the Fokker Planck equation of the positive-$P$ quasi-distribution function. 
\begin{equation}
\Bigl(\frac{d\alpha_r}{dt}\Bigr)_G=-j\alpha_r+\varepsilon_r+\sqrt{\frac{j_1}{2}n_j}\xi_{C,r}+i\sqrt{j_1m_j}\xi_{R,r}, 
\end{equation}
\begin{equation}
\Bigl(\frac{d\alpha_r^+}{dt}\Bigr)_G=-j\alpha_r^+ +\varepsilon_r+\sqrt{\frac{j_1}{2}n_j}\xi_{C,r}^*-i\sqrt{j_1m_j}\xi_{R,r}^+. 
\end{equation}
Here, $\xi_{R,r}$ and $\xi_{R,r}^+$ are independent real random numbers, and 
$\xi_{C,r}$ are complex random numbers satisfying $\langle \xi_{C,r}^*(t)\xi_{C,r'}(t')\rangle=2\delta_{rr'}\delta (t-t')$ 
and $\langle \xi_{C,r}(t)\xi_{C,r'}(t')\rangle=0$. 
We obtain the following equations for $\mu_r=\langle \alpha_r\rangle$, 
$m_r=\langle \delta \alpha_r^2\rangle$, $n_r=\langle \delta \alpha_r^+ \delta \alpha_r\rangle$, 
$\gamma_r=\langle \delta \alpha_r^3\rangle$, and $\kappa_r=\langle \delta \alpha_r^+\delta \alpha_r^2\rangle$. 
\begin{equation}
\label{skg1}
\Bigl(\frac{d\mu_r}{dt}\Bigr)_{G}=-j\mu_r+\varepsilon_r, 
\end{equation}
\begin{equation}
\label{skg2}
\Bigl(\frac{dm_r}{dt}\Bigr)_{G}=-2jm_r+\frac{j_1}{4}(G_j^{-1}-G_j), 
\end{equation}
\begin{equation}
\label{skg3}
\Bigl(\frac{dn_r}{dt}\Bigr)_{G}=-2jn_r+\frac{j_1}{4}(G_j+G_j^{-1}-2), 
\end{equation}
\begin{equation}
\label{skg4}
\Bigl(\frac{d\gamma_r}{dt}\Bigr)_{G}=-3j\gamma_r, 
\end{equation}

\begin{equation}
\label{skg5}
\Bigl(\frac{d\kappa_r}{dt}\Bigr)_{G}=-3j\kappa_r. 
\end{equation}

Next, we derive the skew-Gaussian equations for $(\frac{\partial \hat{\rho}}{\partial t})_{WM}$. 
We obtain the time-development equation of $\mu_r=\langle \hat{a}_r\rangle={\rm Tr}\hat{\rho}\hat{a}_r$ from Eq.(\ref{wmqme}), as 
\begin{equation}
\label{skwm1}
\Bigl(\frac{d\mu_r}{dt}\Bigr)_{WM}=\sqrt{j_1G_j}\dot{W}_r V_r'. 
\end{equation}
Here, $V_r'=\langle \delta \hat{X}_r^2\rangle-\frac{1}{2G_j}$. 
This represents the mean amplitude shift by the indirect homodyne measurement. 
Eq.(\ref{ctsk1}) is obtained by combining Eqs.(\ref{sk1}), (\ref{skg1}), and (\ref{skwm1}). 
Assuming $\dot{W}_r(t)\dot{W}_{r'}(t')\sim \delta_{rr'}\delta(t-t')$, the products of the mean amplitudes follow 
\begin{equation}
\Bigl(\frac{d\mu_r^2}{dt}\Bigr)_{WM}=2\sqrt{j_1G_j}\dot{W}_r V_r'\mu_r+j_1G_jV_r'^2, 
\end{equation}
\begin{equation}
\label{a1WM3}
\Bigl(\frac{d\mu_r^3}{dt}\Bigr)_{WM}=3\sqrt{j_1G_j}\dot{W}_r V_r'\mu_r^2+3j_1G_j\mu_rV_r'^2. 
\end{equation}
On the other hand, the mean of the squared amplitude $\langle \alpha_r^2\rangle={\rm Tr} \hat{\rho}\hat{a}_r^2$ follows 
\begin{widetext}
\begin{eqnarray}
\Bigl(\frac{d\langle \alpha_r^2\rangle}{dt}\Bigr)_{WM}&=&\sqrt{j_1G_j}\dot{W}_r (\langle \alpha_r^3\rangle+\langle \alpha_r^{+}\alpha_r^2\rangle-2\mu_r\langle \alpha_r^2\rangle+\mu_r(1-G_j^{-1}))\nonumber \\
&=&\sqrt{j_1G_j}\dot{W}_r (2\mu_r V_r'+\langle \alpha_r^3\rangle+\langle \alpha_r^{+}\alpha_r^2\rangle+4\mu_r^3-4\mu_r\langle \alpha_r^2\rangle-2\mu_r\langle \alpha_r^{+}\alpha_r\rangle). 
\end{eqnarray}
Here, we have used $(1-G_j^{-1})=2 V_r'-2\langle \alpha_r^2\rangle-2\langle \alpha_r^{+}\alpha_r\rangle+4\mu_r^2$. 
The equation of the mean photon number $\langle \alpha_r^+\alpha_r\rangle={\rm Tr}\hat{\rho} \hat{a}_r^{\dagger}\hat{a}_r$ is 
\begin{eqnarray}
\Bigl(\frac{d\langle \alpha_r^{+}\alpha_r\rangle}{dt}\Bigr)_{WM}&=&\sqrt{j_1G_j}\dot{W}_r (2\langle \alpha_r^{+}\alpha_r^2\rangle-2\mu_r\langle \alpha_r^{+}\alpha_r\rangle+\mu_r(1-G_j^{-1}))\nonumber \\
&=&\sqrt{j_1G_j}\dot{W}_r (2\mu_r V_r'+2\langle \alpha_r^{+}\alpha_r^2\rangle+4\mu_r^3-2\mu_r\langle \alpha_r^2\rangle-4\mu_r\langle \alpha_r^{+}\alpha_r\rangle). 
\end{eqnarray}
From these equations, the Gaussian variances $\langle \delta \alpha_r^2\rangle=\langle \alpha_r^2\rangle-\langle \alpha_r\rangle^2$ 
and $\langle \delta \alpha_r^+ \delta \alpha_r\rangle=\langle \alpha_r^+ \alpha_r\rangle-\langle \alpha_r\rangle^2$, 
are affected by the homodyne measurement as follows, 
\begin{equation}
\Bigl(\frac{d\langle \delta \alpha_r^2\rangle}{dt}\Bigr)_{WM}=\sqrt{j_1G_j}\dot{W}_r (\langle \alpha_r^3\rangle+\langle \alpha_r^{+}\alpha_r^2\rangle+4\mu_r^3-4\mu_r\langle \alpha_r^2\rangle-2\mu_r\langle \alpha_r^{+}\alpha_r\rangle)-j_1G_jV_r'^2, 
\end{equation}
\begin{equation}
\Bigl(\frac{d\langle \delta \alpha_r^{+}\delta \alpha_r \rangle}{dt}\Bigr)_{WM}=\sqrt{j_1G_j}\dot{W}_r (2\langle \alpha_r^{+}\alpha_r^2\rangle+4\mu_r^3-2\mu_r\langle \alpha_r^2\rangle-4\mu_r\langle \alpha_r^{+}\alpha_r\rangle)-j_1G_jV_r'^2. 
\end{equation}
By decomposing the mean of the amplitude products using $\alpha_r=\mu_r+\delta \alpha_r$ and $\alpha_r^{+}=\mu_r+\delta \alpha_r^{+}$, 
the variances are affected by the indirect homodyne measurement as 
\begin{equation}
\label{skwm2}
\Bigl(\frac{d m_r}{dt}\Bigr)_{WM}=-j_1G_j V_r'^2+\sqrt{j_1G_j} \dot{W}_r (\gamma_r+\kappa_r), 
\end{equation}
\begin{equation}
\label{skwm3}
\Bigl(\frac{d n_r}{dt}\Bigr)_{WM}=-j_1G_j V_r'^2+\sqrt{j_1G_j} \dot{W}_r (2\kappa_r). 
\end{equation}
Eqs.(\ref{ctsk2}) and (\ref{ctsk3}) are obtained by using Eqs.(\ref{sk2}), (\ref{skg2}), and (\ref{skwm2}) 
and Eqs.(\ref{sk3}), (\ref{skg3}), and (\ref{skwm3}), respectively. 

Next, we derive the equations of the third-order fluctuation products. 
The third-order moment $\langle \alpha_r^3\rangle$ follows 
\begin{eqnarray}
\label{a3WM}
\Bigl(\frac{d\langle \alpha_r^3\rangle}{dt}\Bigr)_{WM}&=&\sqrt{j_1G_j}\dot{W}_r (\langle \alpha_r^4\rangle+\langle \alpha_r^{+}\alpha_r^3\rangle-2\mu_r\langle \alpha_r^3\rangle+\frac{3}{2}\langle \alpha_r^2\rangle (1-G_j^{-1}))\nonumber \\
&=&\sqrt{j_1G_j}\dot{W}_r (3\langle \alpha_r^2\rangle V_r'+\langle \alpha_r^4\rangle+\langle \alpha_r^{+}\alpha_r^3\rangle+6\mu_r^2\langle \alpha_r^2\rangle-2\mu_r\langle \alpha_r^3\rangle-3\langle \alpha_r^2\rangle^2-3\langle \alpha_r^2\rangle \langle \alpha_r^{+}\alpha_r\rangle). 
\end{eqnarray}
The skew of the amplitude is $\langle \delta \alpha_r^3\rangle=\langle \alpha_r^3\rangle-3\langle \alpha_r^2\rangle \langle \alpha_r\rangle+2\langle \alpha_r\rangle^3$, 
which is obtained from Eq.(\ref{a1WM3}), Eq.(\ref{a3WM}), and 
\begin{eqnarray}
\label{a2a1WM}
\Bigl(\frac{d\langle \alpha_r^2\rangle \langle \alpha_r\rangle}{dt}\Bigr)_{WM}&=&\sqrt{j_1G_j}\dot{W}_r (\langle \alpha_r^2\rangle V_r'+2\mu_r^2 V_r'+\mu_r \langle \alpha_r^3\rangle+\mu_r \langle \alpha_r^{+}\alpha_r^2\rangle+4\mu_r^4-4\mu_r^2\langle \alpha_r^2\rangle-2\mu_r^2\langle \alpha_r^{+}\alpha_r\rangle)\nonumber \\
&+&j_1G_jV_r' (2\mu_r V_r'+\langle \alpha_r^3\rangle+\langle \alpha_r^{+}\alpha_r^2\rangle+4\mu_r^3-4\mu_r\langle \alpha_r^2\rangle-2\mu_r\langle \alpha_r^{+}\alpha_r\rangle), 
\end{eqnarray}
as 
\begin{eqnarray}
\Bigl(\frac{d\langle \delta \alpha_r^3\rangle }{dt}\Bigr)_{WM}&=&\sqrt{j_1G_j}\dot{W}_r (\langle \alpha_r^4\rangle+\langle \alpha_r^{+}\alpha_r^3\rangle-5\mu_r\langle \alpha_r^3\rangle-3\mu_r \langle \alpha_r^{+}\alpha_r^2\rangle\nonumber \\
&&-12\mu_r^4+18\mu_r^2\langle \alpha_r^2\rangle+6\mu_r^2\langle \alpha_r^{+}\alpha_r\rangle-3\langle \alpha_r^2\rangle^2-3\langle \alpha_r^2\rangle \langle \alpha_r^{+}\alpha_r\rangle)\nonumber \\
&-&3j_1G_jV_r' (\langle \alpha_r^3\rangle+\langle \alpha_r^{+}\alpha_r^2\rangle+4\mu_r^3-4\mu_r\langle \alpha_r^2\rangle-2\mu_r\langle \alpha_r^{+}\alpha_r\rangle). 
\end{eqnarray}
By decomposing the moments on the R.H.S. using $\alpha_r=\mu_r+\delta\alpha_r$ and $\alpha_r^+=\mu_r+\delta \alpha_r^+$, 
the equation of the skew variable $\gamma_r=\langle \delta \alpha_r^3\rangle$ due to the homodyne measurement is 
\begin{equation}
\label{skwm4}
\Bigl(\frac{d \gamma_r}{dt}\Bigr)_{WM}=-3j_1G_j V_r'(\kappa_r+\gamma_r). 
\end{equation}
We obtain Eq.(\ref{ctsk4}) from Eqs.(\ref{sk4})(\ref{skg4}), and (\ref{skwm4}). 

The cross skew of $\alpha_r$ and $\alpha^{+}_r$ is 
$\langle \delta \alpha_r^{+}\delta \alpha_r^2\rangle=\langle \alpha_r^{+}\alpha_r^2\rangle-2\langle \alpha_r^{+}\alpha_r\rangle \langle \alpha_r\rangle-\langle \alpha_r^2\rangle \langle \alpha_r\rangle+2\langle \alpha_r\rangle^3$. 
We use Eq.(\ref{a1WM3}), Eq.(\ref{a2a1WM}), 
\begin{eqnarray}
\Bigl(\frac{d\langle \alpha_r^{+} \alpha_r^2\rangle}{dt}\Bigr)_{WM}&=&\sqrt{j_1G_j}\dot{W}_r (\langle \alpha_r^{+} \alpha_r^3\rangle+\langle \alpha_r^{+ 2}\alpha_r^2\rangle-2\mu_r\langle \alpha_r^{+} \alpha_r^2\rangle+\frac{1}{2}(2\langle \alpha_r^{+}\alpha_r\rangle+\langle \alpha_r^2\rangle) (1-G_j^{-1})) \\
&=&\sqrt{j_1G_j}\dot{W}_r (2\langle \alpha_r^{+}\alpha_r\rangle V_r'+\langle \alpha_r^2\rangle V_r'+\langle \alpha_r^{+} \alpha_r^3\rangle+\langle \alpha_r^{+ 2}\alpha_r^2\rangle+4\mu_r^2\langle \alpha_r^{+}\alpha_r\rangle+2\mu_r^2\langle \alpha_r^2\rangle\nonumber \\
&&-2\mu_r\langle \alpha_r^{+}\alpha_r^2\rangle-\langle \alpha_r^2\rangle^2-3\langle \alpha_r^2\rangle \langle \alpha_r^{+}\alpha_r\rangle-2\langle \alpha_r^{+}\alpha_r\rangle^2), \nonumber
\end{eqnarray}
and 
\begin{eqnarray}
\Bigl(\frac{d\langle \alpha_r^{+}\alpha_r\rangle \langle \alpha_r \rangle }{dt}\Bigr)_{WM}&=&\sqrt{j_1G_j}\dot{W}_r (\langle \alpha_r^{+}\alpha_r\rangle V_r'+2\mu_r^2 V_r'+2\mu_r \langle \alpha_r^{+}\alpha_r^2\rangle+4\mu_r^4-2\mu_r^2\langle \alpha_r^2\rangle-4\mu_r^2\langle \alpha_r^{+}\alpha_r\rangle) \nonumber \\
&+&j_1G_jV_r' (2\mu_r V_r'+2\langle \alpha_r^{+}\alpha_r^2\rangle+4\mu_r^3-2\mu_r\langle \alpha_r^2\rangle-4\mu_r\langle \alpha_r^{+}\alpha_r\rangle), 
\end{eqnarray}
to obtain the equation of motion for the third-order moment $\kappa_r=\langle \delta \alpha_r^{+} \delta \alpha^2_r\rangle$: 
 \begin{eqnarray}
\Bigl(\frac{d\langle \delta \alpha_r^{+} \delta \alpha_r^2\rangle}{dt}\Bigr)_{WM}&=&\sqrt{j_1G_j}\dot{W}_r (\langle \alpha_r^{+} \alpha_r^3\rangle+\langle \alpha_r^{+ 2}\alpha_r^2\rangle-\mu_r\langle \alpha_r^3\rangle-7\mu_r \langle \alpha_r^{+}\alpha_r^2\rangle \nonumber \\
&&-12\mu_r^4+10\mu_r^2\langle \alpha_r^2\rangle+14\mu_r^2\langle \alpha_r^{+}\alpha_r\rangle-\langle \alpha_r^2\rangle^2-3\langle \alpha_r^2\rangle \langle \alpha_r^{+}\alpha_r\rangle-2\langle \alpha_r^{+}\alpha_r\rangle^2) \nonumber \\
&-& j_1G_jV_r' (\langle \alpha_r^3\rangle+5\langle \alpha_r^{+}\alpha_r^2\rangle+12\mu_r^3-8\mu_r\langle \alpha_r^2\rangle-10\mu_r\langle \alpha_r^{+}\alpha_r\rangle). 
\end{eqnarray}
\end{widetext}
By decomposing the moments on the R.H.S. using $\alpha_r=\mu_r+\delta \alpha_r$ and $\alpha_r^{+}=\mu_r+\delta \alpha_r^{+}$, we obtain 
\begin{equation}
\label{skwm5}
\Bigl(\frac{d \kappa_r}{dt}\Bigr)_{WM}=-j_1G_j V_r'(5\kappa_r+\gamma_r). 
\end{equation} 
We obtain Eq.(\ref{ctsk5}) by combining Eqs.(\ref{sk5}), (\ref{skg5}), and (\ref{skwm5}). 
Equations (\ref{skwm4}) and (\ref{skwm5}) represent the effects of the measurement-induced state-reduction on the skew variables. 

\setcounter{equation}{0}
\renewcommand{\theequation}{C\arabic{equation}}

\section{Analytical details on the signs of the skew variables}

Here, we discuss the signs of the skew variables in mean-field coupled DOPOs (Eq.(\ref{qmemfc})), 
where the coherent injection $\varepsilon_r$ depends on the mean amplitude $\mu_r$, as $\varepsilon_r=jG_F\mu_r=(1-p_{thr}+j) \mu_r$. 
We start with the skew-Gaussian equations of a DOPO (Eqs.(\ref{sk1}-\ref{sk5})), 
where the skew variables $\gamma_r=\langle \delta \alpha_r^3\rangle$ and $\kappa_r=\langle \delta \alpha_r^+ \alpha_r^2\rangle$ 
follow Eq.(\ref{sk4}) and Eq.(\ref{sk5}), respectively. 
In Eqs.(\ref{sk4}) and (\ref{sk5}), we assume that $-3g^2\gamma_r$ and $-g^2\kappa_r$ appearing in the first term on the R.H.S. is negligible. 
By introducing the coupling loss $j$ terms (as $-3j\gamma_r$ and $-3j\kappa_r$) and neglecting the products between one of variances 
and one of skew variables $(m_r \gamma_r, m_r \kappa_r, n_r \gamma_r, n_r \kappa_r)$ in Eqs.(\ref{sk4}) and (\ref{sk5}), 
the steady-state skew variables satisfy 
\begin{widetext}
\begin{equation}
\begin{bmatrix} 1+j+2g^2\mu_r^2 & -p+g^2\mu_r^2 \\ -p+g^2\mu_r^2 & 3-2p+3j+8g^2\mu_r^2 \end{bmatrix} \begin{bmatrix} \gamma_r \\ \kappa_r \end{bmatrix}
=-2g^2 \mu_r \begin{bmatrix} m_r+m_r^2+2m_rn_r \\ n_r+2m_r^2+4m_rn_r+3n_r^2 \end{bmatrix}. 
\end{equation}
The steady-state self-skewness $\langle \delta \hat{X}_r^3\rangle=\frac{\gamma_r+3\kappa_r}{\sqrt{2}}$, 
and the cross-skewness $\langle \delta \hat{X}_r\delta \hat{P}_r^2\rangle=\frac{-\gamma_r+\kappa_r}{\sqrt{2}}$ are obtained as 
\begin{equation}
\label{sk_theory}
\langle \delta \hat{X}_r^3\rangle=-\frac{\sqrt{2}g^2\mu_r}{3g^2\mu_r^2+1+j-p}\langle :\delta \hat{X}_r^2:\rangle\Bigl(3\langle :\delta \hat{X}_r^2:\rangle+1\Bigr), 
\end{equation}
\begin{equation}
\label{sk_theory2}
\langle \delta \hat{X}_r\delta \hat{P}_r^2\rangle=-\frac{\sqrt{2}g^2\mu_r}{5g^2\mu_r^2+3+3j+p}\Bigl[2\langle :\delta \hat{P}_r^2:\rangle-\langle:\delta \hat{X}_r^2:\rangle+\langle :\delta \hat{P}_r^2:\rangle\Bigl(2\langle :\delta \hat{X}_r^2:\rangle+\langle :\delta \hat{P}_r^2:\rangle\Bigr)\Bigr]. 
\end{equation}
\end{widetext}
When $\langle \delta \hat{X}_r^2\rangle$ is greater than $\frac{1}{6}$, 
the sign-adjusted self-skewness ${\rm sgn}(\langle \hat{X}_r\rangle)\langle \delta \hat{X}_r^3\rangle$ has 
the opposite sign from the normally ordered variance $\langle :\delta \hat{X}_r^2:\rangle$. 

We checked these analytical results when $p_{thr}=1$ ($G_F=1$), 
where the coherent injection $\varepsilon_r=j\mu_r$ compensates for the coupling loss $-j\mu_r$. 
The steady-state amplitude is $\mu_r=\frac{\sqrt{p-1}}{g}$ 
and the variances are $\langle :\delta \hat{X}_r^2:\rangle=\frac{1}{2(2p-2+j)}$, 
$\langle :\delta \hat{P}_r^2:\rangle=-\frac{1}{2(2p+j)}$\cite{Inui19}. 
The mean amplitude $\mu_r=\langle \hat{a}_r\rangle$ can be calculated with the analytical exact solution of a single DOPO\cite{Drummond81,Milburn81} 
and the self-consistent equation\cite{Boite13}, 
\begin{equation}
\label{scl}
\mu_r=\frac{\sum_{k=0}^{\infty}\frac{-2^k c^{2k+1} {}_2F_1(-k,x+e;2x;2) _2F_1(-k-1,x+e;2x;2)}{k!} }{\sum_{k=0}^{\infty}\frac{2^k c^{2k}{}_2F_1(-k,x+e;2x;2) _2F_1(-k,x+e;2x;2)}{k!} },
\end{equation}
where $c=\frac{\sqrt{p}}{g}$, $x=\frac{1+j}{g^2}$, and $e=\frac{\varepsilon_r}{\sqrt{pg^2}}$.
The generalized moment is obtained from the steady-state $\mu_r$ and 
\begin{widetext}
\begin{equation}
\label{scl2}
\langle \hat{a}_r^{\dagger m}\hat{a}_r^n\rangle=\frac{\sum_{k=0}^{\infty}\frac{2^k (-c)^{k'+k''} {}_2F_1(-k',x+e;2x;2) _2F_1(-k'',x+e;2x;2)}{k!} }{\sum_{k=0}^{\infty}\frac{2^k c^{2k}{}_2F_1(-k,x+e;2x;2) _2F_1(-k,x+e;2x;2)}{k!} }, 
\end{equation}
\end{widetext}
where $k'=k+m$ and $k''=k+n$. 
Figure 12(a) compares the results from Eqs.(\ref{sk_theory}) and (\ref{sk_theory2}) with those from the self-consistent equations Eqs.(\ref{scl}) and (\ref{scl2}), 
for $g^2=0.02$, $j=3$, and $\varepsilon_r=j\mu_r$. 
Starting from $\mu_r=1$, we iterated Eq.(\ref{scl}) $10^3$ times. 
Using the final value of $\mu_r$, and $\varepsilon_r=j\mu_r$, we obtained the skew variables from Eq.(\ref{scl2}). 
The mean canonical coordinate $\langle \hat{X}_r\rangle=\sqrt{2}\mu_r$ and skew variables obtained from Eq.(\ref{scl2}) are shown as circles in Figure 12(a). 
The sign of $\langle \hat{X}_r\rangle$ is always positive and at $p=2$ it approaches $\langle \hat{X}_r\rangle \sim 10$. 
The self-skewness $\langle \delta \hat{X}_r^3\rangle$ has the opposite sign from the mean amplitude, whereas 
the cross-skewness $\langle \delta \hat{X}_r\delta \hat{P}_r^2\rangle$ has the same sign as the mean amplitude, as we saw in Figure 3(a). 
The peak absolute values are reached near $p-1 \sim 0.4$ and $0.7$, for the self- and cross-skewness, respectively. 

In an amplitude-controlled CIM with $\varepsilon_r=je_r \sum_{r'}\tilde{J}_{rr'}\tilde{\mu}_{r'}$, 
even when the internal state experiences a large loss because of a large two-photon absorption $g^2$, 
the squared mean amplitude $\mu_r^2$ is stabilized to a positive value $\tau_0$ (defined in Eq.(\ref{tau0def2}) or Eq.(\ref{tau0def})) by a coherent injection. 
In the large $g^2$ limit, the large $g^2\mu_r^2$ terms determine the fluctuation characteristics. 
By taking the large $g^2\mu_r^2$ limit in Eqs.(\ref{ctsk2}) and (\ref{ctsk3}), we obtain 
$\langle :\delta \hat{X}_r^2:\rangle\rightarrow -\frac{1}{6}$, and $\langle :\delta \hat{P}_r^2:\rangle\rightarrow \frac{1}{2}$ \cite{Chaturvedi77}. 
Here, the signs of the self-skewness $\langle \delta \hat{X}_r^3\rangle$ and cross-skewness $\langle \delta \hat{X}_r\delta \hat{P}_r^2\rangle$ 
are respectively the same as and opposite from the mean amplitude $\mu_r$, following Eqs.(\ref{sk_theory}) and (\ref{sk_theory2}). 
Such a regime in an amplitude controlled CIM is assumed to have a large injection gain $G_F> 1$, or negative $p_{thr}$. 
When $p>p_{thr}$ in a system with $p_{thr}\ne 1$, 
the steady-state amplitude is bifurcated to $\mu_r=\pm\frac{\sqrt{p-p_{thr}}}{g}$ 
and the variances are $\langle :\delta \hat{X}_r^2:\rangle=\frac{p_{thr}}{2(2p+1-3p_{thr}+j)}$, 
$\langle :\delta \hat{P}_r^2:\rangle=-\frac{p_{thr}}{2(2p+1-p_{thr}+j)}$. 
Figure 12(b) shows the skew variables and the mean amplitude for $p_{thr}=-0.5, g^2=0.02$, and $j=3$. 
The self-skewness $\langle \delta \hat{X}_r^3\rangle$ has the same sign as the mean amplitude, whereas 
the cross-skewness $\langle \delta \hat{X}_r\delta \hat{P}_r^2\rangle$ has the opposite sign from the mean amplitude, 
as we saw in Figure 3(b) and Figure 4(d). 
The peak absolute values are reached near $p-p_{thr}\sim 0.8$ and $0.6$, for the self- and cross-skewness, respectively. 

\begin{figure}
\begin{center}
\includegraphics[width=9.0cm]{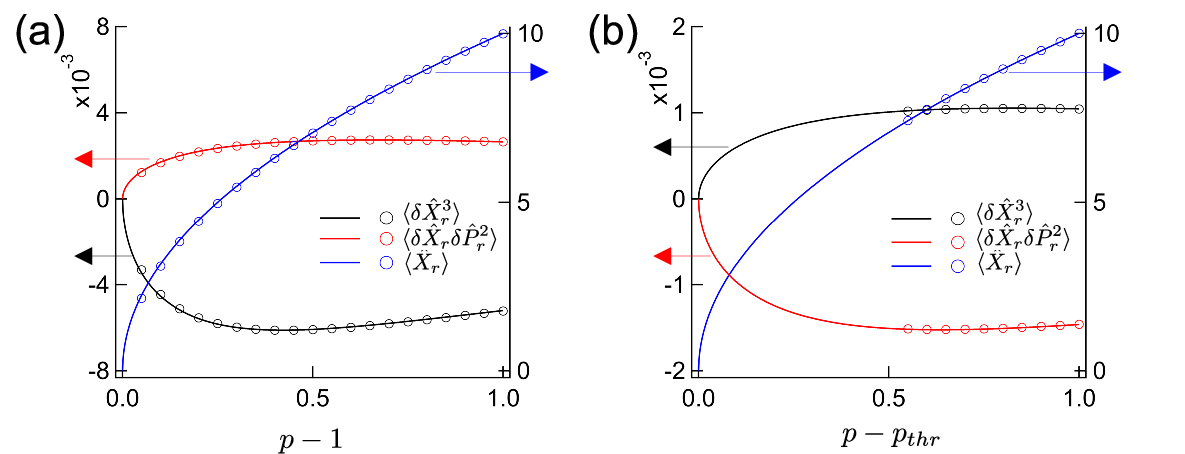}
\caption{Steady-state self-skewness $\langle \delta \hat{X}_r^3\rangle$ and cross-skewness $\langle \delta \hat{X}_r \delta \hat{P}_r^2\rangle$ 
in the mean-field coupled CIM above the threshold ($p>p_{thr}$). 
(a) The parametric-gain- ($p$-) dependent 
self-skewness $\langle \delta \hat{X}_r^3\rangle$, cross-skewness $\langle \delta \hat{X}_r \delta \hat{P}_r^2\rangle$ (the left axis), 
and mean amplitude $\langle \hat{X}_r\rangle$ (the right axis), 
under the mean field coupling $\varepsilon_r=j\mu_r (p_{thr}=1)$. 
Analytical results obtained by the skew-Gaussian model 
Eq.(\ref{sk_theory}), Eq.(\ref{sk_theory2}), and 
$\langle \hat{X}_r\rangle=\sqrt{\frac{2(p-p_{thr})}{g^2}}$ are shown as lines. 
The results from the self-consistent equations (Eqs.(\ref{scl}) and (\ref{scl2})) are shown as circles. 
(b) The same $p$-dependent characteristics with a large coherent injection 
$\varepsilon_r=(1.5+j)\mu_r$ where the linear loss and gain are in balance at $p_{thr}=-0.5$. }
\end{center}
\end{figure}

\setcounter{equation}{0}
\renewcommand{\theequation}{D\arabic{equation}}

\section{Discrete-component models}

\subsection{Discrete-component model of quantum master equation}

We present the discrete-component model (Figure 1(a)) using the density-matrix $\hat{\rho}$, 
which we call discrete-component quantum master equation (QME). 
Here, the density matrix $\hat{\rho}$ is a direct product of the density matrices representing pulses, 
$\hat{\rho}=\otimes_{r=1}^N \hat{\rho}^{(r)}$, where $\hat{\rho}^{(r)}$ is the density-matrix of the $r$-th pulse. 
The time-development of the signal pulse is described 
using the photon-number state expansion $\hat{\rho}^{(r)}=\sum_{N_rN_r'}\rho_{N_rN_r'}^{(r)}|N_r\rangle \langle N_r'|$. 
Here, the initial state of the first round trip has only one non-zero component, $\rho_{00}^{(r)}=1$. 
The density-matrix element $\rho_{N_rN_r'}^{(r)}$ obeys the master equation of the $\chi^{(2)}$ DOPO (Eq.(\ref{qmedopo})) \cite{Kinsler91}: 
\begin{widetext}
\begin{eqnarray}
\label{qmetd}
\frac{\partial \rho^{(r)}_{N_r,N_r'}}{\partial t}&=& 2\sqrt{(N_r+1)(N_r'+1)}\rho^{(r)}_{N_r+1,N_r'+1}-(N_r+N_r')\rho^{(r)}_{N_r,N_r'} \nonumber \\
&+& g^2 \sqrt{(N_r+1)(N_r+2)(N_r'+1)(N_r'+2)}\rho^{(r)}_{N_r+2,N_r'+2}-\frac{g^2}{2}[N_r(N_r-1)+N_r'(N_r'-1)]\rho^{(r)}_{N_r,N_r'} \nonumber \\
&+& \frac{p}{2}(\sqrt{N_r(N_r-1)}\rho^{(r)}_{N_r-2,N_r'}+\sqrt{N_r'(N_r'-1)}\rho^{(r)}_{N_r,N_r'-2}) \nonumber \\
&-&\frac{p}{2}(\sqrt{(N_r+1)(N_r+2)}\rho^{(r)}_{N_r+2,N_r'}+\sqrt{(N_r'+1)(N_r'+2)}\rho^{(r)}_{N_r,N_r'+2}). 
\end{eqnarray}
\end{widetext}
We denote the state after the time-development for a period $\Delta t$ at the DOPO, as $\hat{\rho}^{(r)'}$. 
To describe the effect of BS1, we introduce the two-mode density-matrix $\hat{\rho}_{AB}^{(r)}$ 
for the $r$-th internal pulse represented by $\hat{\rho}^{(r)'}$ and the input probe field $|\psi_{B,r}\rangle$, 
\begin{equation}
\hat{\rho}_{AB}^{(r)}=\hat{\rho}^{(r)'}\otimes |\psi_{B,r}\rangle \langle \psi_{B,r}|. 
\end{equation}
The input probe field of BS1 is prepared in a squeezed state and expressed by Trotterization with $N_{T0}$ divisions, 
$|\psi_{B,r}\rangle=e^{\frac{S}{2}(\hat{b}_{r}^{\dagger2}-\hat{b}_{r}^2)}|0\rangle=\Bigl[1+\frac{S}{2N_{T0}}(\hat{b}_{r}^{\dagger2}-\hat{b}_{r}^2)\Bigr]^{N_{T0}}|0\rangle$, 
where $S=-\ln \sqrt{G_j}$. 
In the two-mode description, BS1 mixes the internal signal state $\hat{a}_r$ with the input probe state $\hat{b}_r$ and 
yields the transmitted state and reflected state. 
In the numerical simulation, the effect of the mode mixing at BS1 is calculated by iterating the equation below $N_{T1}$ times: 
\begin{equation}
\label{discre1}
\hat{\rho}_{AB}^{(r)}\rightarrow \hat{\rho}^{(r)}_{AB}+\frac{\theta_1}{N_{T1}}(\hat{a}_{r}^{\dagger}\hat{b}_{r}-\hat{b}_{r}^{\dagger}\hat{a}_{r})\hat{\rho}^{(r)}_{AB}+\frac{\theta_1}{N_{T1}}\hat{\rho}^{(r)}_{AB}(\hat{b}_{r}^{\dagger}\hat{a}_{r}-\hat{a}_{r}^{\dagger}\hat{b}_{r}). 
\end{equation}
The characteristics of the reflected mode are obtained from $\hat{\rho}^{(r)}_B={\rm Tr}_A \hat{\rho}^{(r)}_{AB}$, 
which is the density-matrix after tracing out the transmitted mode. 
The measured values obtained by the homodyne detector are approximated to follow a skew-Gaussian distribution. 
We calculate the moments of the reflected field $\langle \hat{X}_{R,r}^n\rangle={\rm Tr}\hat{\rho}^{(r)}_B \hat{X}_{R,r}^n (n=1,2,3)$\cite{Braginsky95}, 
where $\hat{X}_{R,r}=\frac{\hat{b}_{r}+\hat{b}_{r}^{\dagger}}{\sqrt{2}}$,
and obtain the variance and self-skewness of the canonical coordinate as 
$V_{R,r}^X=\langle \hat{X}_{R,r}^2\rangle-\langle \hat{X}_{R,r}\rangle^2$, 
and $\mathfrak{S}_{R,r}=\langle \hat{X}_{R,r}^3\rangle -3\langle \hat{X}_{R,r}^2\rangle \langle \hat{X}_{R,r}\rangle+2\langle \hat{X}_{R,r}\rangle^3$. 
There have been attempts in quantum optics that have used the square root of a Gaussian random number 
to describe third-order fluctuations\cite{Plimak01}. 
Here, we generate the measured value $X_{B,r}$ with a normal random variable $\mathcal{N}_r$, as follows: 
\begin{equation}
\label{XRmeas}
X_{B,r}=\langle \hat{X}_{R,r}\rangle-\frac{\mathfrak{S}_{R,r}}{6V_{R,r}^X}+\sqrt{V_{R,r}^X+\frac{\mathfrak{S}_{R,r}}{3\sqrt{V_{R,r}^X}}\mathcal{N}_r}\mathcal{N}_r. 
\end{equation}
By using the measured value $X_{B,r}$, the transmitted state after the homodyne measurement is written as \cite{Braginsky95}, 
\begin{equation}
\hat{\rho}^{(r)''}=\frac{\langle X_{B,r}|\hat{\rho}^{(r)}_{AB}|X_{B,r}\rangle}{\langle X_{B,r}|\hat{\rho}_B^{(r)}|X_{B,r}\rangle}. 
\end{equation}
We calculate $\hat{\rho}^{(r)''}$ by using the photon-number state expansion, and $\langle N|X_{B,r}\rangle=H_{N}(X_{B,r})\frac{e^{-\frac{X_{B,r}^2}{2}}}{\pi^{1/4}\sqrt{2^{N} N!}}$. 
Here, $H_N(X)$ represents Hermite polynomials, $H_N(X)=(-1)^Ne^{X^2}\frac{d^N}{dX^N}e^{-X^2}$. 
In the FPGA feedback circuit, the inferred internal amplitude $\tilde{\mu}_r=X_{B,r}/\sqrt{2R_1}$ is used to calculate the feedback injection signal, 
$\zeta_r=\frac{j\Delta t}{\sqrt{R_2}}e_r\sum_{r'}\tilde{J}_{rr'}\tilde{\mu}_{r'}$, 
and to calculate the dynamics of the auxiliary variable $e_r$ (Eq.(\ref{tdeve})). 
Using the intensity modulator and phase modulator, the coherent amplitude of the feedback signal is set to $\zeta_{r}$. 
We again prepare a two-mode density-matrix $\hat{\rho}_{AZ}^{(r)}$, 
\begin{equation}
\hat{\rho}_{AZ}^{(r)}=\hat{\rho}^{(r)''}\otimes |\zeta_{r}\rangle \langle \zeta_{r}|, 
\end{equation}
for the internal mode $\hat{a}_{r}$ and the feedback signal $\hat{z}_{r}$. 
The mode mixing at BS2 is calculated by discretizing $\theta_2$ in Eq.(\ref{rhotf}) with $N_{T2}$ divisions, 
that is, iterating the following equation $N_{T2}$ times, 
\begin{equation}
\hat{\rho}_{AZ}^{(r)}\rightarrow \hat{\rho}^{(r)}_{AZ}+\frac{\theta_2}{N_{T2}}(\hat{a}_{r}^{\dagger}\hat{z}_{r}-\hat{z}_{r}^{\dagger}\hat{a}_{r})\hat{\rho}^{(r)}_{AZ}+\frac{\theta_2}{N_{T2}}\hat{\rho}^{(r)}_{AZ}(\hat{z}_{r}^{\dagger}\hat{a}_{r}-\hat{a}_{r}^{\dagger}\hat{z}_{r}).
\end{equation} 
The state after the round trip is obtained by tracing out the transmitted feedback signal, $\hat{\rho}^{(r)}={\rm Tr}_Z \hat{\rho}_{AZ}^{(r)}$. 
At the end of the round trip, the auxiliary variable $e_r$ is incremented in accordance with Eq.(\ref{tdeve}). 

Although we assume that the measured values follow a skew-Gaussian distribution, 
the QME simulation using the density-matrix is accurate in its description of the nonlinear saturation for large $g^2$ and measurement-induced state-reduction. 

\subsection{Discrete-component skew-Gaussian model}

The assumption of a skew-Gaussian distribution was introduced into QME 
to reduce the numerical cost of calculating the 
distribution $\langle X|\hat{\rho}^{(r)}_B|X\rangle$ for a sufficiently high resolution of $X$, 
which is used to determine the measured value $X_{B,r}$ in a more rigorous way. 
However, even without that the computational cost of the discrete-component QME 
is much higher than that of discrete-component Gaussian models\cite{Clements17,Ng22}, 
because of the photon-number state expansion and Trotterization of the beam splitter couplings. 

Here, we develop a discrete-component skew-Gaussian model of the CIM as 
a skew-Gaussian approximation of the QME. 
We introduce the self-skewness $\mathfrak{S}_r:=\langle \delta \hat{X}_r^3\rangle$ 
and the cross-skewness $\mathfrak{C}_r:=\langle \delta \hat{X}_r\delta \hat{P}_r^2\rangle$, 
in addition to the mean amplitude $\langle \hat{X}_r\rangle$, and 
variances $V_r^X=\langle \delta \hat{X}_r^2\rangle$, $V_r^P=\langle \delta \hat{P}_r^2\rangle$ that are used in the Gaussian model\cite{Ng22}. 
Figure 6 shows the variables we used in the discrete-component skew-Gaussian model. 
The first round trip starts from the vacuum state, 
$\langle \hat{X}_r\rangle=0$, $V_r^X=\frac{1}{2}$, $V_r^P=\frac{1}{2}$, $\mathfrak{S}_r=0$, and $\mathfrak{C}_r=0$. 
The time-development at the DOPO obeys the equations (\ref{sk1}-{\ref{sk5}), which are derived in Appendix B.1. 
Here, we convert variables $(\langle \hat{X}_r\rangle, V_r^X, V_r^P, \mathfrak{S}_r, \mathfrak{C}_r)$ 
into $(\mu_r, m_r, n_r, \gamma_r, \kappa_r)$ following 
$\mu_r=\frac{\langle \hat{X}_r\rangle}{\sqrt{2}}$, $m_r=\frac{V_r^X-V_r^P}{2}$, $n_r=\frac{V_r^X+V_r^P-1}{2}$, 
$\gamma_r=\frac{\mathfrak{S}_r-3\mathfrak{C}_r}{2\sqrt{2}}$, and $\kappa_r=\frac{\mathfrak{S}_r+\mathfrak{C}_r}{2\sqrt{2}}$. 
The state after the DOPO $\hat{\rho}^{(r)'}$, 
represented by $(\langle \hat{X}_r\rangle', V_r^{X'}, V_r^{P'}, \mathfrak{S}_r', \mathfrak{C}_r')$, 
is obtained by the inverse transform of variables $\langle \hat{X}_r\rangle'=\sqrt{2}\mu_r$, $V_r^{X'}=n_r+m_r+\frac{1}{2}$,$V_r^{P'}=n_r-m_r+\frac{1}{2}$, 
$\mathfrak{S}_r'=\frac{\gamma_r+3\kappa_r}{\sqrt{2}}$, $\mathfrak{C}_r'=\frac{-\gamma_r+\kappa_r}{\sqrt{2}}$. 
Here, $(\mu_r, m_r, n_r, \gamma_r, \kappa_r)$ are values after the time-development (Eqs.(\ref{sk1}-\ref{sk5})) for a period $\Delta t$. 

After the DOPO, the signal pulses are partially extracted at BS1 with a power reflection rate $R_1=j_1 \Delta t$, 
where the internal signals are mixed with the squeezed vacuum probe whose variances are $V_{B,r}^X=\frac{1}{2G_j}$, $V_{B,r}^P=\frac{G_j}{2}$. 
The transmitted mode has 
\begin{equation}
\langle \hat{X}_{T,r}\rangle=\sqrt{1-R_1}\langle \hat{X}_r\rangle', 
\end{equation}
\begin{equation}
V_{T,r}^a =(1-R_1) V_r^{a'} +R_1 V_{B,r}^a (a=X,P), 
\end{equation}
\begin{equation}
\mathfrak{S}_{T,r}= (\sqrt{1-R_1})^3\mathfrak{S}_r', 
\end{equation}
\begin{equation}
\mathfrak{C}_{T,r}= (\sqrt{1-R_1})^3\mathfrak{C}_r'.
\end{equation}
The moments of the canonical coordinate in the reflected mode follow 
\begin{equation}
\langle \hat{X}_{R,r}\rangle=\sqrt{R_1}\langle \hat{X}_r\rangle', 
\end{equation}
\begin{equation}
V_{R,r}^X=(1-R_1)V_{B,r}^X+R_1 V_{r}^{X'}, 
\end{equation}
\begin{equation}
\mathfrak{S}_{R,r}=(\sqrt{R_1})^3 \mathfrak{S}_r'. 
\end{equation}
Using Eq.(\ref{XRmeas}), we obtain the measured value $X_{B,r}$, and 
the inferred internal amplitude $\tilde{\mu}_{r}=X_{B,r}/\sqrt{2R_1}$ is calculated in an FPGA. 
The transmitted signals are modified by the measurement-induced state-reduction\cite{Wiseman93a,Eisert02,Adesso14}, 
\begin{equation}
\langle \hat{X}_{r}\rangle'' =\langle \hat{X}_{T,r}\rangle+\frac{V_{TR,r}^X}{\sqrt{V_{R,r}^X}}\mathcal{N}_r,
\end{equation}
\begin{equation}
\label{stred1}
V_{r}^{X''} =V_{T,r}^X-\frac{V_{TR,r}^{X2}}{V_{R,r}^X}+\sqrt{2j_1G_j \Delta t}\mathcal{N}_r\mathfrak{S}_r'. 
\end{equation}
Here, $\mathcal{N}_r$ is the same Gaussian random variable as was used to calculate $X_{B,r}$. 
$V_{TR,r}^{X}$ is the quantum correlation of the canonical coordinates between transmitted and reflected modes, 
\begin{equation}
V_{TR,r}^X:=\langle \delta \hat{X}_{T,r}\delta \hat{X}_{R,r}\rangle=\sqrt{R_1(1-R_1)}(V_r^{X'}-V_{B,r}^X).
\end{equation}
The last term in Eq.(\ref{stred1}) is a correction due to the self-skewness (Appendix B.2). 
The transmitted canonical momenta have the following correction due to the cross-skewness, 
\begin{equation}
\label{stred2}
V_{r}^{P''}= V_{T,r}^{P}+\sqrt{2j_1G_j \Delta t}\mathcal{N}_r\mathfrak{C}_r'. 
\end{equation}
The skew variables are affected by the homodyne measurement: 
\begin{equation}
\label{stred3}
\mathfrak{S}_{r}''=\mathfrak{S}_{T,r}-6j_1G_j \Delta t \mathfrak{S}_r'\Bigl(V_r^{X'}-\frac{1}{2G_j}\Bigr),
\end{equation}
\begin{equation}
\label{stred4}
\mathfrak{C}_{r}''= \mathfrak{C}_{T,r}-2j_1G_j \Delta t \mathfrak{C}_r'\Bigl(V_r^{X'}-\frac{1}{2G_j}\Bigr).
\end{equation}
The above descriptions of the measurement-induced state-reduction 
assume that the reflection rate $R_1(=j_1\Delta t)$, 
self-skewness $\mathfrak{S}_r$, and cross-skewness $\mathfrak{C}_r$ are small but non-negligible 
compared with the mean amplitude and variance, 
but that the products of these small values are negligible. 

The feedback signal is prepared in a coherent state with 
variances of $V_{Z,r}^a=\frac{1}{2}(a=X,P)$. 
The feedback mean amplitude is expressed as $\langle \hat{X}_{Z,r}\rangle=\sqrt{2}\zeta_{r}$, 
where $\zeta_{r}$ is calculated in an FPGA digital circuit. 
The feedback signal is injected into the cavity via BS2. 
The state after BS2 is described using a power reflection rate $R_2=j_2 \Delta t$: 
\begin{equation}
\label{XF_BS2}
\langle \hat{X}_r\rangle=\sqrt{1-R_2}\langle \hat{X}_{r}\rangle''+\sqrt{R_2}\langle \hat{X}_{Z,r}\rangle, 
\end{equation}
\begin{equation}
V^a_{r}=(1-R_2)V_{r}^{a''}+R_2 V_{Z,r}^a (a=X,P), 
\end{equation}
\begin{equation}
\mathfrak{S}_r= (\sqrt{1-R_2})^3\mathfrak{S}_{r}'', 
\end{equation}
\begin{equation}
\mathfrak{C}_r= (\sqrt{1-R_2})^3\mathfrak{C}_{r}''.
\end{equation} 
The auxiliary variable $e_r$ for the chaotic amplitude control is incremented in accordance with Eq.(\ref{tdeve}). 

\subsection{Comparison of time-developments}

We compared time-developments of discrete-component models for an $N=6$ CIM. 
The parameters were $\Delta t=3.2\times 10^{-3}, p=2, g^2=5, G_j=1.1, j=10, R=9, \beta=1, \tau_0=1$, and $e_r(t=0)=0.4$, 
and the instance was 
\begin{equation}
\label{jijN6}
\tilde{J}_{rr'}=
\begin{bmatrix}
0 & -0.5 & 0 & -0.6 & 0.2 & -0.1 \\
-0.5 & 0 & -0.4 & 0.4 & 0.2 & -0.2 \\
0 & -0.4 & 0 & 1 & 0.8 & -0.6 \\
-0.6 & 0.4 & 1 & 0 & -0.6 & -0.4 \\
0.2 & 0.2 & 0.8 & -0.6 & 0 & 0.5 \\
-0.1 & -0.2 & -0.6 & -0.4 & 0.5 & 0
\end{bmatrix}
.
\end{equation}
The ground state $(\uparrow,\downarrow,\downarrow,\downarrow,\uparrow,\uparrow)$ has energy $E=-3.5$, 
and the first excitation state $(\uparrow,\downarrow,\downarrow,\downarrow,\downarrow,\uparrow)$ has energy $E=-2.9$. 
The maximum photon number used in the QME was 35 for the $\hat{a}_r$ and $\hat{b}_r$ modes and 45 for the $\hat{z}_r$ mode. 
The sub-divisions of the time period $\Delta t$ in Eq.(\ref{qmetd}) were 64. 
The number of Trotterizations $N_{T0}$, $N_{T1}$ and $N_{T2}$ were $10^8$, 256 and 64 respectively. 
Figure 13(a-c) show the time-dependent mean canonical coordinate $\langle \hat{X}_r\rangle'$ 
obtained using the discrete-component skew-Gaussian, QME, and Gaussian models, respectively, for the 
same sequence of random numbers $\mathcal{N}_r$ representing measurements. 
At $t\sim 7$, the trajectories of amplitudes $\langle \hat{X}_r\rangle'$ obtained by QME model differed 
from the trajectories obtained by the other two, although they converged again at $t\sim 8$. 
On the other hand, the trajectories of the Gaussian model diverged from those of the skew-Gaussian and QME models at $t\sim 8.5$. 
Finally at $t=10$, the signs ${\rm sgn}(\langle \hat{X}_r\rangle')$ had one of the ground-state configurations, 
$(\uparrow,\downarrow,\downarrow,\downarrow,\uparrow,\uparrow)$, for the skew-Gaussian and QME models, 
and they had one of the first excitation state configurations, 
$(\downarrow,\uparrow,\uparrow,\uparrow,\uparrow,\downarrow)$, for the Gaussian model. 
Figures 13(d) and (e) show the self-skewness and cross-skewness for the $r=1$ mode, 
obtained by the skew-Gaussian and QME models. 
The self-skewness $\mathfrak{S}_1'$ and the cross-skewness $\mathfrak{C}_1'$ 
primarily had the same sign as and opposite sign from the mean amplitude $\langle \hat{X}_1\rangle'$, respectively. 
Figure 13(f) shows the Wigner function of $r=1$ mode obtained by the discrete-component QME model at $t=10$ (after the BS2), 
which shows a slightly $\hat{X}$-squeezed state with $\langle \hat{X}_1\rangle\sim 1.04$, $V_1^{X}\sim 0.465$, $V_1^P\sim 0.537$, 
$\mathfrak{S}_1\sim 0.0066$, and $\mathfrak{C}_1\sim -0.0080$. 

\begin{figure*}
\begin{center}
\includegraphics[width=15.0cm]{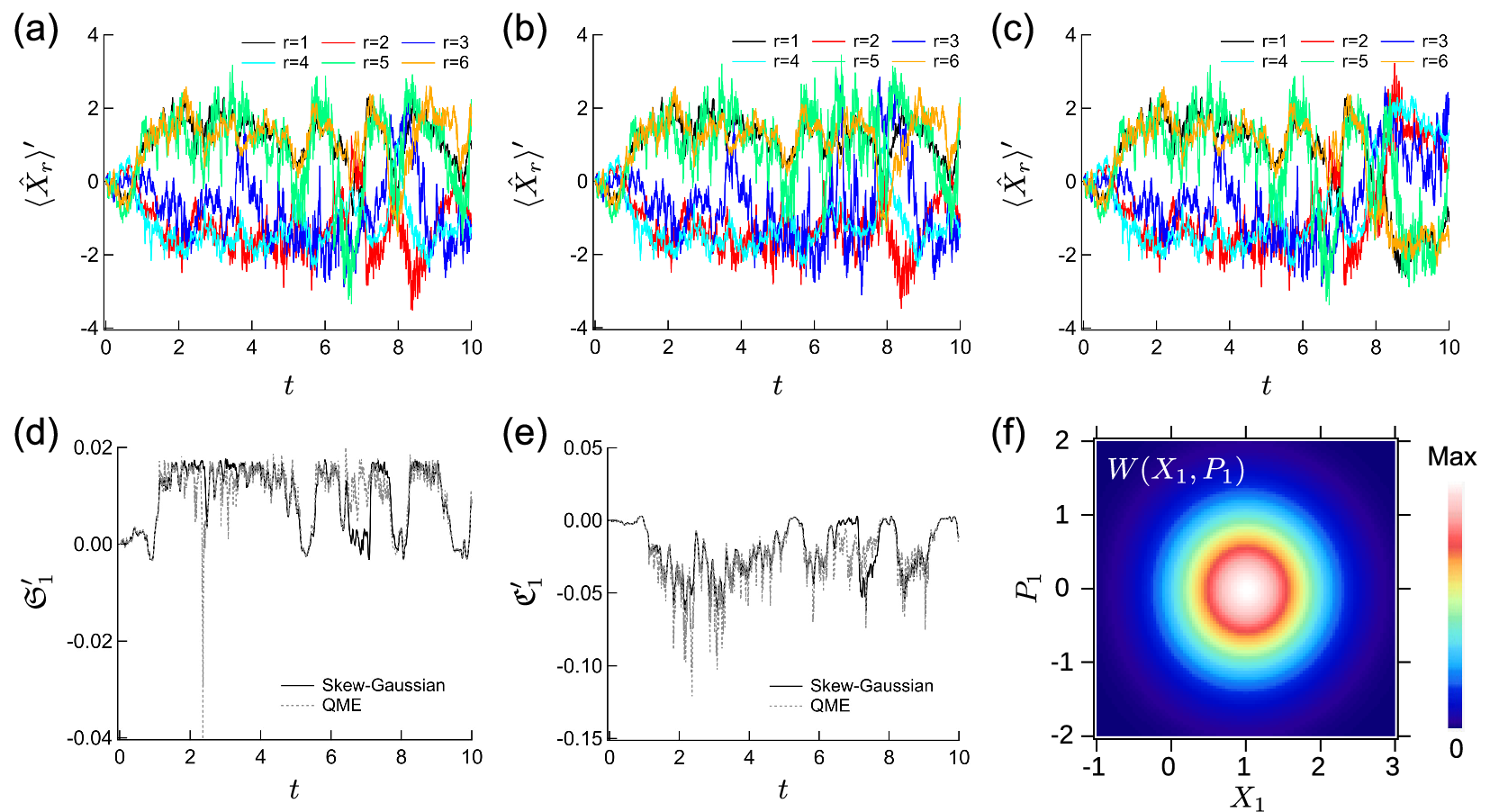}
\caption{Comparison of discrete-component models. 
(a)(b)(c) Time-development of mean canonical coordinate $\langle \hat{X}_r\rangle'$ 
obtained by the skew-Gaussian, QME, and Gaussian models respectively. 
(d) Time-dependent self-skewness $\mathfrak{S}_1'$ for skew-Gaussian and QME models. 
(e) Time-dependent cross-skewness $\mathfrak{C}_1'$ for skew-Gaussian and QME models. 
(f) Wigner function $W(X_1,P_1)$ at $t=10$ obtained by the QME model. }
\end{center}
\end{figure*}

\setcounter{equation}{0}
\renewcommand{\theequation}{E\arabic{equation}}

\section{Optimization and site-number-dependent characteristics of small-photon-number CIM}

Figure 14 shows information about the parameter optimization of small-photon-number CIM. 
It presents the cumulative success probability $P_s'(t)$ 
(the probability of finding the ground state at least once until time $t$) 
and minimum number of matrix-vector-multiplications 
needed to reach a solution $\min {\rm MVMTS}$ for $N=100$, $\alpha_{WP}=0.8$ Wishart planted instances. 
$P_s'$ and $\min_t {\rm MVMTS}(t)$ are defined in the same way as in Section VI. 
The three parameters $j$, $g^2$ and $\beta$ are varied in figures, 
and the other parameters are fixed to $\Delta t=3.2\times 10^{-3}, p=0, G_j=2, R=9, \tau_0=1.5$, and $e_r(t=0)=2$. 
The number of simulation runs was $10^5$. 
Figure 14(a) plots the $P_s'(t=10)$ and $\min_{t\le 10}{\rm MVMTS}(t)$ versus $j$ when $g^2=15$ and $\beta=0.3$. 
The success probability $P_s'$ peaks at $j\sim 100$. 
Figure 14(b) is a plot versus $g^2$ when $j=100$ and $\beta=0.3$, 
and Figure 14(c) is one versus $\beta$ when $g^2=15$ and $j=100$. 
The success probability $P_s'(t=10)$ peaks at $g^2\sim 13$ and $\beta\sim 0.25$, respectively. 

Figure 15 shows details on the $N$ dependence of small-photon-number CIM in Figure 10 for $j=100$, $g^2=1.5\sqrt{N}$ and $\beta=3000/N^2$. 
In particular, the plotted results are for the discrete-component skew-Gaussian model. 
Figures 15(a) and (b) show the time-dependent cumulative success probability $P_s'(t)$ and 
matrix-vector-multiplications to solution ${\rm MVMTS}(t)=\frac{t}{\Delta t}\frac{\ln(0.01)}{\ln(1-P_s'(t))}$. 
Figure 15(c) shows the two types of fitting, $e^{a+b\sqrt{N}}$ and $e^{a+bN}$, 
for the minimum number of matrix-vector-multiplications needed to reach a solution $\min_{t\le 10}{\rm MVMTS}(t)$. 
The numerically obtained $\min_{t\le 10}{\rm MVMTS}(t)$ was closer to the $e^{a+b\sqrt{N}}$ fitting \cite{Hamerly19,Mohseni22} than the $e^{a+bN}$ fitting. 
The values of $R^2$ for the fittings were 0.999, and 0.994, respectively. 

\begin{figure*}
\begin{center}
\includegraphics[width=15.0cm]{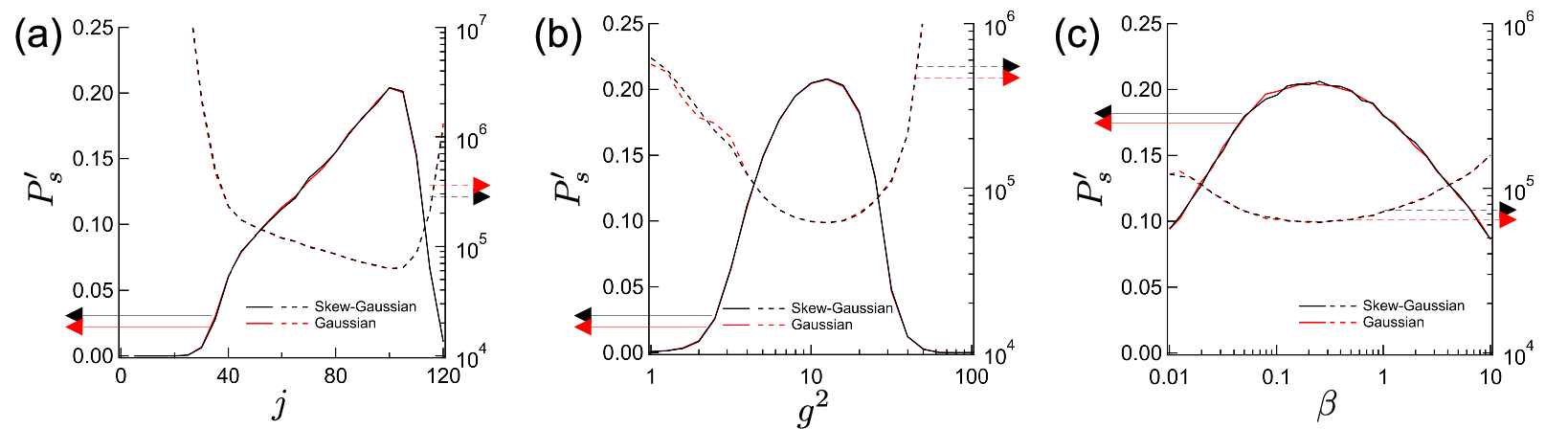}
\caption{Success probability $P_s'$ at $t=10$ (left axis) 
and minimum number of matrix-vector-multiplications needed to reach a solution $\min_{t\le 10} {\rm MVMTS}(t)$ (right axis) for small-photon-number $N=100$ CIM. 
(a) $j$ dependence ($g^2=15, \beta=0.3$). 
(b) $g^2$ dependence ($j=100, \beta=0.3$). 
(c) $\beta$ dependence ($g^2=15, j=100$). }
\end{center}
\end{figure*}

\begin{figure*}
\begin{center}
\includegraphics[width=15.0cm]{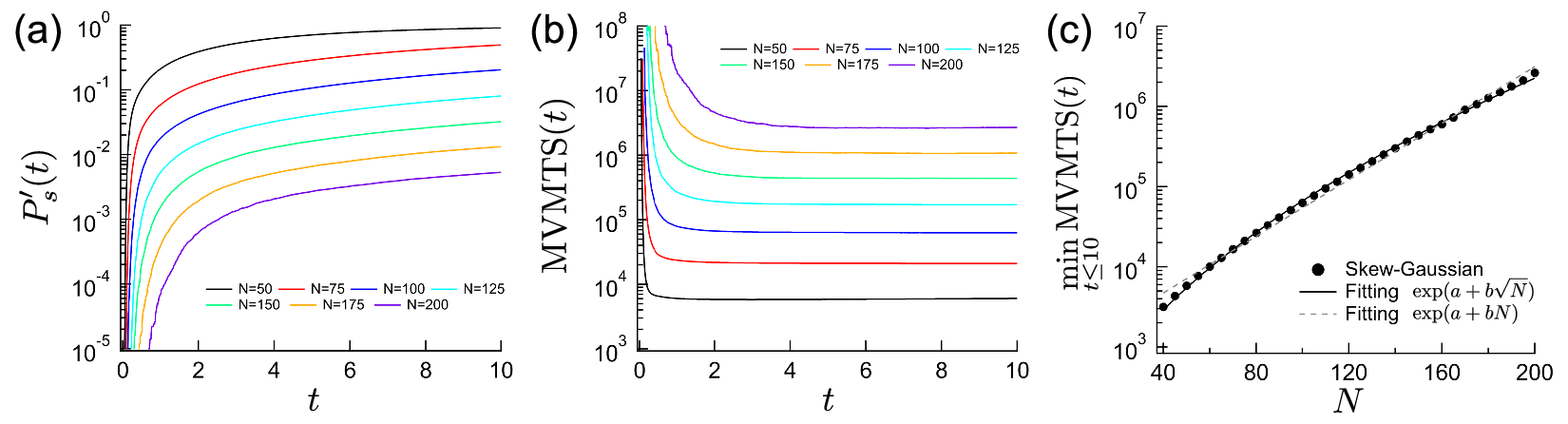}
\caption{Details of success probability $P_s'$ and $\min_t {\rm MVMTS}(t)$ of small-photon-number CIM, obtained by the skew-Gaussian model. 
(a) Time-dependent cumulative success probability $P_s'(t)$ for various site numbers $N$. 
(b) Time-dependent matrix-vector-multiplications to solution ${\rm MVMTS}(t)$ for various site numbers $N$. 
(c) Site-number-dependent $\min_{t\le 10} {\rm MVMTS}(t)$ and its $\exp(a+b\sqrt{N})$ and $\exp(a+bN)$ fittings. }
\end{center}
\end{figure*}

\setcounter{equation}{0}
\renewcommand{\theequation}{F\arabic{equation}}

\section{Large-photon-number CIM}

Here, we give the detailed information on the large-photon-number CIM model in Figures 10 and 11, 
which was compared with the small-photon-number CIM. 
The large-photon-number CIM was simulated using a discrete-component skew-Gaussian model (Appendix D.2), 
and spin was defined in the same way $\sigma_r={\rm sgn}(\tilde{\mu}_r)$ as in the small-photon-number CIM. 
However, it had a different equation of the auxiliary variable $e_r$ from Eq.(\ref{tdeve}) \cite{Kako20,Inui22}, 
\begin{equation}
\frac{de_r}{dt}=-\beta (g^2\tilde{\mu}_r^2-\tau)e_r. 
\end{equation}
Moreover, we did not use Eq.(\ref{tau0def}) to redefine the target value. 
Figures 16(a-c) show the time-development of an $N=6$ large-photon-number CIM 
for the instance shown in Eq.(\ref{jijN6}). 
The parameters are $\Delta t=3.2\times 10^{-3}, p=-40, g^2=10^{-3}, G_j=1, j=100, R=9, \beta=4, \tau=3$, and $e_r(t=0)=2$. 
In Figures 16(a) and (b), the mean amplitudes $\langle \hat{X}_r\rangle'$ and auxiliary variables $e_r$ continued oscillating after 
the transient period up to $t\sim 0.6$. 
The signs of the mean amplitude ${\rm sgn}(\langle \hat{X}_r\rangle')$ had one of the ground-state configurations, 
$(\downarrow,\uparrow,\uparrow,\uparrow,\downarrow,\downarrow)$, at $t=2$. 
Figure 16(c) plots the variances $V_1^{X'},V_1^{P'}$, self-skewness $\mathfrak{S}_1'$, and cross-skewness $\mathfrak{C}_1'$. 
The canonical coordinate $\hat{X}_1$ was squeezed ($V_1^{X'}<0.5$) due to the coherent injection, 
and the self-skewness $\mathfrak{S}_1'$ had the same sign (negative sign) as the mean amplitude $\langle \hat{X}_1\rangle'$. 
The absolute values of skew variables are much smaller than those of small-photon-number CIMs because of small $g^2$. 

Figures 16(d-f) show information about the parameter optimization of large-photon-number CIM. 
These plot the cumulative success probability $P_s'$ and the 
minimum number of matrix-vector-multiplications needed to reach a solution $\min_t {\rm MVMTS}(t)$, 
for $N=100$, $\alpha_{WP}=0.8$ Wishart planted instances. 
The $P_s'$ and $\min_t {\rm MVMTS}(t)$ are defined in the same way as were done for small-photon-number CIM in Section VI. 
The three parameters $j$, $g^2$, and $\beta$ are varied in figures, 
and the other parameters were fixed to $\Delta t=3.2\times 10^{-3}, p=-40, G_j=1, R=9, \tau=3$, and $e_r(t=0)=2$. 
The number of simulation runs was $10^5$. 
Figure 16(d) plots the $j$-dependent $P_s'(t=10)$ and $\min_{t\le 10}{\rm MVMTS}(t)$ for $g^2=10^{-3}$, and $\beta=0.08$; 
$P_s'(t=10)$ peaks around $j\sim 95$. 
Figure 16(e) shows the $g^2$ dependence for $j=100$, and $\beta=0.08$, 
where $P_s'$ is almost constant when $g^2$ is sufficiently smaller than 1. 
Figure 16(f) shows the $\beta$ dependence for $g^2=10^{-3}$, and $j=100$. 
$P_s'$ has a peak around $\beta\sim 0.08$. 
In the blue dashed line in Figure 10 representing the large-photon-number CIM, $\beta=800/N^2$ is the only parameter that depended on $N$. 
The other parameters used to plot Figure 10 were $\Delta t=3.2\times 10^{-3}, p=-40, g^2=10^{-3}, G_j=1, j=100, R=9, \tau=3$, and $e_r(t=0)=2$.  
The parameters in Figure 11 were the same where $\beta$ was fixed to $0.08$. 

\begin{figure*}
\begin{center}
\includegraphics[width=15.0cm]{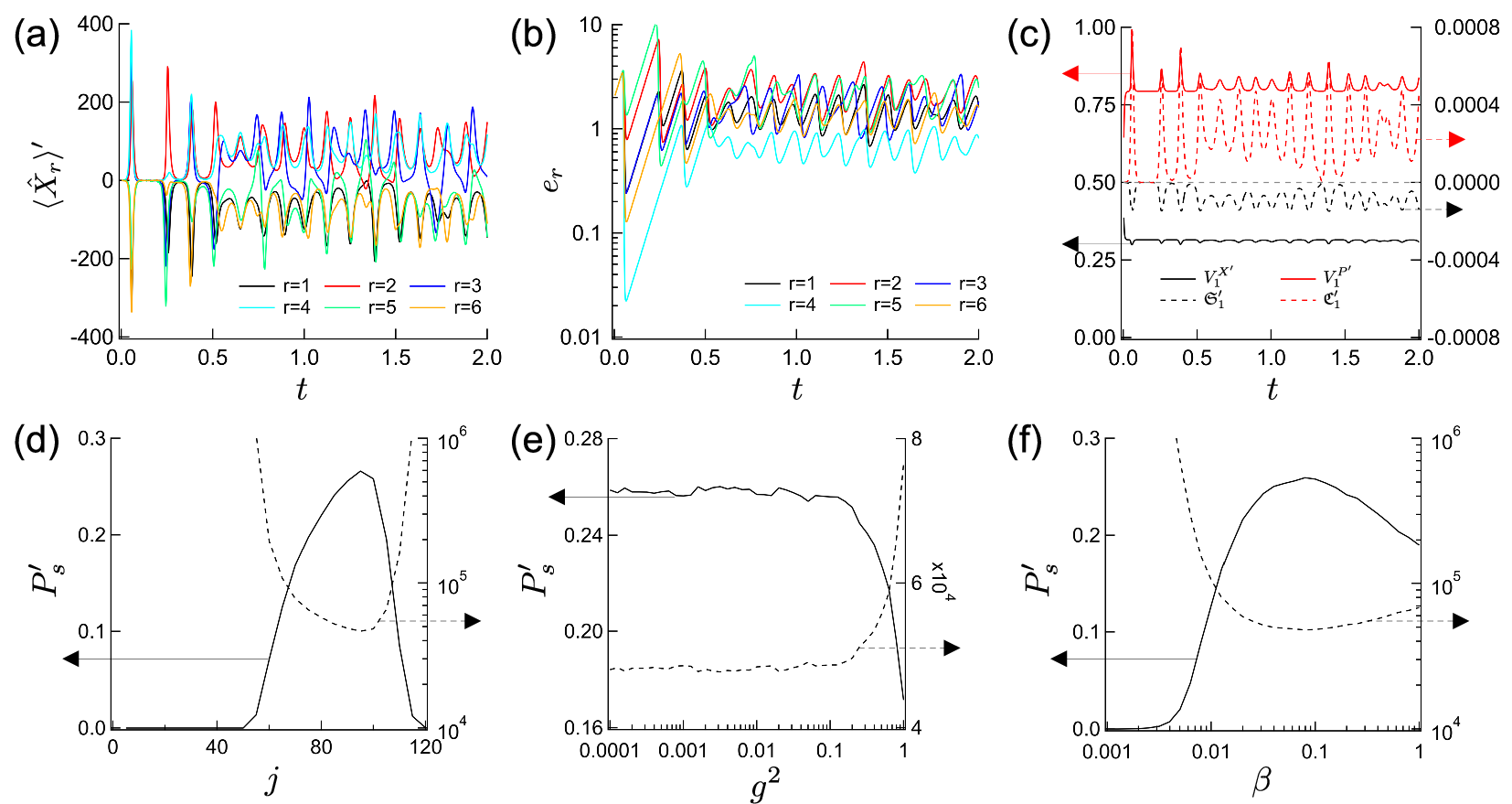}
\caption{Typical time-development and optimization of large-photon-number CIM. 
(a-c) Time development for an $N=6$ instance (Eq.(\ref{jijN6})). 
(a) Mean canonical coordinate $\langle \hat{X}_r\rangle'$. 
(b) Auxiliary variables $e_r$. 
(c) Variances $V_r^{X'}$, $V_r^{P'}$ (left axis), 
and skew variables $\mathfrak{S}_r'$, $\mathfrak{C}_r'$ (right axis). 
(d-f) Instance-averaged success probability $P_s'$ at $t=10$ (left axis), 
and minimum number of matrix-vector-multiplications needed to reach a solution $\min_{t\le 10}{\rm MVMTS}(t)$ (right axis), for $N=100$, $\alpha_{WP}=0.8$ Wishart planted instances. 
(d) $j$ dependence ($g^2=10^{-3}, \beta=0.08$). 
(e) $g^2$ dependence ($j=100, \beta=0.08$). 
(f) $\beta$ dependence ($g^2=10^{-3}, j=100$). }
\end{center}
\end{figure*}

\end{document}